%% file: pg_dust.tex
\documentclass{aastex61}
\usepackage{graphicx}
\usepackage{subfigure}
\usepackage{multirow}
\usepackage{comment}
\usepackage{natbib}
\usepackage{hyperref}
\usepackage{mathtools}
\usepackage{mathrsfs}
\usepackage{fontenc}
\usepackage{color}
\usepackage{url}
\usepackage{hyperref}
\usepackage{gensymb}
\usepackage{pifont}
\newcommand{\MH}{$\mathrm{H}_2$}
\newcommand{\hi}{H{\sevenrm\,I}}

\newcommand{\hbeta}{H{$\beta$}}
\newcommand{\halpha}{H{$\alpha$}}

\newcommand{\OIII}{[O{\sevenrm\,III}]}

\newcommand{\NII}{[N{\sevenrm\,II}]}

\newcommand{\mcl}{\multicolumn}
\newcommand{\ph}{\phantom}
\newcommand{\tnm}{\tablenotemark}
\newcommand{\cohii}{CO-to-$\mathrm{H}_2$}
\newcommand{\acounit}{$M_\odot \, \mathrm{(K\,km\,s^{-1}\,pc^2)}^{-1}$}
 
 \font\sevenrm=cmr7 scaled 1000

\def\spitzer{{\it Spitzer}}
\def\irs{{IRS}}
\def\mips{{MIPS}}
\def\herschel{{\it Herschel}}
\def\pacs{{PACS}}
\def\spire{{SPIRE}}

\def\wise{{\it WISE}}
\def\w{{\it W}}
\def\tmass{{2MASS}}

\def\hst{{\it HST}}
\def\galfit{{\tt GALFIT}}
\def\mc{{\tt emcee}}

\newcommand{\mzr}{$M_*$--$Z$}
\newcommand{\gzr}{$\delta_\mathrm{GDR}$--$Z$}
\newcommand{\gdr}{$\delta_\mathrm{GDR}$}

\submitjournal{ApJ}

\begin{document}


\title{On the Gas Content and Efficiency of AGN Feedback in Low-redshift Quasars}

\shorttitle{Gas and Feedback in Quasars}

\shortauthors{SHANGGUAN ET AL.}

\author{Jinyi Shangguan}
\affil{Kavli Institute for Astronomy and Astrophysics, Peking University,
Beijing 100871, China}
\affil{Department of Astronomy, School of Physics, Peking University,
Beijing 100871, China}

\author{Luis C. Ho}
\affil{Kavli Institute for Astronomy and Astrophysics, Peking University,
Beijing 100871, China}
\affil{Department of Astronomy, School of Physics, Peking University,
Beijing 100871, China}

\author{Yanxia Xie}
\affil{Kavli Institute for Astronomy and Astrophysics, Peking University,
Beijing 100871, China}

\begin{abstract}
The interstellar medium is crucial to understanding the physics of active galaxies 
and the coevolution between supermassive black holes and their host galaxies.  
However, direct gas measurements are limited by sensitivity and other uncertainties.  
Dust provides an efficient indirect probe of the total gas.  We apply this technique 
to a large sample of quasars, whose total gas content would be prohibitively 
expensive to measure.  We present a comprehensive study of the full (1 to 500 
\micron) infrared spectral energy distributions of 87 redshift $< 0.5$ quasars selected 
from the Palomar-Green sample, using photometric measurements from 2MASS, 
{\it WISE}, and {\it Herschel}, combined with {\it Spitzer} mid-infrared (5--40 \micron) 
spectra. With a newly developed Bayesian Markov Chain Monte Carlo fitting method, 
we decompose various overlapping contributions to the integrated spectral energy 
distribution, including starlight, warm dust from the torus, and cooler dust on galaxy 
scales.  This procedure yields a robust dust mass, which we use to infer the gas mass, 
using a gas-to-dust ratio constrained by the host galaxy stellar mass.  Most ($90\%$) 
quasar hosts have gas fractions similar to those of massive, star-forming galaxies, 
although a minority ($10\%$) seem genuinely gas-deficient, resembling present-day 
massive early-type galaxies.  This result indicates that ``quasar mode'' feedback does 
not occur or is ineffective in the host galaxies of low-redshift quasars.  We also find 
that quasars can boost the interstellar radiation field and heat dust on galactic scales.  
This cautions against the common practice of using the far-infrared luminosity to 
estimate the host galaxy star formation rate.
\end{abstract}


\section{Introduction}\label{sec:intro}

The tight correlation between the mass of supermassive black holes (BHs) and
the bulge properties of their host galaxies \citep{Magorrian1998AJ,
Ferrarese2000ApJ,Gebhardt2000ApJ} implicates a strong connection between
BH growth and galaxy evolution \citep{Kormendy2013ARAA,
Heckman2014ARAA}.  However, the physical mechanisms behind this apparent
BH--galaxy coevolution are still unclear.  Energy feedback from active galactic
nuclei (AGNs) is widely invoked to regulate galactic-scale star formation
\citep{Fabian2012ARAA}.  When accretion onto the BH reaches sufficiently high
levels, such that the AGN is powerful enough to be regarded as a quasar,
radiative or mechanical energy may drive a strong outflow that can blow the cold
gas out of the galaxy \citep{Silk1998AA}.  ``Quasar mode'' feedback may also
play a central role in the popular gas-rich, major merger-driven evolutionary
scenario for AGNs \citep{Sanders1988ApJ}, as they transform from an initially
dust-enshrouded stage to their final unobscured quasar stage
\citep{Hopkins2008ApJS}.  Many modern cosmological simulations frequently
invoke AGN feedback to effectively quench star formation in massive galaxies
(e.g., \citealt{Dubois2016MNRAS,Weinberger2017MNRAS}).

From an observational perspective, however, it is still elusive when, where, and
how AGNs influence their host galaxies.  Is AGN feedback actually as pervasive
as commonly assumed? Is it really as effective as we hope?  Does AGN feedback
suppress or, in fact, enhance star formation?

Recent studies offer a variety of promising, albeit ambiguous, clues.  AGN
outflows appear to be common at various redshifts \citep{Perna2015AA,
Woo2016ApJ,Nesvadba2017AA}, but their contribution to feedback is unclear
\citep{Woo2017ApJ}.  Spatially resolved optical spectroscopy show that AGN
winds may suppress star formation within the outflow, but they can also enhance
star formation along the edges of the flow (e.g., \citealt{Cresci2015ApJ,
Carniani2016AA}).  \cite{Maiolino2017Natur} argue that considerable star
formation can be driven by outflows, which may also affect the overall
morphology and kinematics of the galaxy.  Submillimeter observations find
strong outflows ($\gtrsim 100\,M_\odot$ yr$^{-1}$) in local ultraluminous
infrared (IR) galaxies and AGNs \citep{Cicone2014AA,Stone2016ApJ,
Alfonso2017ApJ}. However, the sample size is limited, and it is not clear
whether the gas in the end actually gets blown out of the galaxy.

Independent of the specific details of the physical processes involved, AGN
feedback, if it is effective enough to influence the host galaxy on large
scales, ought to leave an imprint on the global cold interstellar medium (ISM)
content of the system \citep{Ho2008ApJ}.   For example, in the merger-driven
scenario realized in hydrodynamical simulations (e.g.,
\citealt{Hopkins2006ApJS}), broad-line (type~1) AGNs  emerge in the aftermath
of dust/gas expulsion by energy feedback, toward the end of the merger
sequence.  In such a scenario, we expect the cold ISM content in type~1
AGNs---especially those powerful enough to be deemed quasars---to be
gas deficient relative to normal galaxies of similar mass.  Is this true?

This basic, robust prediction has been difficult to test in practice because
direct gas measurements are still lacking for large, well-defined samples of
AGNs, particular those of sufficient luminosity to expect feedback processes
to operate.  \cite{Ho2008ApJS} conducted the first systematic survey for \hi\
gas in a large sample of nearby broad-line AGNs using the Arecibo telescope.
Surprisingly, there is no evidence for gas deficit, casting doubt on the role
of AGN feedback in these systems \citep{Ho2008ApJ}.  The sample of Ho et al.,
however, restricted to very low redshifts ($z\lesssim0.1$) because of the
limitations of current \hi\ facilities, largely comprises relatively
low-luminosity AGNs (Seyfert~1 galaxies), hardly powerful enough to qualify as
bona fide quasars.  Observations of the CO molecule can probe molecular gas in
AGNs over a wide range of redshifts and luminosities, from relatively nearby lower
luminosity sources \citep{Scoville2003ApJ,Evans2006AJ,Bertram2007AA,
Husemann2017MNRAS} to powerful quasars out to $z\gtrsim6$ (e.g.,
\citealt{Walter2004ApJ,Wang2013ApJ,Cicone2014AA,Wang2016ApJ}).  However, CO
observations are still relatively time consuming, precluding studies of large,
statistically meaningful samples.  Moreover, even when detected, the
interpretation of the observations is still plagued by the uncertainty of the
\cohii\ conversion factor $\alpha_\mathrm{CO}$ \citep{Bolatto2013ARAA}.

An alternative, independent strategy to probe the gas content of galaxies is
to measure the dust mass, since these two constituents of the ISM are tightly
linked through the gas-to-dust ratio (\gdr).  This approach has been commonly
and effectively exploited in a variety of studies, especially with the advent
of the {\it Herschel Space Observatory} \citep{Pilbratt2010AA}, whose
unprecedented sensitivity and angular resolution have furnished a wealth of
far-IR (FIR) data for local and distant galaxies (e.g., \citealt{Leroy2011ApJ,
Dale2012ApJ,Eales2012ApJ,Berta2013AA,Berta2016AA}) and AGNs (e.g.,
\citealt{Leipski2014ApJ,Vito2014MNRAS,Podigachoski2015AA,Westhues2016AJ,
Shimizu2017MNRAS}).

This paper analyzes IR spectral energy distributions (SEDs) of a large sample
of bright, low-redshift quasars, using complete (1--500 \micron), high-quality
photometric measurements obtained from 2MASS, {\it WISE}, and {\it Herschel},
supplemented by mid-IR (MIR) spectroscopy over the wavelength range 5--40
\micron\ from the {\it Spitzer}\ Infrared Spectrometer (IRS).  The primary goal
of this paper is to derive robust total dust masses for the sample, with
well-understood uncertainties, carefully taking into account all known sources
of systematic effects.  To this end, we must decompose the IR SED into its three
main constituents: stellar emission, AGN-heated dust emission, and host galaxy
dust emission.  We use the widely applied (e.g., \citealt{Draine2007ApJ,
Magdis2012ApJ,Ciesla2014AA}) dust emission templates from
\citet[hereafter DL07]{DraineLi2007ApJ} to model the galactic dust emission.
One of the major uncertainties comes from the treatment of the AGN dust torus
emission, since it dominates the MIR and extends into the FIR
\citep{Nenkova2008ApJa,Honig2010AA,Honig2017ApJL,Siebenmorgen2015AA,
Xie2017ApJS}.  Our analysis takes full advantage of the important constraints
on the torus emission provided by the IRS spectra.  Many works have tried to 
decouple the galactic dust emission by decomposing the torus component from 
the observed IR SED.  However, none of the current widely used codes (e.g.,
{\tt DecompIR}, \citealt{Mullaney2011MNRAS}; {\tt BayeSED}, \citealt{Han2014ApJS};
{\tt CIGALE}, \citealt{Noll2009AA,Ciesla2015AA}; AGN{\it fitter},
\citealt{CalistroRivera2016ApJ};  but see \citealt{Sales2015ApJ} and
\citealt{HerreroIllana2017MNRAS}), properly fits spectroscopic data
simultaneously with photometric data.  Some study the spectra and the
photometric SED separately (e.g., \citealt{Marshall2007ApJ, Kirkpatrick2015ApJ}).
This approach, although practical, is not optimal, as it cannot provide a global,
self-consistent solution with properly constrained uncertainties.  We develop a
new Bayesian Markov Chain Monte Carlo (MCMC) method\footnote{We make the code
publicly available at \url{https://github.com/jyshangguan/Fitter}.} that
simultaneously incorporates photometric and spectral data in the fitting.  We
extensively evaluate a number of potential systematic uncertainties by comparing
various methods to fit the SED.

We find evidence that quasars can heat dust on galactic scales.  This implies
that star formation rates traditionally estimated from the FIR may be biased
by the AGN, even after accounting for the contribution from the torus emission.
We derive robust dust masses for the host galaxies and use them to estimate
the total mass of the cold gas.  We show that the widely adopted method (e.g.,
\citealt{Magdis2012ApJ,Santini2014AA,Berta2016AA}) of estimating \gdr\ from the
galaxy stellar mass, in combination with other well-established galaxy scaling
relations, provides reliable total gas masses within the main body of the galaxy
(i.e., $\lesssim R_{25}$).\footnote{$R_{25}$ is the isophotal radius of the 
galaxy at a surface brightness of 25 $B$ $\mathrm{mag\,arcsec}^{-2}$.}  We also
present an empirical formalism to estimate the global gas content of the galaxy.
We find that most quasar host galaxies have similar cold gas content to massive
star-forming galaxies, although a minority are as gas poor as quenched
elliptical galaxies.  We argue that ``quasar mode'' feedback does not operate
effectively in all quasar host galaxies.

The paper is organized as follows.  We introduce the quasar and galaxy samples
used in our study in Section \ref{sec:sample}. Section \ref{sec:data} describes
the data reduction and construction of the SEDs for the quasar sample.  Our
method to model the SEDs with a newly developed Bayesian MCMC fitting algorithm
is explained in Section \ref{sec:method}, and the results of our measurements
are presented in Section \ref{sec:result}.  Finally, in Section
\ref{sec:discuss} we evaluate different methods to measure dust masses and
discuss the implications of our results for AGN feedback.  This work adopts the
following parameters for a $\Lambda$CDM cosmology: $\Omega_m = 0.308$,
$\Omega_\Lambda = 0.692$, and $H_{0}=67.8$ km s$^{-1}$ Mpc$^{-1}$
\citep{Planck2016AA}.

\section{Quasar and Galaxy Samples}\label{sec:sample}

We study the lower redshift ($z<0.5$) subset of 87 bright, UV/optically selected
quasars from the Palomar-Green (PG) survey \citep{Schmidt1983ApJ}, as
summarized in \cite{Boroson1992ApJS}.  Although the PG quasar sample is not
complete because of large photometric errors and its simple color selection
criterion (e.g., \citealt{Goldschmidt1992MNRAS}), this representative sample of
bright, nearby quasars has been extensively studied for decades, allowing us to
take advantage of a wealth of archival and literature multiwavelength data.
As a major motivation of this study is to try to quantify, in as much detail
as practical, various sources of systematic uncertainties in the derived
dust properties, the availability of high-quality data across the entire IR
(1 to 500 \micron) region is crucial.  The PG sample has the best and most
complete set of IR observations for quasars or AGNs to date, encompassing not
only six bands of \herschel\ photometry but also \spitzer\ IRS spectroscopy,
and, of course, the full complement of shorter-wavelength measurements from
the all-sky surveys of 2MASS and \wise\ (Section \ref{sec:data}).

Equally importantly, the PG sample has available a rich repository of additional
ancillary data from which critical physical properties of the central engine and
host galaxy can be derived, including BH masses and Eddington ratios (optical
spectra: \citealt{Boroson1992ApJS,Ho2009ApJS}), accretion disk (X-ray spectra:
\citealt{Reeves2000MNRAS,Bianchi2009AA}), jets (radio continuum:
\citealt{Kellermann1989AJ,Kellermann1994AJ}), and host galaxy stellar
morphology [{\it Hubble Space Telescope (\hst)}\ images: \citealt{Kim2008ApJ,
Kim2017ApJS}].

The physical properties of PG quasars are summarized in Table \ref{tab:result}.
Apart from properties related to the dust and ISM of the hosts, we also
include information on the optical AGN luminosity, broad H$\beta$ line width,
BH mass, and stellar mass of the host galaxies.  Direct estimates of total
stellar mass ($M_*$) are available for 55 objects for which \cite{Zhang2016ApJ}
were able to analyze high-resolution optical and near-IR (NIR) images.  For
the remaining 32 objects that do not have direct estimates of stellar masses,
we provide an indirect estimate of the lower limit for the total stellar mass
from the bulge mass ($M_{\rm bulge}$), adopting the tight
$M_\mathrm{BH}$--$M_\mathrm{bulge}$ relation of local inactive galaxies
(\citealt{Kormendy2013ARAA}; Equation (10))\footnote{\cite{Kormendy2013ARAA}
calculate the bulge mass based on the $K$-band mass-to-light ratio 
($M/L_K$) constrained by the optical color ($B-V$).  They use the $M/L_K$--color
relation from \cite{Into2013MNRAS} but modify its intercept according to 
dynamical measurements.  Therefore, our bulge mass obtained from the 
$M_\mathrm{BH}$--$M_\mathrm{bulge}$ relation should be close to that based 
on Kroupa-like initial mass functions (IMFs), such as \cite{Kroupa1998ASPC,Kroupa2001MNRAS}
and \cite{Kroupa1993MNRAS}, that are relevant to our work.  Since the \cite{Chabrier2003PASP} 
and Kroupa-like IMFs will only introduce very little difference ($\lesssim 10\%$) to the stellar 
mass \citep{Madau2014ARAA}, we do not differentiate between the two kinds of IMFs 
throughout the paper.}.  We apply the recent calibration of \citet[Equation (4)]{Ho2015ApJ} 
to calculate single-epoch virial BH masses ($M_\mathrm{BH}$) using the 5100 \AA\ monochromatic 
luminosity [$\lambda L_\lambda$(5100 \AA)], adjusted to our cosmology, and the full width
at half maximum (FWHM) of the broad H$\beta$ emission line (FWHM$_\mathrm{H\beta}$),
as listed in \citet[Tables 1 and 7]{Vestergaard2006ApJ}\footnote{Their values of 
FWHM$_\mathrm{H\beta}$ for PG 0923$+$129 and PG 0923$+$201 appear to have been 
interchanged by mistake; the correct values are listed in Table \ref{tab:result}.}.

An integral part of our analysis will compare the ISM properties of PG quasars
with those of local inactive galaxies (Section \ref{ssec:ism}).  We choose
three samples of inactive galaxies.

\begin{enumerate}
\item{
KINGFISH \citep{Kennicutt2011PASP} consists of 61 representative local
star-forming galaxies, with stellar masses measured using optical-to-NIR
color and $H$-band luminosity \citep{Skibba2011ApJ}, assuming 
a \cite{Kroupa2001MNRAS} stellar IMF.  The IR SEDs of the galaxies
have been studied by \cite{Draine2007ApJ} and, more recently, \cite{Dale2012ApJ,
Dale2017ApJ}, using the DL07 model.  The dust properties for most of the
galaxies are reported in \cite{Draine2007ApJ}, which we adopt.
}

\item{
The Herschel Reference Survey (HRS; \citealt{Boselli2010PASP}) comprises 322
$K$-band selected galaxies within a distance of $D_{L} \approx$ 15--25 Mpc.
The stellar masses were determined from the $i$-band luminosity with $g-i$
color-dependent stellar mass-to-light ratio from \cite{Zibetti2009MNRAS}, assuming 
the \cite{Chabrier2003PASP} stellar IMF.  The ISM properties of HRS galaxies 
have been extensively studied \citep{Cortese2012AA,Cortese2016MNRAS,Boselli2014AA2,
Ciesla2014AA}.  \cite{Ciesla2014AA} measured dust properties by fitting DL07 
models to the 8--500 \micron\ SED using {\tt CIGALE}.  \cite{Boselli2014AA1} reported 
\hi\ measurements, mainly from the Arecibo ALFALFA survey, and various CO(1--0) 
observations whereby the CO line fluxes were corrected according to the galaxy optical 
size.  We adopt the molecular gas masses converted with a luminosity-dependent 
$\alpha_\mathrm{CO}$ conversion factor, considering that the stellar masses of the 
HRS galaxies span a wide range and the conversion factor varies with the gas-phase 
metallicity (and hence stellar mass; \citealt{Boselli2002AA}), although using a constant 
conversion factor only affects the molecular gas masses by, on average, $<0.1$ dex and 
makes essentially no difference in our results.
}

\item{
The COLD GASS \citep{Saintonge2011MNRAS} sample includes 366 nearby
($D_{L} \approx$ 100--200 Mpc) massive ($M_* > 10^{10}\,M_\odot$;
\citealt{Saintonge2012ApJ}) galaxies.  The stellar masses come from SED fitting
using photometric data from the Sloan Digital Sky Survey (SDSS;
\citealt{Stoughton2002AJ}) assuming the \cite{Chabrier2003PASP} stellar IMF.
\hi\ gas masses come from Arecibo data, and molecular gas masses
were converted from CO(1--0) line luminosities measured using the IRAM 30~m
telescope, assuming $\alpha_\mathrm{CO}=4.35$ \acounit.  The gas masses for the
COLD GASS and HRS samples account for elements heavier than hydrogen.
}
\end{enumerate}

\input{tab1.tex}

\section{Data Analysis and Compilation}\label{sec:data}

\subsection{2MASS and WISE}
The \tmass\ \citep{Skrutskie2006AJ} $J$ (1.235 $\mu$m), $H$ (1.662 $\mu$m), and
$K_s$ (2.159 $\mu$m) bands \citep{Cohen2003AJ} are dominated by emission from
the old stellar population of the host galaxy.  Since the quasar host galaxies may be 
resolved, the measurements from the \tmass\ Point Source Catalog are not accurate.  
At the same time, only a small fraction of the PG quasars are included in the \tmass\ 
Extended Source Catalog.  Therefore, we reanalyze the \tmass\ data for the entire 
sample.  We collect the \tmass\ images from the NASA/IPAC Infrared Science Archive 
(IRSA)\footnote{\url{irsa.ipac.caltech.edu/frontpage/}} by matching each source
with a search radius of $4\arcsec$ with respect to the optical position of the quasar 
and performing aperture photometry using the Python
package {\tt photutils}\footnote{\url{http://photutils.readthedocs.io/en/stable/}}.
To measure the integrated flux, we use the default aperture radius of 7\arcsec\
\citep{Jarrett2003AJ} with the sky annulus set to a radius of 25\arcsec\ to
35\arcsec.  For the nearest ($z\lesssim0.1$) quasars having more extended host
galaxies, we use a larger aperture radius of $20\arcsec$ but the same sky
annulus.  To determine the uncertainty, we perform 500 random aperture
measurements of the sky, in exactly the same way as the quasar, with all
sources masked, and use the standard deviation of the spatial variation of the
sky to be the uncertainty of our measurement.  We do not apply any aperture
correction, which is found to be very
small\footnote{\url{www.astro.caltech.edu/\~jmc/2mass/v3/images/}}.  The
apertures of five targets (PG 0921$+$525, PG 1115$+$407, PG 1216$+$069,
PG 1534$+$580, and PG 1612+261) are affected by projected close companions.
As all the companions are $\gtrsim 4\arcsec$ away from the quasars, we first
use \galfit\ \citep{Peng2002AJ,Peng2010AJ} to fit and remove them from the
images.  The point-spread function (PSF) of each image is derived from the stars
in the field using {\tt DAOPHOT} in IRAF\footnote{IRAF is distributed by the
National Optical Astronomy Observatories, which are operated by the Association
of Universities for Research in Astronomy, Inc., under cooperative agreement
with the National Science Foundation.} \citep{Tody1986SPIE}.  The residual
images are measured using the same method described above.  For PG 1216$+$069,
its companion is a very bright foreground star, and hence its \galfit\ residual
image suffers from exceptionally large uncertainty.

In order to obtain accurate measurements that avoid the influence of projected
companions, we also decide to perform our own aperture photometry on the \wise\
images.  We similarly collect \wise\ \citep{Wright2010AJ,Jarrett2011ApJ} \w1
(3.353 \micron), \w2 (4.603 \micron), \w3 (11.561 \micron), and \w4 (22.088
\micron) data of the PG sample from IRSA.  As the effective wavelengths of the
\w3 and \w4 bands overlap with the bandpass of the \spitzer\ IRS spectra, we
use them to check for possible systematic zeropoint offsets between these two
data sets (Appendix \ref{apd:sys}).  We choose not to include these two \wise\
bands in the final SED fitting, because they are known to suffer from systematic
(though correctable) uncertainties due to the red color of the targets (Appendix
\ref{apd:sys}).  Our method to measure the \wise\ data is similar to that used for
\tmass\ data.  We adopt ``standard'' aperture radii \citep{Cutri2012wise},
8\farcs25 for the \w1, \w2, and \w3 bands, and 16\farcs5 for the \w4 band, along
with a sky annulus of 50\arcsec--70\arcsec.  We use coadded PSFs
\citep{Cutri2012wise} of the four \wise\ bands to calculate the aperture
correction factors from the PSF curves of growth.  The uncertainty is also
estimated by making 500 random measurements throughout the sky region.  Visual
examination shows that the source apertures of seven objects (PG 1048$-$090,
PG 1103$-$006, PG 1119$+$120, PG 1216$+$069, PG 1448$+$273, PG 1612$+$261, and
PG 1626$+$554) are contaminated by projected companions.  Due to the differences
in wavelength and resolution, the projected companions in \wise\ images are not
necessarily the same as those in the \tmass\ images.  As with the \tmass\ images,
we use \galfit\ to subtract the companions and then perform aperture photometry
on the residual images.  The \tmass\ and \wise\ measurements are listed in Table
\ref{tab:nir}.  The 3\% calibration uncertainties for both \tmass\
\citep{Jarrett2003AJ} and \wise\ \citep{Jarrett2011ApJ} are not included.  The
objects with companions are marked; we note that our main statistical results
are not affected by whether or not we include these objects.

\input{tab2.tex}

\subsection{Spitzer}
The entire sample of $z < 0.5$ PG quasars has been uniformly observed by
\spitzer\ \irs.  We utilize the data as processed by \cite{Shi2014ApJS}, who
scaled the short-low ($\sim$ 5--14 \micron) spectra to match the long-low
($\sim$ 14--40 \micron) spectra, and the overall flux of the spectra was scaled
to match the \mips\ 24 \micron\ photometry.  The flux scale of the spectra is
also well-matched to the \wise\ data (Appendix \ref{apd:sys}), and thus no
further normalization is applied to the \spitzer\ data.  PG 0003$+$199 only has
short-low spectra, and we supplement it with a high-resolution spectrum
($\sim$ 10--37 \micron; AORKey=25814528) from the CASSIS database
\citep{Lebouteiller2015ApJS}.  The high-resolution spectrum of PG 0003$+$199
is resampled to match the low-resolution spectra, binning the spectrum by
taking the median value of the wavelength and flux density for every 10 points.
The uncertainty is the median uncertainty in each bin divided by $\sqrt{10}$.
The spectra are combined by scaling the short-low spectrum to the
high-resolution spectrum at 13 \micron.  We do not scale the combined spectrum
further because there is no reference \spitzer\ photometric observation of this
source, and the spectrum already seems to match the photometric data reasonably
well. However, we caution that the SED of PG 0003$+$199 may suffer larger
systematic uncertainties than the rest of the targets.

\subsection{Herschel}

We observed nearly the entire PG sample with the Photodetector Array Camera
and Spectrometer (\pacs; \citealt{Poglitsch2010AA}) and the Spectral and
Photometric Imaging Receiver (\spire; \citealt{Griffin2010AA}) instruments
on board \herschel\ (program OT1\_lho\_1; PI: L. Ho).  PG 1351$+$640 was
observed only with \pacs\ in our observation.  A few targets were excluded from
our program because they had already been observed by other programs.  We
retrieved these data from the {\it Herschel Science Archive} (HSA).
PG 1226$+$023 was observed only with \spire\ (PI: D. Farrah).  PG 1426$+$015 is
located in one of the fields of the \herschel Thousand Degree
Survey\footnote{http://www.h-atlas.org/} (PI: S. Eales), and we use the \spire\
data from that project.  No \herschel\ observations exist for PG 1444$+$407.
Thus, in total, 86 out of the 87 PG quasars have \herschel\ observations, with 84
having both \pacs\ and \spire\ data.

We quote monochromatic flux densities at 70, 100, and 160 \micron\ for \pacs,
and at 250, 350, and 500 \micron\ for \spire\ (Table \ref{tab:herschel}).  The
objects possibly affected by confusion from close companions are marked in
Table \ref{tab:herschel}; they likely have larger uncertainties.  Our results,
however, are not affected by whether or not these objects are included in the
analysis.  The  standard pipeline assumes a spectral shape $\nu f_\nu \sim$ 
constant.  We provide $3\,\sigma$ upper limits for non-detections.  The calibration
uncertainties for \pacs\ and \spire\ photometry are both $5\%$, which are not
included in the uncertainties quoted in Table \ref{tab:herschel}.  We do not
apply a color correction but do consider the instrument spectral response
functions in the SED modeling.  As documented in Appendix \ref{apd:sys}, our
\pacs\ 70 and 160 \micron\ measurements are generally consistent with \spitzer\
\mips\ measurements.  The \herschel\ data for the PG sample were analyzed
independently by \cite{Petric2015ApJS}; we compare our measurements with theirs
in Appendix \ref{apd:herdata}.

\subsubsection{PACS}

The PACS observations were conducted in mini-scan mode with scan angles
70\degree\ and 110\degree\ at a scanning speed of $20\arcsec \, \mathrm{s}^{-1}$.
PACS simultaneously scans each source in two bands, 70 \micron\ or 100 \micron\
and 160 \micron, over a field of view of $1\farcm75 \times 3\farcm5$.  The
integration time for each scan angle was 180 s.

The data were processed within the Herschel Interactive Processing Environment
(HIPE; \citealt{Ott2010ASPC}) version 14.1.0 (calibration tree version 72).
We use the standard HIPE script for point-source photometry to reduce the
{\tt level1} data into science images.  We first generated a mask based on
signal-to-noise ratio.  All pixels above the $3\,\sigma$ threshold are masked.
Then, a circular mask with $\mathtt{radius}=25\arcsec$ is added at the nominal
position of the target.  The scan maps with different scan directions are
drizzle-combined with the {\tt photProject} function, using the default pixel
fraction ($\mathtt{pixfrac}=1.0$) and reduced output pixel sizes of
$1\farcs1$, $1\farcs4$, and $2\farcs1$ for the 70, 100, and 160 \micron\ bands,
respectively.  A smaller pixel fraction can, in principle, reduce the covariant
noise, but we find that the noise does not significantly change when we set
$\mathtt{pixfrac}=0.6$.  The above-described key parameters follow those used by
\citet[Section 4.1]{Balog2014ExA}.

We perform point-source aperture photometry using aperture sizes and annular
radii for background subtraction as recommended by Paladini's \herschel\ Webinar
``Photometry Guidelines for PACS Data''\footnote{\url{https://nhscsci.ipac.caltech.edu/workshop/Workshop\_Oct2014/Photometry/PACS/PACS\_phot\_Oct2014\_photometry.pdf}}.
The aperture radii for bright sources are $12\arcsec$, $12\arcsec$, and
$22\arcsec$ for the 70, 100, and 160 \micron\ bands, respectively, whereas for
faint sources they are $5\farcs5$, $5\farcs6$, and $10\farcs5$. For concreteness,
we set the division between bright and faint sources as 200 mJy at 100 \micron,
although in practice we find little difference between the flux densities
measured with the large and small apertures for objects with 100 \micron\ flux
densities of $\sim$ 150--200 mJy.  We measure the curves of growth and the
variation of the aperture-corrected fluxes to study the effect of aperture size.
We find that the aperture radius we are using is large enough to measure
accurately even the partially resolved targets with $z<0.05$, at the same time
being small enough to avoid contaminating sources and minimize the noise.

The sky annulus covers the radial range 35\arcsec--45\arcsec, out to which the
sky measurements are affected by the PSF wings by less than $0.1\%$
\citep{Balog2014ExA}.  Aperture correction is always necessary because the
\herschel\ PSFs are very extended (see Table 2 of \citealt{Balog2014ExA}).  For
PG 0923$+$129, whose host galaxy is very extended, we use an aperture radius of
18\arcsec, 18\arcsec, and 30\arcsec\ for the 70, 100, and 160 $\mu$m bands,
respectively.  Some objects with close companions require the companions to be
subtracted first before performing aperture photometry (see below).

To determine the uncertainties of the flux densities, we perform 20 measurements
on the image without background subtraction, centered evenly on the background
annulus (with radius $45\arcsec$).  The aperture sizes are exactly the same as
those used to measure the sources.  We take the standard deviation of the 20
measurements as the $1\,\sigma$ uncertainty of the aperture photometry of the
source \citep{Balog2014ExA}.  The median uncertainties of the 70, 100, and 160
\micron\ bands are 2.96, 3.80, and 11.27 mJy, respectively, for the entire
sample.  Measured flux densities $<3\,\sigma$ are quoted as $3\,\sigma$ upper
limits.  The method of \cite{Leipski2014ApJ} to estimate the uncertainty by
randomly sampling the sky is not applicable here, because in our images, the
region with good exposure coverage ($>75\%$) is too small compared with the
aperture size.

Five objects (PG 0043$+$039, PG 0947$+$396, PG 1048$+$342, PG 1114$+$445, 
and PG 1322$+$659) show close companions that are bright and close enough to 
affect the aperture photometry.  These companions need to be removed prior to 
measuring the source.  In order to generate the PSF, we use observations of $\alpha$ 
Tau (obsid: 1342183538 and 1342183541; \citealt{Balog2014ExA}), reprocessed with 
the same parameters as the PG quasars.  \galfit\ is used to simultaneously fit the
sources and the companions. Visual inspection of the residual images shows that
the companions are very well removed.  Therefore, we perform the aperture
photometry for the targets on the residual images with their companions removed,
using a small aperture size.  The companions of PG 0043$+$039 and PG 0947$+$396
are exceptionally heavily blended in the 160 \micron\ band.  After the
companions are subtracted, PG 0043$+$039 cannot be measured above the
$3\,\sigma$ level.  PG 0947$+$396 can still be measured, but the flux
uncertainty may be larger than the nominal sky error.  Six objects
have faint companions. For all but PG 0844+349, the companions affect 
the measurements by at most 10\%.  We decide not to remove them because the
uncertainties induced by \galfit\ fitting may be even larger, and, for some
companions without optical counterparts, we are not sure whether they actually
belong to the host galaxies or not.  PG 0844+349 is in a merger system and the
ISM of the two galaxies are likely highly disturbed (e.g., \citealt{Kim2017ApJS}),
so our standard small aperture is good to avoid the contamination from the
companion.  However, removing the extended companion galaxy will lead to a 
much larger uncertainty than the usual compact source, and so we decide to 
keep our standard measurements.  The uncertainties of this object are likely
$\lesssim 25\%$ for the three \pacs\ bands.

\subsubsection{SPIRE}

The \spire\ imaging photometer covers a field of view of
$4^{\prime} \times 8^{\prime}$ with an FWHM resolution of $18\farcs1$,
$25\farcs2$, and $36\farcs6$ for the 250, 350, and 500 \micron\ bands,
respectively (\citealt{Griffin2010AA}).  The observations were conducted in the
small-scan-map mode, with a single repetition scan for each object and a total
on-source integration time of 37 s.

The data reduction was performed using HIPE (version 14.1.0; calibration tree
{\tt spire\_cal\_14\_3}) following standard procedures, using a script
dedicated for small maps provided by HIPE. Although our sample contains a
number of bright objects, many of our sources are faint ($< 30$ mJy), and even
undetectable.  Following the suggested strategy for photometry for SPIRE, we
choose the HIPE built-in source extractor {\tt sourceExtractorSussextractor}
\citep{Savage2007ApJ} to measure the locations and fluxes of the sources, with
the error map generated from the pipeline and adopting a $3\,\sigma$ threshold
for the detection limit.  We measure the source within the FWHM of the beam
around the nominal position of the quasar.

Among the sources found with a bright companion in \pacs\ images, PG 0043$+$039,
PG 1114$+$445, and PG 1322$+$659 are undetected with \spire.  For the objects 
with faint companions, the emission is likely dominated by the target whenever they 
are detected in \spire\ maps.  We visually checked all of the images to identify possible 
false detections.  If a target is not detected at 250 \micron, which has the best resolution 
among the three \spire\ bands, but is detected at the longer wavelengths, we check 
whether there is a source detected near the target in the 250 \micron\ map.  If so, the 
detection in the other band(s) is considered false.  As a result of this procedure, we 
consider the detections at 350 and/or 500 \micron\ for PG 0947$+$396, PG 1048$+$342,
PG 1048$-$090, and PG 1626$+$554 to be spurious; Table \ref{tab:herschel} only 
reports upper limits for these four sources.

Following \cite{Leipski2014ApJ}, we use the pixel-to-pixel fluctuations of the
source-subtracted residual map to determine the uncertainty of the flux
measurements.  The residual map is created by subtracting all sources found by
the source extractor from the observed map.  We then calculate the
pixel-to-pixel RMS in a box of size eight times the beam FWHM of each band.  The box
size is large enough to include a sufficient number of pixels for robust
statistics, but small enough to avoid the low-sensitivity area at the edges of
the map.  The median RMS from our measurements are 10.57, 8.98, and 11.52 mJy at
250, 350, and 500 \micron.  \cite{Leipski2014ApJ} found that this method tends
to obtain the uncertainties very close to, but a bit smaller than, that
calculated from the quadrature sum of the confusion noise limits and the
instrument noise \citep{Nguyen2010AA}.  For our sample with one repetition scan,
the expected noise levels are 10.71, 9.79, and 12.76 mJy, respectively, very
close to our measurements.  We provide $3\,\sigma$ upper limits for all
non-detections.  Sources with flux densities below three times the RMS, even if
detected by the source extractor, are considered non-detections.

\input{tab3.tex}

\subsection{Archival Data}

There are no \pacs\ data for PG 1226$+$023 and no \herschel\ data of any kind
for PG 1444$+$407.  Therefore, we use \mips\ 70 and 160 $\mu$m data
\citep{Shang2011ApJS} for these two objects.  For the 16 radio-loud objects in
the sample, we use additional radio data from
NED\footnote{\url{http://ned.ipac.caltech.edu/}} to constrain the nonthermal
jet emission at FIR and submillimeter wavelengths.  Table \ref{tab:arXiv}
lists the archival data used in our analysis.

\subsection{Presentation of the SEDs}

\begin{figure*}
\begin{center}
\begin{tabular}{c c}
\includegraphics[height=0.35\textheight]{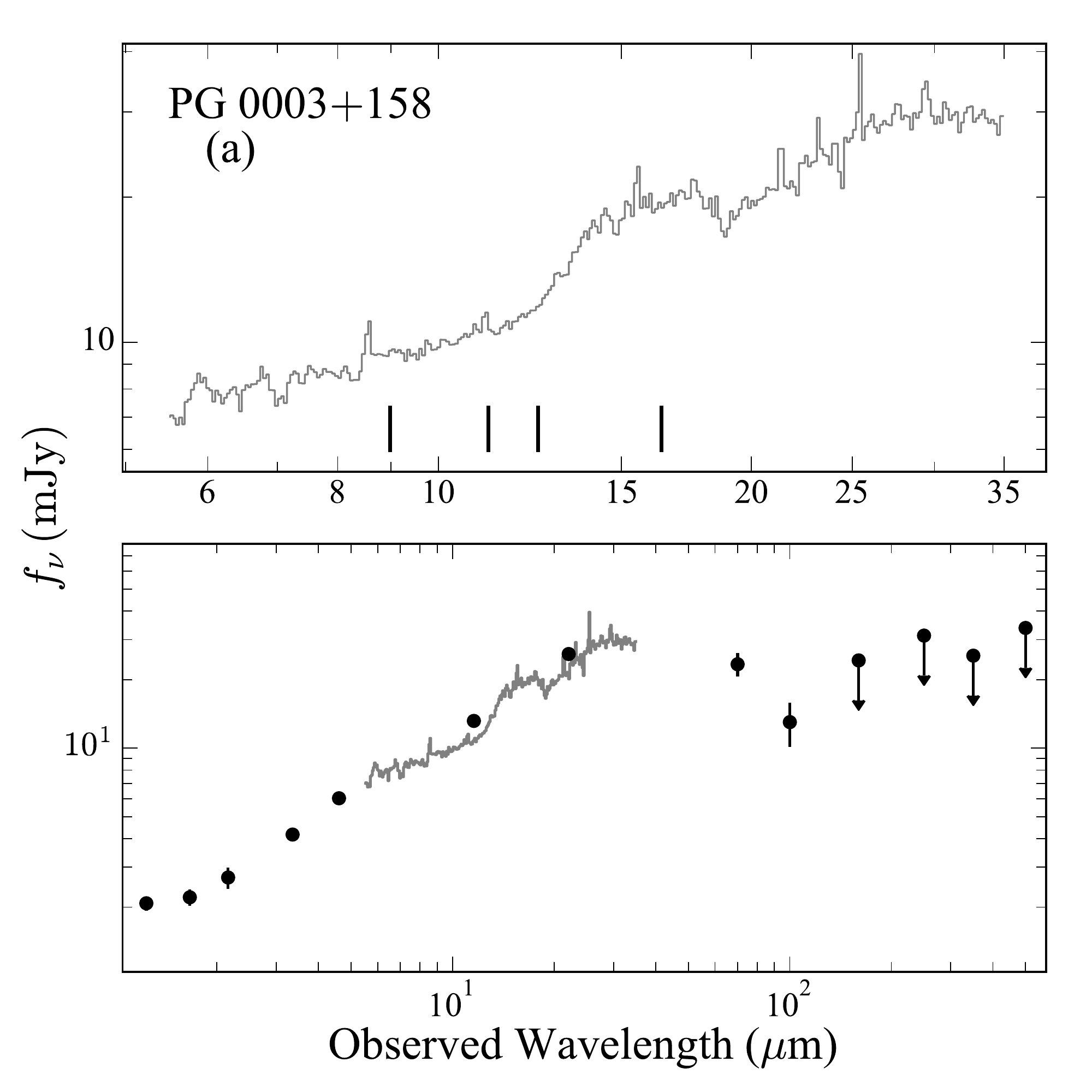} &
\includegraphics[height=0.35\textheight]{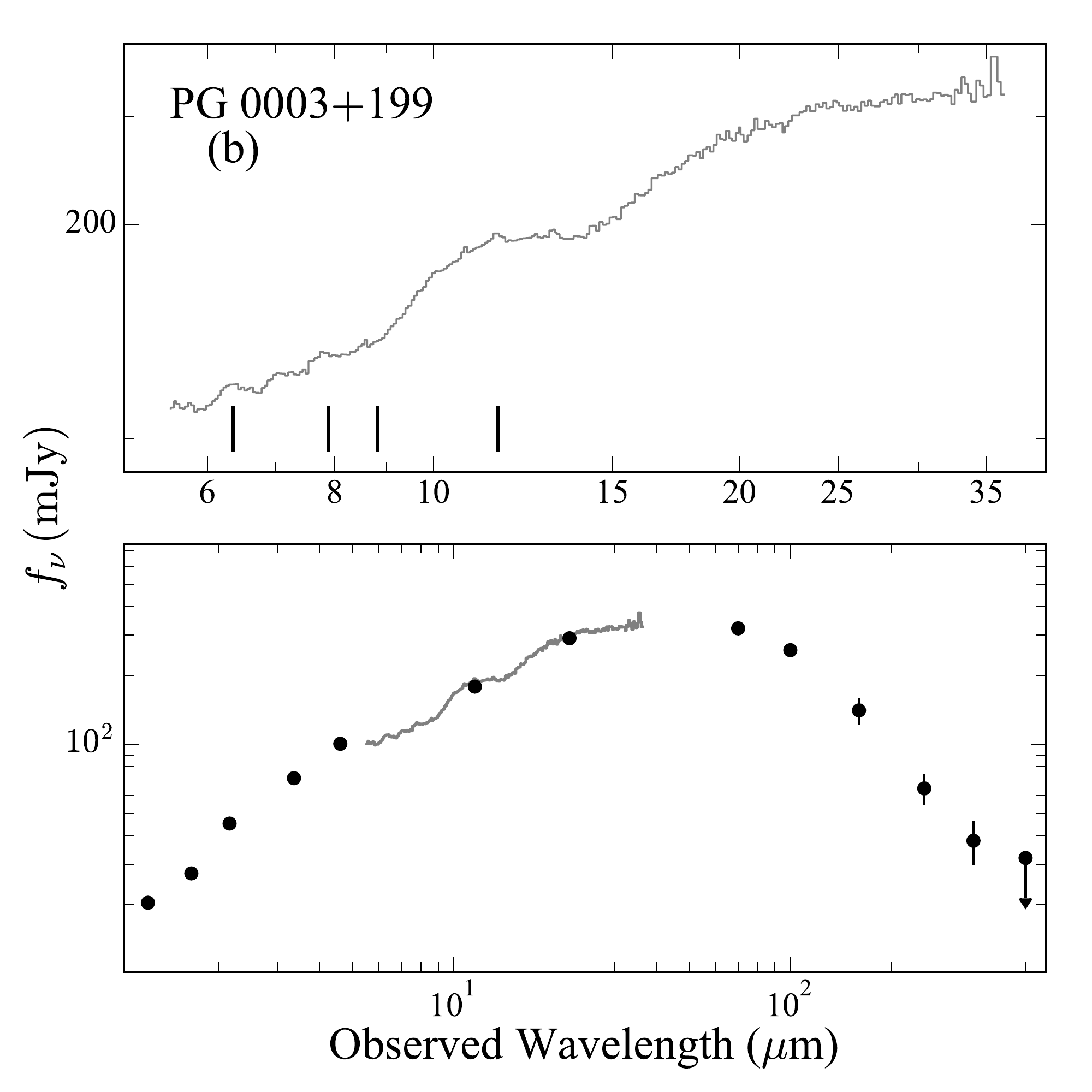} \\
\includegraphics[height=0.35\textheight]{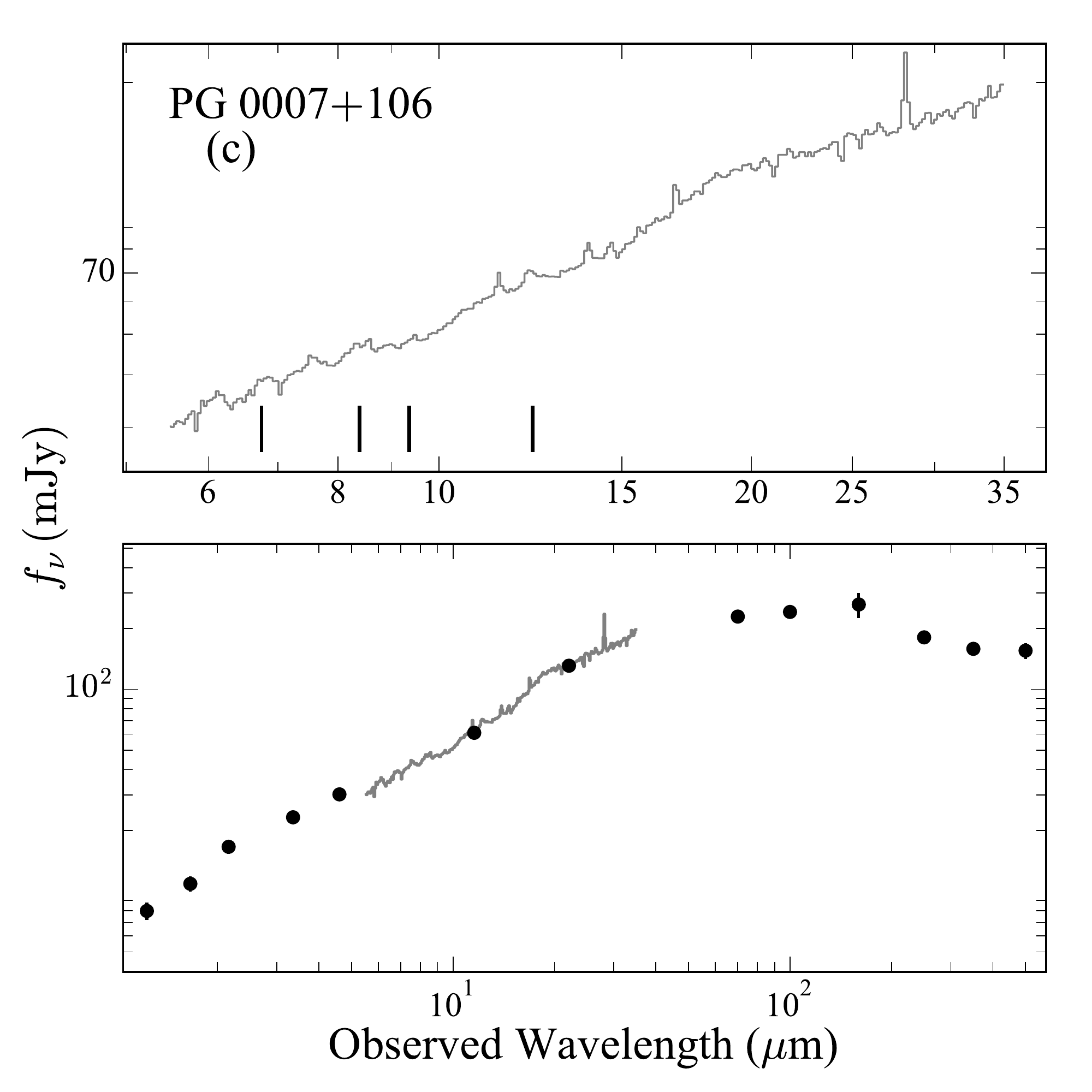} &
\includegraphics[height=0.35\textheight]{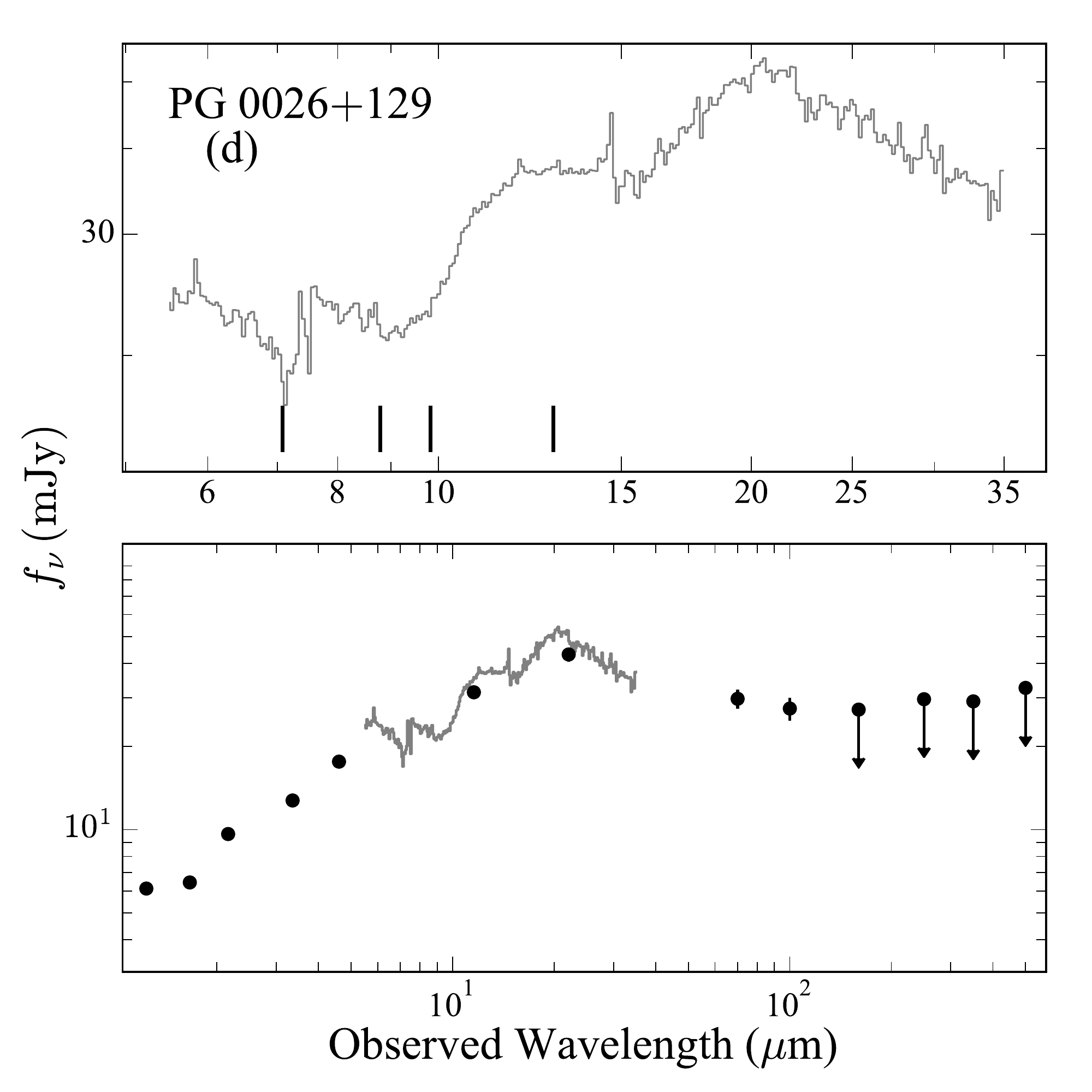}
\end{tabular}
\caption{IR SEDs of four representative PG quasars.  For each object, the
upper panel shows the details of the IRS spectra from $\sim$ 5 to 40 \micron\
(gray), and the lower panel shows the full SED from $\sim$ 1 to 500 \micron.
The black vertical lines in the upper panel highlights the location of the most
prominent PAH features at 6.2, 7.7, 8.6, and 11.3 \micron.  {\it The SEDs of the
entire sample (87 images) can be found in the online version.}}
\label{fig:sed}
\end{center}
\end{figure*}

The IR SEDs of the entire PG sample of 87 low-redshift quasars are displayed in
Figure \ref{fig:sed}.  Two panels are plotted for each object, one highlighting
the \spitzer\ IRS spectrum from $\sim$ 5 to 40 \micron, and the other showing
the entire IR band from $\sim$ 1 to 500 \micron.  Black vertical lines in the
upper panel demarcate the wavelengths of the most prominent features of
polycyclic aromatic hydrocarbons (PAHs) at 6.2, 7.7, 8.6, and 11.3 \micron.

\input{tab4.tex}

\section{SED Fitting Methods}\label{sec:method}
\subsection{SED Models}\label{ssec:model}

\input{tab5.tex}

The IR SED consists of emission from various physical sources inside a galaxy.
The stellar emission usually mainly contributes to the NIR, the bands covered
by \tmass.  The emission from the AGN dust torus dominates the quasar SED up to
$\sim 30$ \micron, covered by \wise\ and \irs.  At longer wavelengths, in the
regime of the \herschel\ bands, dust emission from the galactic-scale ISM
becomes brighter than the torus emission.  If the quasar is radio-loud, the jet
contributes strong synchrotron radiation, which usually dominates the GHz
radio bands but may extend to and sometimes even dominate the submillimeter
regime.  Since the emission from all of these physical components overlap, we
must fit the entire IR SED by simultaneously modeling all of the emission
components, in order to get an unbiased measurement of the host galaxy dust
properties.  The models we consider and their associated parameters are
summarized in Table \ref{tab:pars}.  The following describes them in detail.

The \emph{stellar emission} is represented by a simple stellar population
model from \citet[hereafter BC03]{Bruzual2003MNRAS} with a \cite{Chabrier2003PASP}
stellar IMF.  We use the Python package {\tt EzGal}
\citep{Mancone2012PASP} to generate the template spectra.  The stellar age is,
in principle, a free parameter, but we fix it to 5 Gyr because we can hardly
solve for the stellar age independently without additional constraints on the stellar
emission of the host galaxy in the optical bands.  This, however, is extremely
challenging because of the dominance of AGN emission at shorter wavelengths.
Moreover, the spectral shape of the NIR stellar emission is governed mostly
by the old stellar population, rendering it relatively insensitive to stellar
age.  Therefore, fixing the stellar age of the BC03 template is expected to have
a negligible effect on the derived dust properties.

For the \emph{dust torus emission}, we incorporate the templates generated by
the radiative transfer model CLUMPY \citep{Nenkova2008ApJa, Nenkova2008ApJb}.
We also test two other dust torus radiative transfer models provided by
\cite{Honig2010AA} and \cite{Siebenmorgen2015AA}.  For all three models, in
order to get a good fit to the MIR data, an additional hot ($T\approx 1000$ K)
blackbody (BB) component is required \citep{Deo2011ApJ,Mor2012MNRAS}.  This is
likely because these models all assume that silicate and carbon dust have the
same temperature distributions (R. Siebenmorgan 2017, private communication).
However, in reality, carbonaceous dust can have higher temperature, such that
the real dust torus displays excess emission at wavelengths $\lesssim 5$ \micron\ 
\citep{GarciaGonzalez2017MNRAS,Honig2017ApJL}.  A detailed analysis 
of how different dust torus templates fit quasar SEDs and how they affect the cold dust
measurements is beyond the scope of the current work.  Nevertheless, we do worry
whether the choice of torus model may introduce model-dependent systematic
uncertainties in our fitting.  In Appendix \ref{apd:xl17}, we demonstrate that
the torus model does not bias the key derived cold dust parameters---especially
the dust mass---as long as the FIR data constrain well the peak and the
Rayleigh-Jeans tail of the dust emission.  One of the advantages of the CLUMPY
model is that there are $\sim$ $10^6$ templates available, more than two orders of
magnitude larger than the other two sets of models.  The higher the density of
the sampled parameter grids, the more robust the model we can reconstruct by
interpolating the model templates (see Appendix \ref{apd:mcmc}).  The CLUMPY
model has seven free parameters: the optical depth of the individual cloud $\tau_V$,
the power-law index $q$ of the cloud radial distribution, the ratio $Y$ of outer
and inner radii of the dust torus,\footnote{The inner radius is set by the dust
sublimation temperature, assumed to be 1500 K.} the average number of clouds on
the equatorial ray $N_0$, the standard deviation $\sigma$ of the Gaussian
distribution of the number of clouds in the polar direction, the observer's
viewing angle $i$ from the torus axis, and the luminosity $L$ normalization
factor.  The complementary BB component,

\begin{equation}
f_\mathrm{\nu, BB} = \Omega_{d} \, B_\nu(T),
\end{equation}

\noindent
has two free parameters, $\Omega_{d}$ the solid angle subtended by the dust
and $B_\nu(T)$ the Planck function with temperature $T$.  Hence, the CLUMPY+BB
model has a total of nine parameters.

The \emph{galactic dust emission} is described by the widely used DL07 model.
The model is based on the dust composition and size distribution observed
in the Milky Way (MW).  The dust emission templates are calculated including
the single-photon heating process, which produces the PAH features.  The
radiation intensity relative to the local interstellar radiation field is
parametrized by $U$.  The DL07 model assumes that most of the dust in a galaxy
is located in the ``diffuse ISM'' and exposed to the radiation field with the
same intensity $U=U_\mathrm{min}$, the minimum radiation field intensity 
of the galaxy.  A small fraction ($\gamma$) of dust is
heated by photons from a power-law distribution of $U$, with
$U_\mathrm{min}<U<U_\mathrm{max}$ ($U_\mathrm{min} \ll U_\mathrm{max}$),
referred to as the ``photodissociation region'' component for normal galaxies,
since the wide range of $U$ may come from photodissociation regions related
to massive stars.  The dust grains are a mixture of amorphous silicate and
graphite, including PAH particles with mass fraction $q_\mathrm{PAH}$. The
rest-frame flux density of the DL07 model is

\begin{equation}
f_\mathrm{\nu,\, DL07} = \frac{M_{d} (1 + z)^2}{4 \pi D_{L}^2}
[(1 - \gamma) p_\nu^{(0)}(q_\mathrm{PAH},\, U_\mathrm{min})
+ \gamma p_\nu(q_\mathrm{PAH},\, U_\mathrm{min},\, U_\mathrm{max},\,
\alpha) ],
\end{equation}

\noindent
where $M_{d}$ is the dust mass, $z$ is the redshift, $D_{L}$ is the luminosity
distance, and $p_\nu^{(0)}(q_\mathrm{PAH},\, U)$ is the power radiated per unit
frequency per unit mass of the dust mixture determined by $q_\mathrm{PAH}$
exposed to the radiation field with intensity $U$.  The specific power of unit
dust mass is

\begin{equation}
p_\nu(q_\mathrm{PAH},\, U_\mathrm{min},\, U_\mathrm{max},\, \alpha) =
\frac{(\alpha - 1)}{U_\mathrm{min}^{1 - \alpha} - U_\mathrm{max}^{1 - \alpha}}
\, \int_{U_\mathrm{min}}^{U_\mathrm{max}} p_\nu^{(0)}(q_\mathrm{PAH},\, U)
\, U^{-\alpha} \, dU,
\end{equation}

\noindent
where $\alpha$ is the power-law index of the interstellar radiation field
intensity distribution.  DL07 provide the precalculated
$p_\nu^{(0)}(q_\mathrm{PAH},\, U_\mathrm{min})$ and
$p_\nu(q_\mathrm{PAH},\, U_\mathrm{min},\, U_\mathrm{max},\, \alpha)$ as
model templates.\footnote{\url{http://www.astro.princeton.edu/\~draine/dust/irem.html}}
By studying the SEDs of normal star-forming galaxies, \cite{Draine2007ApJ}
found that, for all situations, we can fix $\alpha = 2$ and
$U_\mathrm{max} = 10^6$.  We adopt this simplification, assuming that quasar
host galaxies have a distribution of radiation field intensity similar to that of 
typical star-forming galaxies.  Therefore, the DL07 model contains four free
parameters: $q_\mathrm{PAH}$ and $U_\mathrm{min}$ are discrete, and $\gamma$ and
$M_d$ are continuous.

The radio-loud objects are defined by the radio-loudness parameter,
$R \equiv f_\nu({\rm 6~cm})/f_\nu$(4400 \AA), such that $R \geq 10$
\citep{Kellermann1989AJ}\footnote{PG 1211+143 was misidentified as radio-loud
\citep{Kellermann1994AJ}.}.  The synchrotron radiation of radio-loud objects may
contaminate considerably the dust thermal emission in the submillimeter.  To
fit the synchrotron component, we adopt a broken power-law model (e.g.,
\citealt{Peer2014SSRv})

\begin{equation}
f_{\nu, \mathrm{syn}} \propto
  \begin{cases}
    \nu^{-\alpha}     & \nu < \nu_\mathrm{c},\\
    \nu^{-\alpha-1/2} & \nu_\mathrm{c} < \nu < \nu_\mathrm{max},\\
    0                 & \nu > \nu_\mathrm{max},
  \end{cases}
\end{equation}

\noindent
where $\nu_{c}=10^{13}$ Hz is the cooling frequency, above which the power-law
slope becomes steeper, and we assume that the highest frequency of the
synchrotron emission is $\nu_\mathrm{max}=10^{14}$ Hz.  The typical power-law
slope for steep-spectrum quasars is $\alpha \approx 0.7$, and we can use the
radio SED to anchor the synchrotron component.  For flat-spectrum quasars
\citep{Urry1995PASP} whose radio emission varies greatly, the archival radio
data, taken at different times, cannot be fitted by the synchrotron radiation model.
Nevertheless, we find that, with the help of submillimeter data, it is possible
to fit three flat-spectrum quasars in our sample (PG 0007$+$106, PG 1226$+$023,
and PG 1302$-$102) with reasonable power-law slopes ($\alpha\approx$ 0.7--1.3).
In the remaining two objects (PG 1309$+$355 and PG 2209$+$184), the synchrotron
emission is not dominant, and so it will only marginally affect, if at all, the global fit.  
The $f_0$ in Table \ref{tab:pars} is the scaling factor of the synchrotron model.

The final model SED is the linear combination of the BC03, BB, CLUMPY, DL07,
and, if necessary, the synchrotron components.  To directly compare with the
observed photometric data, we need to fold the model SED through the response
functions of the respective photometric bands \citep{Bessell2012PASP}. For the
\tmass\ and \wise\ bands,\footnote{The response curves of the \tmass\ and \wise\
filters can be downloaded from
\url{http://www.ipac.caltech.edu/2mass/releases/allsky/doc/sec6_4a.html} and
\url{http://wise2.ipac.caltech.edu/docs/release/allsky/expsup/sec4_4h.html}.}
our quoted flux density is the band-averaged flux density,

\begin{equation}
\langle f_\nu \rangle = \frac{\int f_\nu(\nu) S(\nu) d\nu/\nu}{\int S(\nu)
d\nu/\nu},
\end{equation}

\noindent
where $S(\nu)$ is the system photon response function.  In the case of the
\herschel\ bands, the data are monochromatic flux densities at the nominal
frequency $\nu_0$,

\begin{equation}
f_{\nu_0} = \frac{1}{\nu_0} \frac{\int S'(\nu)d\nu}{\int S'(\nu)d\nu/\nu}
\langle f_\nu \rangle,
\end{equation}

\noindent
where $S'(\nu)$ is the system energy response function,\footnote{The response
curves of \pacs\ are combined from the filter transmission functions and the
detector absorption, while those of \spire\ are combined from the filter
transmission functions for point sources with the aperture efficiency. All the
information are obtained from HIPE.} and the band-averaged flux density becomes
(Section 5.2.4 of \spire\ handbook)

\begin{equation}
\langle f_\nu \rangle = \frac{\int f_\nu(\nu) S'(\nu) d\nu}{\int S'(\nu) d\nu}.
\end{equation}

No additional reprocessing is necessary to mimic the observations of \spitzer\
because the PAH features in DL07 are already designed to match the
low-resolution \irs\ spectra.

\subsection{Fitting Method}\label{ssec:code}

In order to simultaneously fit the photometric and spectroscopic data, we
develop a Bayesian MCMC fitting algorithm.  The code can incorporate an
arbitrary number of models to obtain a combined SED model.  The Bayesian 
method (\citealt{Gregory2005}) implies that the posterior probability density 
function (PDF) of model parameters, $\Theta=[\theta_1, \theta_2, ...]$, given 
the prior knowledge $I$ and data $D$, is

\begin{equation}
p(\Theta | D, I) = \frac{p(\Theta | I) p(D | \Theta, I)}{p(D | I)}.
\end{equation}

\noindent
The {\it prior}, $p(\Theta | I)$, is provided by our prior knowledge about the
probability distribution of the model parameters.  The {\it evidence},
$p(D | I)$, is a normalization factor that does not affect the fitting with a given
model.  It may be important when we need to compare different models, but this
is beyond the scope of the current work, and we do not consider it further.

The {\it likelihood} of the data, $\mathscr{L} = p(D|\Theta,I)$, being observed
with the given prior knowledge and model parameters is assumed to be

\begin{equation}
  \mathrm{ln}\,\mathscr{L} = \mathrm{ln}\,\mathscr{L}_{p, d} +
  \mathrm{ln}\,\mathscr{L}_{p, u} + \mathrm{ln}\,\mathscr{L}_{s},
\end{equation}

\noindent
where $\mathrm{ln}\,\mathscr{L}_{p, d}$ and $\mathrm{ln}\,\mathscr{L}_{p, u}$
are the ln-likelihoods of the photometric data with detection and upper limits,
respectively, while $\mathrm{ln}\,\mathscr{L}_{s}$ is the ln-likelihood of the
spectra.  We adopt

\begin{eqnarray}\label{eq:lnlp}
&&\mathrm{ln}\,\mathscr{L}_{p, d} = -\frac{1}{2}\left(\sum_i^n \,
\frac{(y_i - \tilde{y}_i(\Theta))^2}{s_i^2} + \sum_i^n \mathrm{ln}\,(2\pi
s_i^2)\right), \\
&&\mathrm{ln}\,\mathscr{L}_{p, u} = \sum_j^m \, \mathrm{ln} \,
\frac{1+\mathrm{erf}(z_j)}{2}, ~ z_i = \frac{y_i - \tilde{y}_i(\Theta)}
{\sqrt{2}s_i},
\end{eqnarray}

\noindent
with

\begin{equation}
s_i^2 = \sigma_i^2 + (f \tilde{y}_i(\Theta))^2, \nonumber
\end{equation}

\noindent
where $y_i$ and $\tilde{y}_i$ are the observed and model synthetic flux
densities,  $\mathrm{erf}(x)$ is the error function, and $\sigma_i$ is the
observational uncertainty, three times which is considered to be the upper limit.
We introduce a parameter $f$ into $s_i$, the square root of the inverse weight,
to consider the systematic uncertainty from the model to the real data.  In
order to balance the weight of the data at different wavelengths, some works
assign a 10\% additional uncertainty to all of the bands (e.g.,
\citealt{Draine2007ApJ}).  Others choose to use a uniform weight for all of the
bands instead of incorporating the observational uncertainty.  Our approach, by
contrast, assumes the typical percentage for the model to deviate from the data
to be $f$ and lets the MCMC algorithm fit for $f$ as a free parameter.  This method
to consider upper limits for the data has been widely used (e.g.,
\citealt{Isobe1986ApJ,Lyu2016ApJ,Shimizu2017MNRAS}).  For spectroscopic data,
the residual between data and model may be highly correlated
\citep{Czekala2015ApJ}, and so we need to model the residual correlation.  For
the spectra, we adopt

\begin{eqnarray}
\mathrm{ln}\,\mathscr{L}_{s} &=& -\frac{1}{2}\left(\textbf{r}^T \,
\textbf{K}^{-1} \, \textbf{r} + \mathrm{ln} \, \det{\textbf{K}} + N \,
\mathrm{ln} \, 2\pi \right), \label{eq:chis} \\
K_{i, j} &=& s_i^2 \, \delta_{i, j} + k_{i, j}, \label{eq:Kij} \nonumber \\
k_{i, j} &=& a^2 \left(1 + \frac{\sqrt{3}|\lambda_i - \lambda_j|}{\tau} \right)
\exp{\left(-\frac{\sqrt{3}|\lambda_i - \lambda_j|}{\tau}\right)}, \label{eq:kij}
\end{eqnarray}

\noindent
where $r_i = y_i - \tilde{y}_i(\Theta)$ is the residual between data and model,
$\textbf{K}$ is the covariance matrix, $N$ is the length of the data,
$\delta_{i, j}$ is the Kronecker delta, and $k_{i, j}$ describes the correlation
between two residuals at wavelengths $\lambda_i$ and $\lambda_j$. We choose
$k_{i, j}$ to be the Mat\'{e}rn $3/2$ function \citep{Rasmussen2006}, where
$a$ is the strength of the correlation and $\tau$ is the characteristic length
of the correlation.  There are, in total, three free parameters ($f$, $a$, and
$\tau$) that enter the fitting to model the uncertainties.  We use the Python
package {\tt George} \citep{Ambikasaran2014} to calculate the matrix inverse and
determinant with Gaussian process regression method.  With more realistic
treatments of the uncertainties and residuals, our likelihood function is
flexible enough to balance the weight of the photometric and spectroscopic data
in the fitting.

Due to the complexity of the model (up to 19 parameters to be fitted), we have
to rely on the MCMC method to sample the parameter 
space.  We develop a Python code to construct the model and use the package 
\mc\ \citep{ForemanMackey2013} to sample the posterior PDF (see Appendix
\ref{apd:mcmc} for details).  In order to ascertain whether the Bayesian MCMC
fitting method can effectively constrain the model parameters, we generate mock
SEDs with the best-fit models of the quasar SEDs and their realistic
uncertainties and upper limits. The details of the test are described in
Appendix \ref{apd:rel}.  We find that the DL07 parameters can be reliably
measured with our fitting strategy. The scatter of the input and best-fit dust
masses is 0.16 dex for the entire sample, with no systematic deviation.  For the
44 objects whose FIR SEDs are good enough to cover the peak and Rayleigh-Jeans
tail of the dust emission, the scatter of the dust mass is only 0.09 dex.
$U_\mathrm{min}$ and $q_\mathrm{PAH}$ are discrete parameters.  Their best-fit
results are typically $\lesssim 2$ grid points away from the input values,
except for some objects with very poor detections in the FIR.  The $\gamma$
parameter controls the amount of dust emission from the power-law part of the
radiation field, which mainly contributes in the MIR, overlapping with the AGN
torus emission.  Therefore, $\gamma$ is mostly affected by the AGN torus model.
The fitting results may be unreliable for objects with $\gamma \lesssim 0.01$.

\section{Results}\label{sec:result}

\subsection{SED Fitting}\label{ssec:fitting}

\begin{figure*}
\begin{center}
\begin{tabular}{c c}
\includegraphics[height=0.35\textheight]{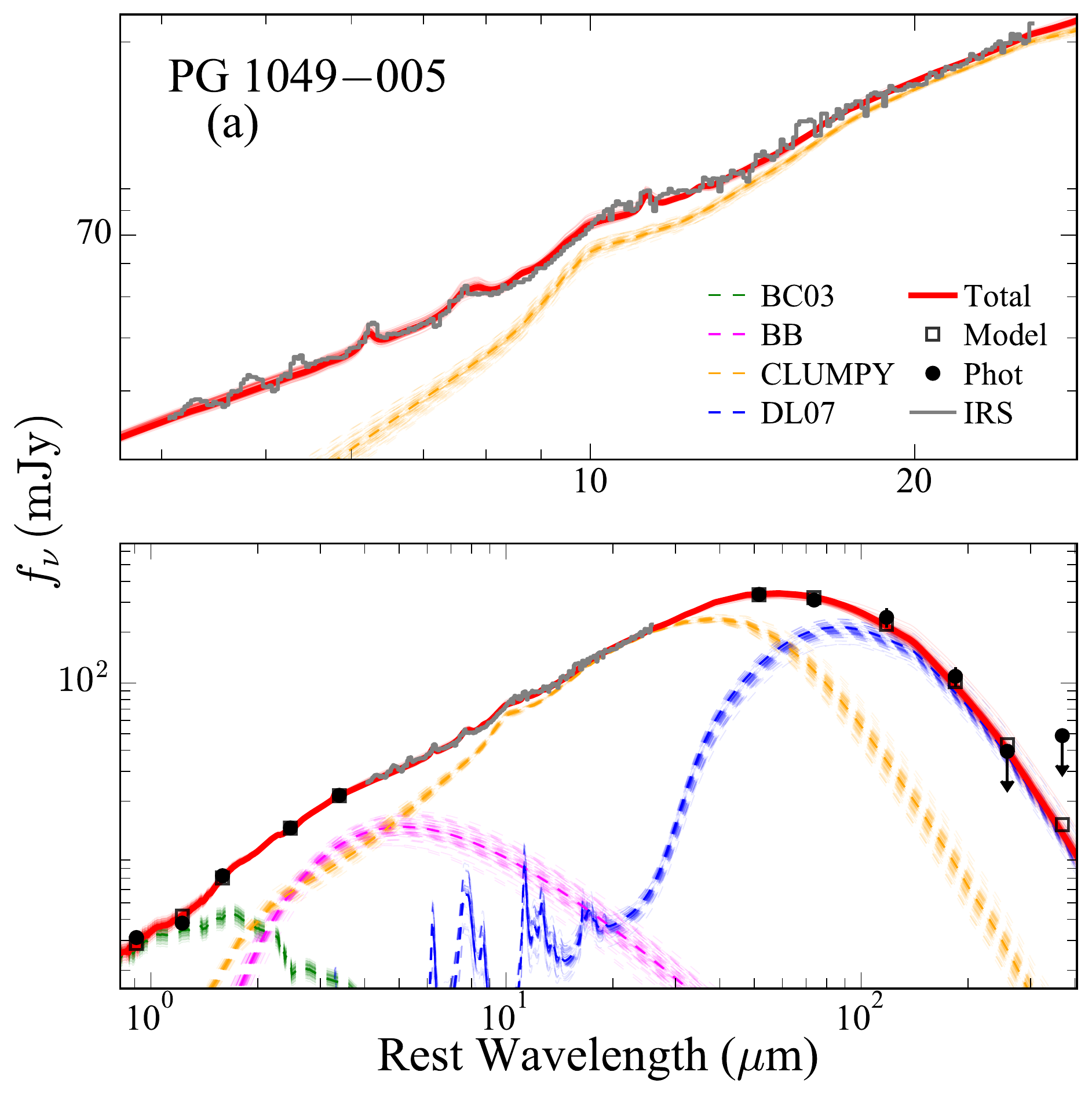} &
\includegraphics[height=0.35\textheight]{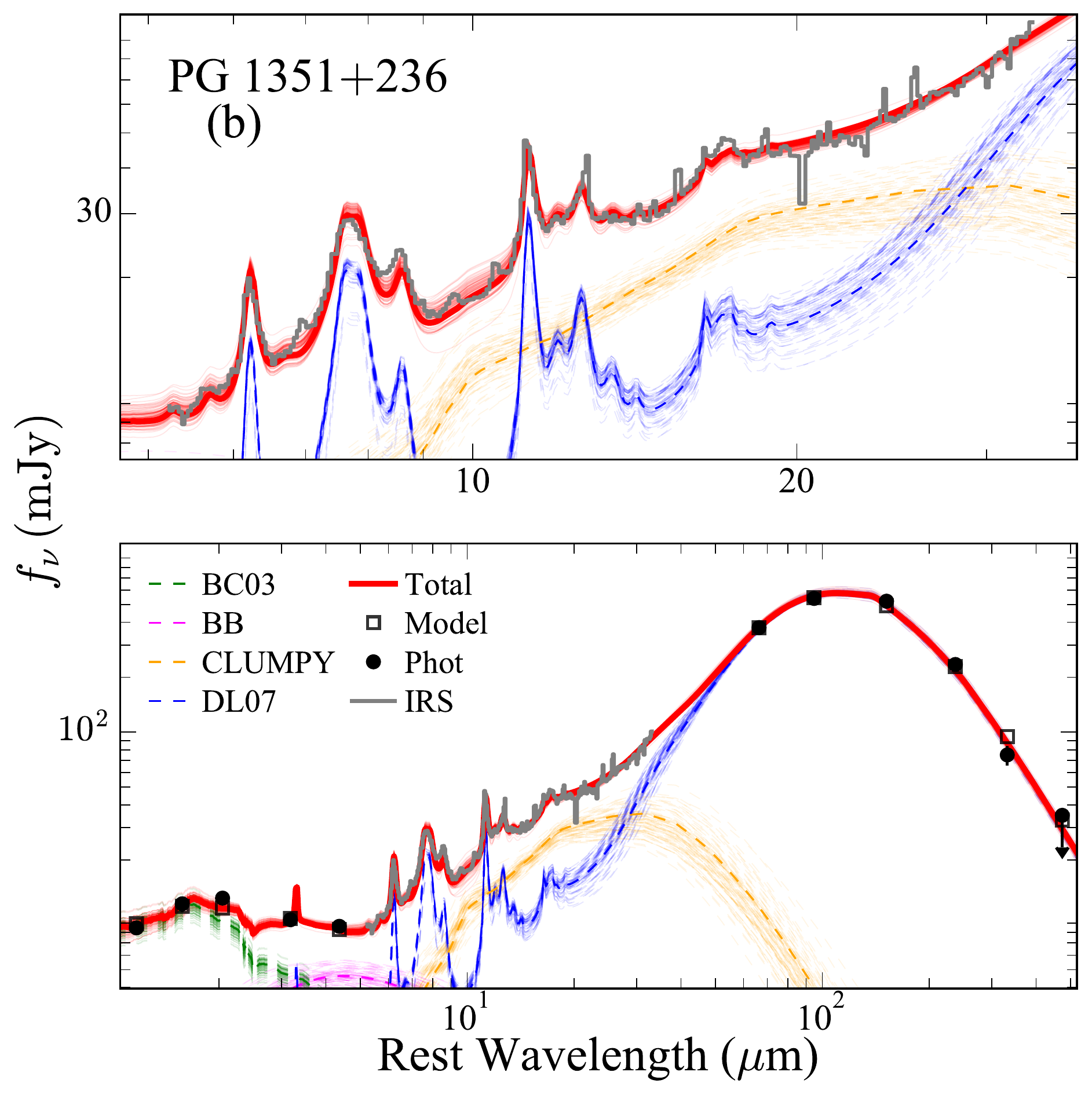}
\end{tabular}
\caption{Best-fit results for (a) PG 1049$-$005 and (b) PG 1351$+$236.  The lower 
panels show the entire IR SEDs while the upper panels zoom in to display the details 
of the IRS spectrum covering $\sim$ 5 to 40 \micron\ (gray line).  The black points 
are the photometric data from \tmass, \wise, and \herschel.  The dashed lines are the 
best-fit models: BC03 (green), BB hot dust  (magenta), CLUMPY torus (orange), and 
DL07 (blue).  The combined total model is the red solid line.  To visualize the model 
uncertainties, the associated thin lines in light color represent 100 sets of models with 
parameters drawn randomly from the space sampled by the MCMC algorithm.  With 
detections in four \herschel\ bands, PG 1049$-$005 can already provide good constraints 
on the model.  In the case of PG 1351$+$236, which has prominent PAH features, the 
best-fit model captures the features of the SED on both large and small scales. {\it The 
best-fit results of the entire sample (87 images) can be found in the online version.}
}
\label{fig:fitgood}
\end{center}
\end{figure*}

Best-fit results are shown in Figure \ref{fig:fitgood} for two objects with \herschel\ 
detections in four or more bands.  The best fit and each component of the models 
are displayed with dashed lines in different colors.  To illustrate the uncertainty of 
the model (components), we randomly choose 100 sets of parameters from the 
MCMC-sampled parameter space and plot them with light thin lines.  
The lower panels show the full SED and the best-fit models while the upper panels 
zoom in to display the details of the spectra in the range 5--40 \micron.  The best-fit 
model not only matches the large-scale structure of the SED but also properly captures 
the detailed PAH features of the spectra.  

\begin{figure*}
\begin{center}
\begin{tabular}{c c}
\includegraphics[height=0.35\textheight]{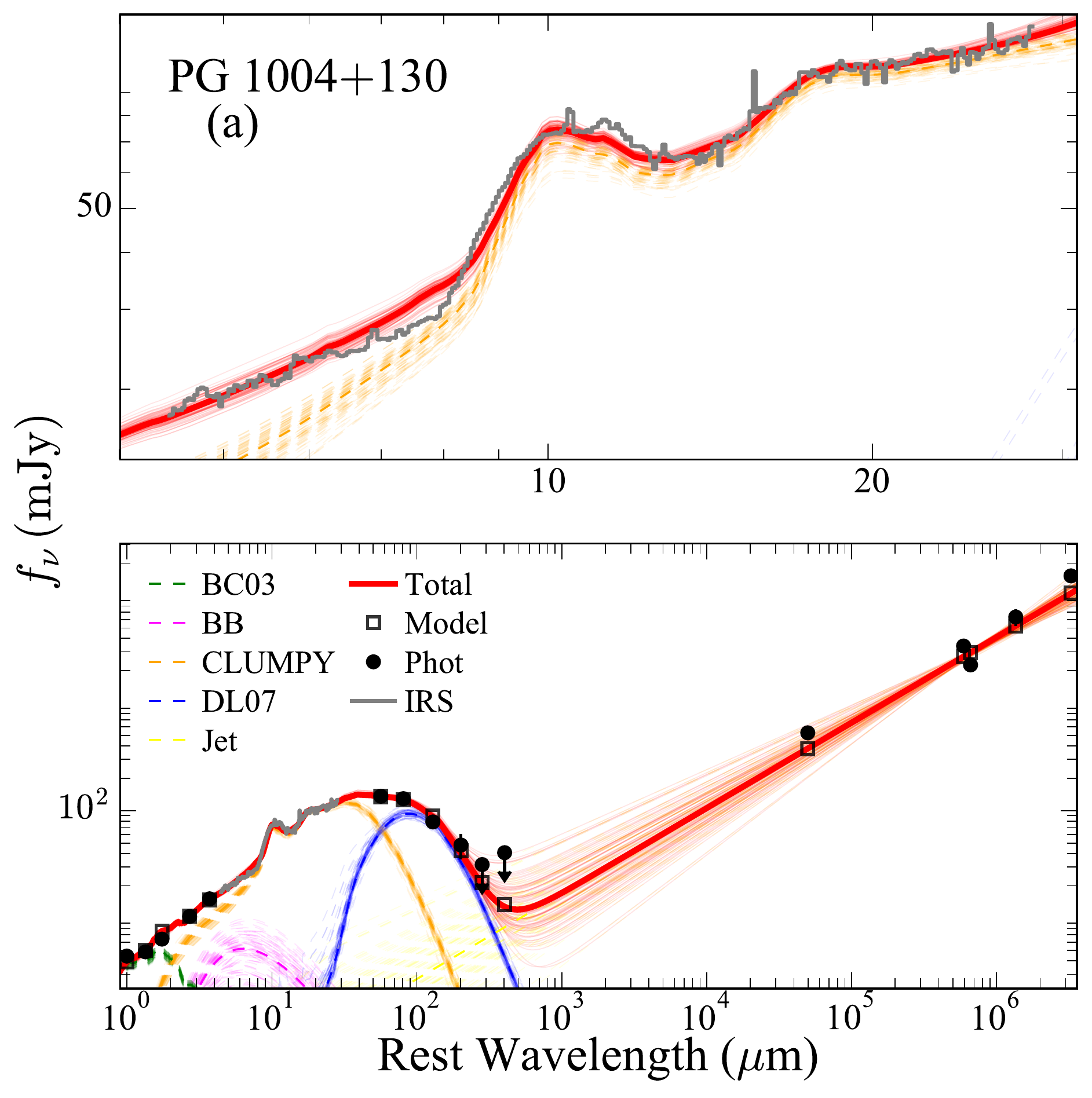} &
\includegraphics[height=0.35\textheight]{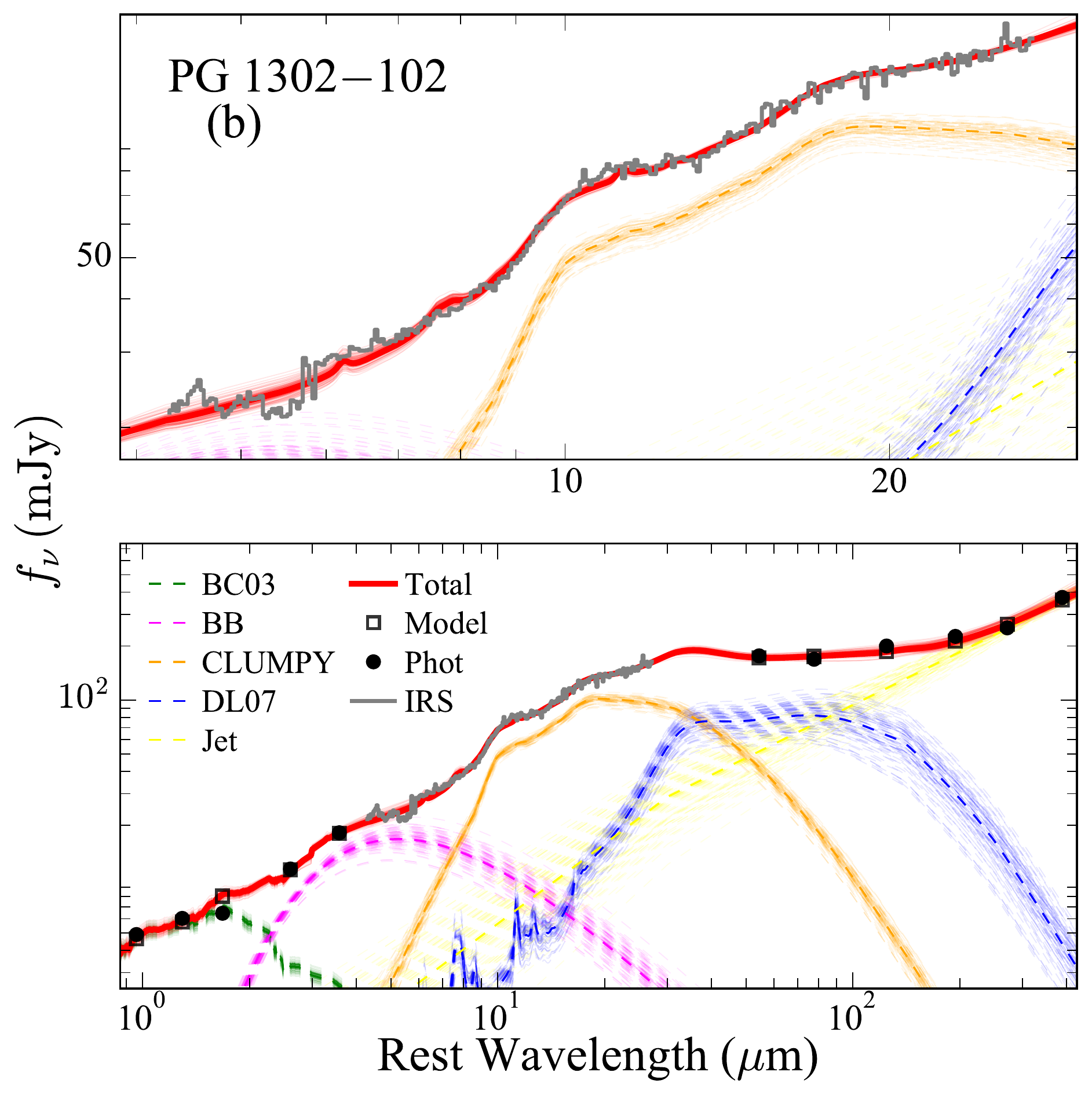}
\end{tabular}
\caption{Best-fit results for (a) the steep-spectrum quasar PG 1004$+$130 and 
(b) the flat-spectrum quasar PG 1302$-$102.  The symbols are the same as in 
Figure \ref{fig:fitgood}.  The synchrotron emission (yellow) is negligible at FIR 
wavelengths for PG 1004+130, whereas it is dominant in PG 1302$-$102, although 
its radio emission varies too much to be used to constrain the synchrotron model.
}
\label{fig:fitradio}
\end{center}
\end{figure*}

Figure \ref{fig:fitradio} shows the fitting results for two radio-loud objects.
PG 1004$+$130 is a steep-spectrum radio quasar.  The synchrotron emission
(yellow), anchored by radio data collected from the archives, contributes
negligibly at FIR wavelengths.  As a flat-spectrum radio quasar, PG 1302$-$102
exhibits too much radio variability to constrain the synchrotron model, and we
resort to fitting the IR SED without additional radio data.  Even though the
synchrotron emission is very strong, all of the dust components are reasonably
well constrained.  The power-law slope is $\sim$ 0.8.  The DL07 component is not
significantly affected by the synchrotron emission for all the radio-loud
objects. The only exception is PG 1226+023, whose synchrotron emission is so
strong that the cold dust emission is totally overwhelmed; its dust mass is very
uncertain, as reflected in its error bar.

\begin{figure*}
\begin{center}
\begin{tabular}{c c}
\includegraphics[height=0.35\textheight]{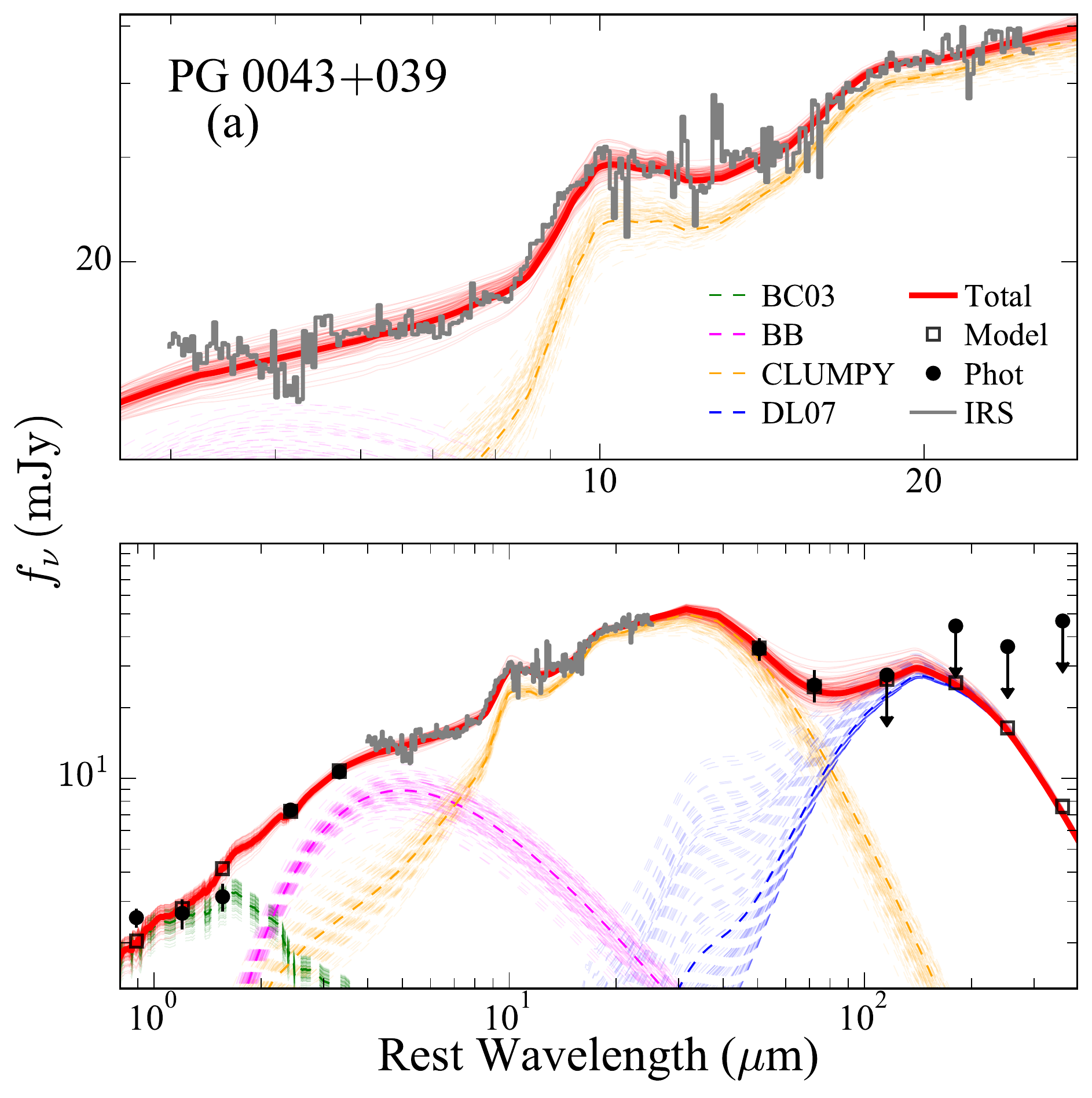} &
\includegraphics[height=0.35\textheight]{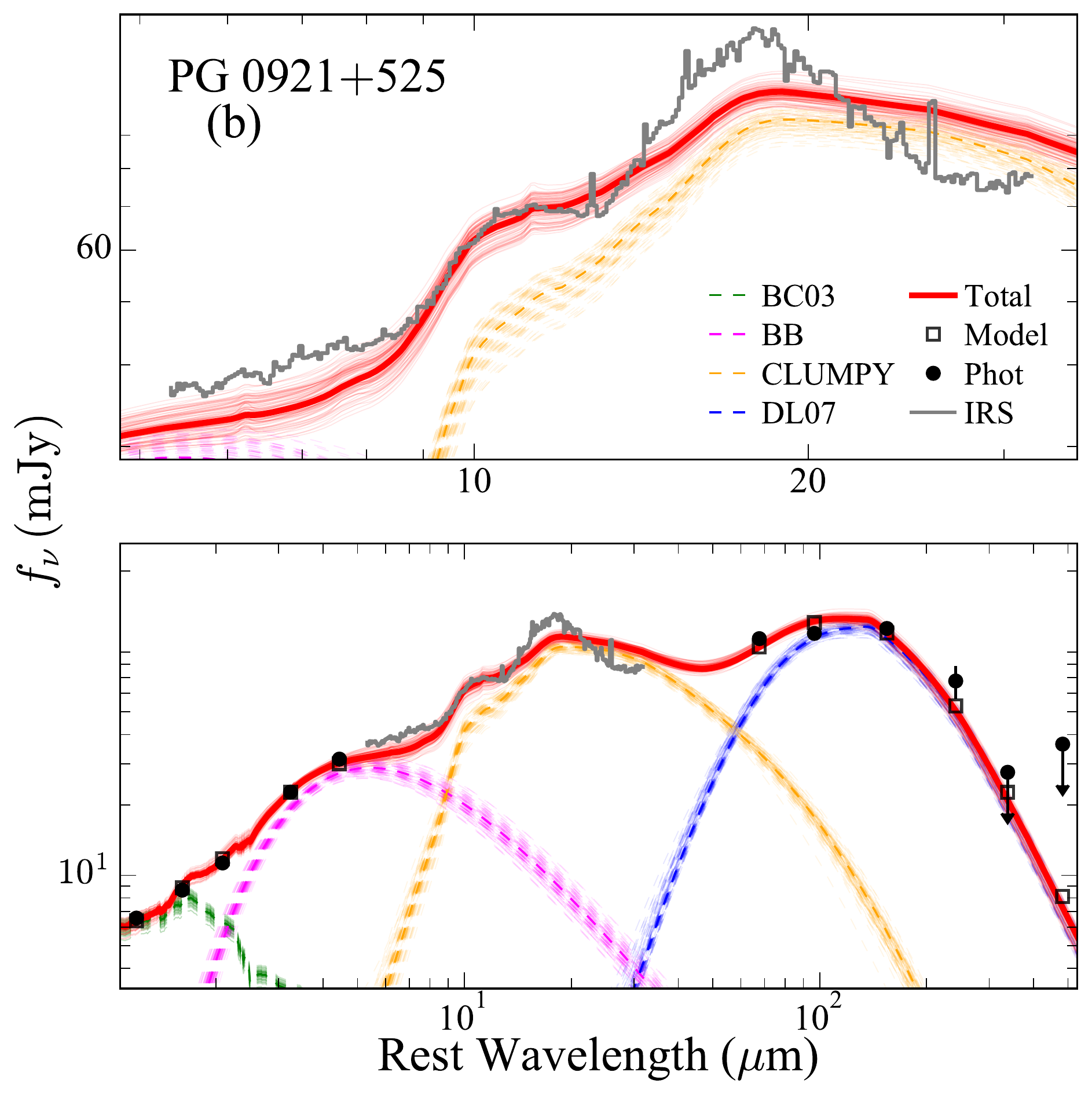}
\end{tabular}
\caption{Best-fit results for (a) PG 0043+039 and (b) PG 0921+525.
The symbols are the same as in Figure \ref{fig:fitgood}.  The DL07 
model is not well constrained for PG 0043$+$039, mainly because 
there are only two \herschel\ bands detected.  Nevertheless, the 
upper limits still provide some useful constraints on the best-fit model, 
albeit with larger uncertainties.  The spectrum of PG 0921$+$525 is 
not well fit by the CLUMPY model, which likely cannot account for the
different dust composition in this object.  However, this mismatch will 
barely affect the measurement of the dust mass (Appendix \ref{apd:xl17}).
}
\label{fig:fitbad}
\end{center}
\end{figure*}

Still, not all fits are reliable.  This applies primarily to some distant
(fainter) objects that are not well detected by \herschel.  As illustrated by
PG 0043$+$039 (Figure \ref{fig:fitbad}(a)), the DL07 model cannot be well
constrained.  However, this only happens when there is no detected \herschel\
band where the DL07 model contributes non-negligible emission.  We visually
check all of the fitting results and find 11 objects whose DL07 model cannot be
well constrained by the FIR SED. If we allow the DL07 parameters to be free,
the model adjusts to mainly fit the mismatch between the data and the CLUMPY
component.  Under these circumstances, we simply attempt to place an upper limit
on the allowed dust mass.  We fix the dust mass in the fit, manually and
iteratively adjusting $M_d$ in increments of 0.1 dex.  Meanwhile, $M_d$ is
degenerate with $U_\mathrm{min}$: lower values of $U_\mathrm{min}$ lead to
higher $M_d$.  For the purposes of obtaining a robust, conservative upper limit
on $M_d$, we fix $U_\mathrm{min}=1.0$ since the diffuse radiation field of
quasar host galaxies is not likely weaker than that of the solar neighborhood.
In normal, star-forming galaxies, $U_\mathrm{min}$ hardly ever reaches below 1
\citep{Draine2007ApJ}.  We also fix $q_\mathrm{PAH}=0.47$, the minimum value of
the model grid, although in practice the actual value of $q_\mathrm{PAH}$ makes
little difference because the DL07 component of the 11 objects is always
negligible at MIR wavelengths compared to the torus component.

Another complication arises when the CLUMPY component cannot fit the \irs\ spectrum 
well (Figure \ref{fig:fitbad}(b)), presumably because the dust torus of some objects has 
an unusual chemical composition \citep[hereafter, XLH17]{Xie2017ApJS} that differs 
from that assumed in the standard CLUMPY model.  In these situations, we usually need 
to limit the amplitude of the covariance ($a$), so that the template is forced to match the 
spectrum, regardless of the detailed features.  This may introduce systematic errors to
the DL07 model. This issue is addressed in Appendix \ref{apd:xl17}, where we investigate 
the impact on the DL07 parameters by replacing the CLUMPY model with the optically thin 
dust emission model proposed by XLH17.  We find that both torus models yield 
consistent values of $U_\mathrm{min}$ and $M_d$, especially for the objects with good FIR 
data.   The $\gamma$ parameter shows some systematic discrepancies, but this is expected 
because it is mostly degenerate with the torus model.  The scatter in $q_\mathrm{PAH}$ 
is large, likely because, for some cases, the XLH17 model poorly matches the 
spectra below $\sim$ 10 \micron\ (see Appendix \ref{apd:xl17} for details).

Furthermore, as we later show (Section \ref{ssec:mbb}), the modified blackbody
(MBB) model, when properly used, can provide dust masses that are quite
consistent with those derived with the DL07 model from full SED fitting.
In summary: our measurements of dust masses in quasar host galaxies from the
DL07 model and full SED fitting are not likely biased compared to those of
normal galaxies.

\subsection{ISM Radiation Field: Evidence for AGN Heating of Dust}\label{ssec:rad}

\begin{figure*}
\begin{center}
\includegraphics[width=0.9\textwidth]{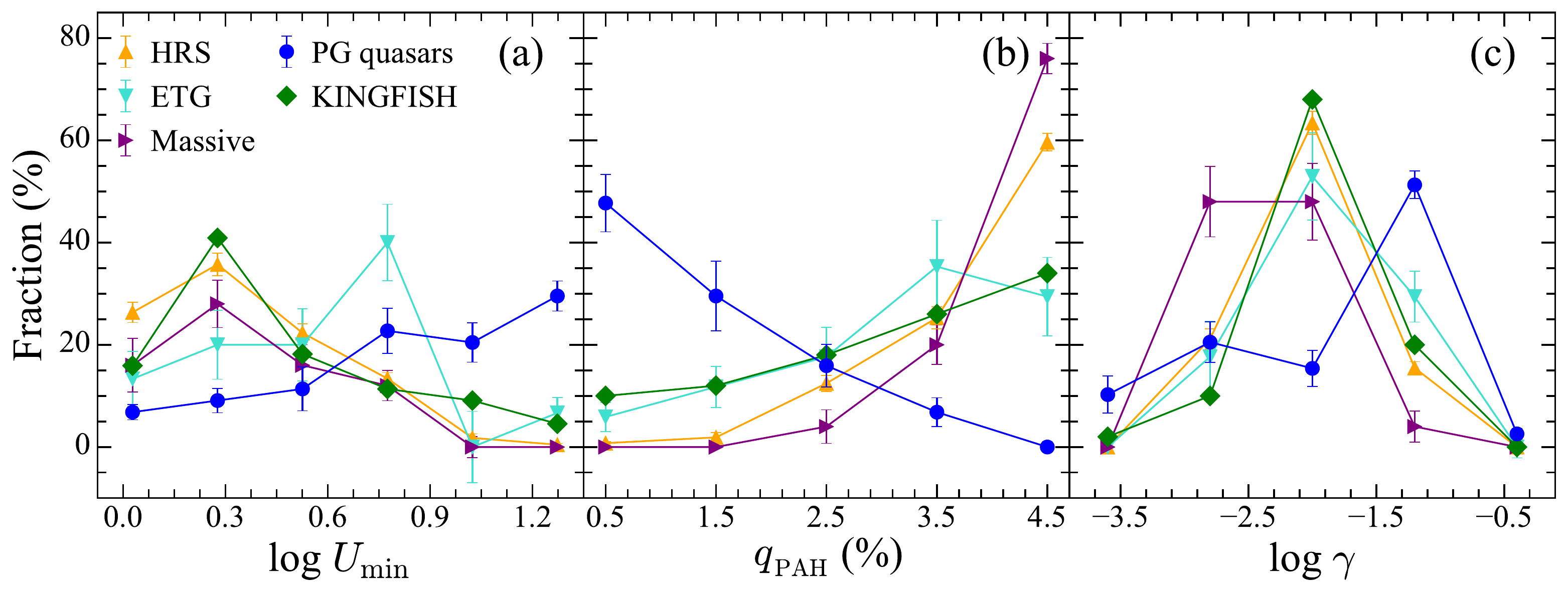}
\caption{Distribution of DL07 parameters (a) $U_\mathrm{min}$, (b)
$q_\mathrm{PAH}$, and (c) $\gamma$ for PG quasars (blue 
circles) and star-forming galaxies from the KINGFISH (green 
diamonds) and HRS (upward orange triangles) samples.  
Two subsamples of HRS galaxies are plotted.  The downward cyan 
triangles are early-type galaxies (S0 and elliptical, according to 
\citealt{Boselli2010PASP}; 17 objects), while the rightward purple triangles 
are massive galaxies with stellar mass $> 10^{10.5}\,M_\odot$ (25 objects).  
The uncertainties for PG quasars and HRS galaxies are estimated with a Monte 
Carlo method, resampling the parameters according to their measured 
uncertainties and calculating the number of galaxies in each bin for 500 
times.  The local star-forming and quenched galaxies in the KINGFISH and HRS 
samples peak at low $U_\mathrm{min}$ but high $q_\mathrm{PAH}$.  By contrast, 
PG quasar host galaxies tend to have higher $U_\mathrm{min}$ and lower 
$q_\mathrm{PAH}$.  The early-type galaxies tend to have higher $U_\mathrm{min}$ 
but mainly peak at intermediate values, $U_\mathrm{min} < 10$.
}
\label{fig:dl07}
\end{center}
\end{figure*}

\begin{figure}
\begin{center}
\includegraphics[width=0.4\textwidth]{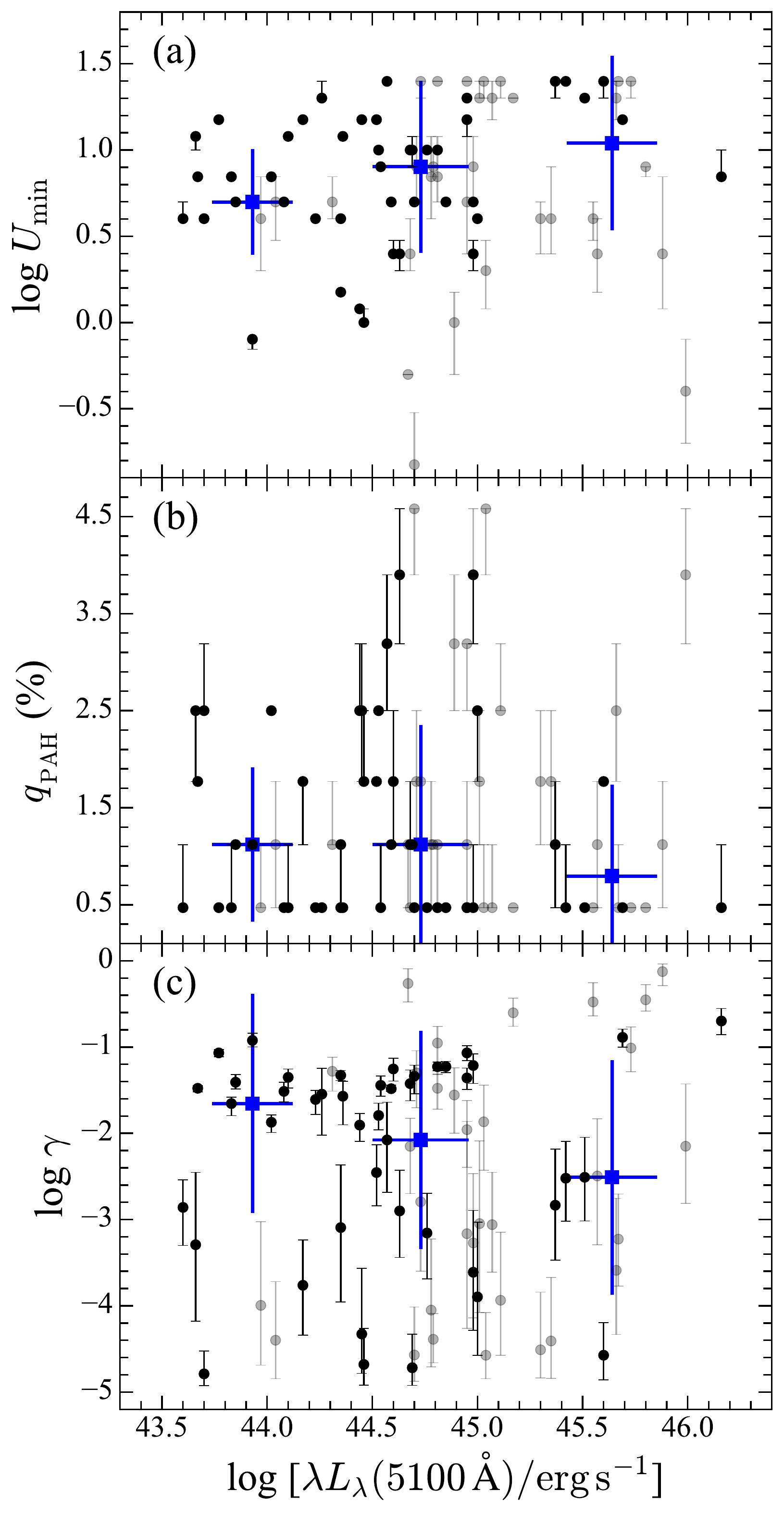}
\caption{Relation between AGN optical luminosity and DL07 parameters
(a) $U_\mathrm{min}$, (b) $q_\mathrm{PAH}$, and (c) $\gamma$.
The dark points represent more robust fitting results than the gray points; we
omitted objects for which only upper limits are available for the dust mass.
The sample is binned according to $\mathrm{log}\,[\lambda L_{\lambda}$(5100 
\AA)/$\mathrm{erg\,s^{-1}}]$: $< 44.25$, 44.25--45.25, and $\geq 45.25$; the 
median and standard deviation of the parameter values in each bin are plotted 
as blue squares with error bars.  Since $U_\mathrm{min}$ and $q_\mathrm{PAH}$ 
are discrete parameters, their errors may not be resolvable if they are smaller 
than the size of the grid.}
\label{fig:dl07agn}
\end{center}
\end{figure}

Our approach to SED fitting using the DL07 model allows us to diagnose some
important properties of the ISM, namely the strength of the ``diffuse'' 
radiation field ($U_\mathrm{min}$), the mass fraction of warm dust ($\gamma$),
and the mass fraction of the dust contained in PAHs ($q_\mathrm{PAH}$).
Although the best-fit parameters for individual objects may have relatively
large uncertainties, the distribution of parameters for the sample may yield
insights into the ensemble properties of quasar host galaxies.  Figure
\ref{fig:dl07} compares the distributions of $U_\mathrm{min}$, $q_\mathrm{PAH}$,
and $\gamma$ for the PG quasars in relation to the sample of normal galaxies
from KINGFISH and HRS.  The distributions of the three parameters for the
KINGFISH and HRS galaxies are very similar, even though the KINGFISH sample
comprises essentially gas rich, star-forming galaxies while more than half of 
the HRS galaxies are gas poor \citep{Ciesla2014AA}.  The uncertainties for 
PG quasars and HRS galaxies are estimated with a Monte Carlo method, 
resampling the parameters according to their measured uncertainties and 
calculating the number of galaxies in each bin for 500 times.\footnote{In order
to provide a conservative confidence level, the discrete parameters of PG
quasars are perturbed around the closest grids around the measured values if
their uncertainties are not resolved.  No uncertainty is provided for the
KINGFISH galaxies \citep{Draine2007ApJ}.}

Relative to the normal galaxies, the quasar hosts display a higher fraction of
$U_\mathrm{min}$ at high values.  A higher $U_\mathrm{min}$ signifies a
stronger ISM radiation field.  What is the source of this enhancement?  One
possibility is that quasar host galaxies may have stronger star formation
activity than normal galaxies.  Quasar host galaxies may have experienced a 
recent starburst, whose magnitude scales with the AGN luminosity 
\citep{Kauffmann2003MNRAS}.  This interpretation, however, is not supported
by the evidence in hand.  Based on the strength of the 11.3 $\mu$m PAH feature,
\cite{Zhang2016ApJ} find that PG quasars have similar star formation rates to 
``main-sequence'' star-forming galaxies of similar stellar mass.  
\cite{Husemann2014MNRAS} come to the same conclusion, for another quasar 
sample.  Our own analysis indicates that quasar hosts, in fact, have lower values 
of $q_\mathrm{PAH}$ compared with normal galaxies (Figure \ref{fig:dl07}(b)).  
In conjunction with the mild reduction of $q_\mathrm{PAH}$ with increasing AGN
luminosity (Figure \ref{fig:dl07agn}(b)), this supports the idea that PAHs
tend to be destroyed by the high-energy photons from the AGN
\citep{Smith2007ApJ,Sales2010ApJ,Wu2010ApJ}. It is unlikely that the reduction
of PAH strength stems from enhanced MIR extinction, as we find no clear evidence
for dust absorption features in the \irs\ spectra.  In this work, we will not
attempt to resolve the inherent ambiguity on the interpretation of the reduced
strength of PAH features in PG quasars (i.e. intrinsic reduction in star
formation rate or AGN destruction of PAHs).  Suffice it to say, there is no
compelling evidence that the star formation rate is enhanced in our sample of
PG quasars.  In support of this conclusion, we note that among the
six objects with the highest values of $U_\mathrm{min}$ and optical AGN
luminosity [$\mathrm{log}\,U_\mathrm{min}>1.2$ and
$\lambda L_\lambda$(5100 \AA) $>10^{45}\,\mathrm{erg s^{-1}}$],\footnote{We
visually check the SED fitting results and find that the $U_\mathrm{min}$ of
PG 1004$+$130, PG 1049$-$005, PG 1116$+$215, PG 1416$-$129, PG 1543$+$489, 
and PG 1704$+$608 are robustly constrained.} three (PG 1004$+$130, PG 1116$+$215,
and PG 1416$-$129) have host galaxies that resemble giant elliptical galaxies in \hst\ 
images (Y. Zhao et al. 2018, in preparation).  Furthermore, PG 1416$-$129 is found to 
be gas poor (Section \ref{ssec:ism}).  Alternatively, perhaps $U_\mathrm{min}$ 
is enhanced by old stars.  An evolved stellar population or enhanced stellar 
surface density may drive the radiation field to a very high intensity level (e.g., 
\citealt{MentuchCooper2012ApJ}), although \cite{Rowlands2015MNRAS} find 
that the cold dust temperature for a small sample of post-starburst galaxies is 
not unusually high compared to normal star-forming galaxies.  In Figure \ref{fig:dl07}, 
we also plot two subsamples of HRS galaxies, with early-type galaxy morphology 
and with stellar masses $>10^{10.5}\,M_\odot$.  The values of $U_\mathrm{min}$ 
of early-type galaxies, dominated by an old stellar population, tend to be higher than 
those of other galaxy samples but are not as high as in quasar host galaxies.

If the elevated radiation intensity of quasar hosts is not due to
an excess of young or old stars, it is likely that the ISM is
heated, at least in part, by the central AGN.  From the spatial extent of the
narrow-line region \citep{Greene2011ApJ,Husemann2014MNRAS}, we know that the
radiation field of the AGN can reach large distances into the host galaxy.  As
the narrow-line region gas is dusty (e.g., \citealt{Kraemer2011ApJ,
Wild2011MNRAS}), it is natural for the associated dust to experience enhanced
heating from the AGN.  Studies of the MIR spectra of quasars
also reveal that AGN-heated silicate emission likely comes from the narrow-line
region \citep{Schweitzer2008ApJ,Mor2009ApJ}.  Figure \ref{fig:dl07agn}(a)
shows that $U_\mathrm{min}$ increases with increasing AGN luminosity, although
the scatter is relatively large for individual objects.  We note that the
distribution of $\gamma$ (Figures \ref{fig:dl07}(c)) further supports the
notion that the dust in the host galaxies of quasars is exposed to a higher
intensity radiation field than star-forming galaxies, while the scatter in
Figure \ref{fig:dl07agn}(c) is large (see Appendix \ref{apd:rel} for caveats
on the interpretation of $\gamma$).  

The ability for the AGN or any sources other than young stars
to heat dust appreciably on galactic scale raises serious doubt for the common
practice of using the FIR luminosity to estimate star formation rates in AGN
host galaxies (e.g., \citealt{Leipski2014ApJ,Podigachoski2015AA,Westhues2016AJ,
Shimizu2017MNRAS}).  Our results suggest that attempts to remove the dust torus
contribution alone from the IR SED may not be enough to guarantee that the FIR
luminosity is uncontaminated by AGN emission.

\subsection{ISM Mass}\label{ssec:ism}

\begin{figure*}[ht]
\begin{center}
\begin{tabular}{c c}
\includegraphics[height=0.35\textheight]{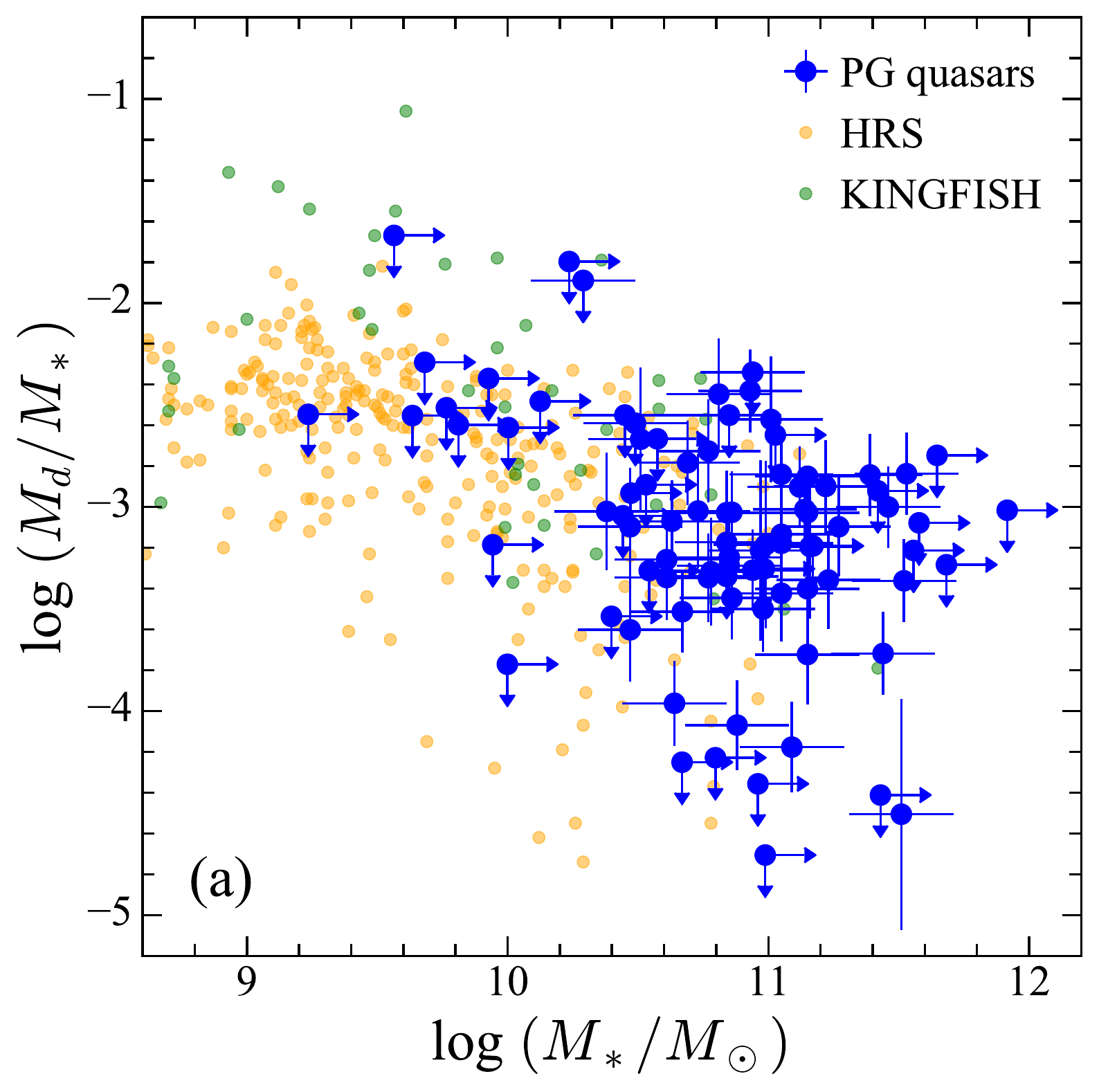} &
\includegraphics[height=0.35\textheight]{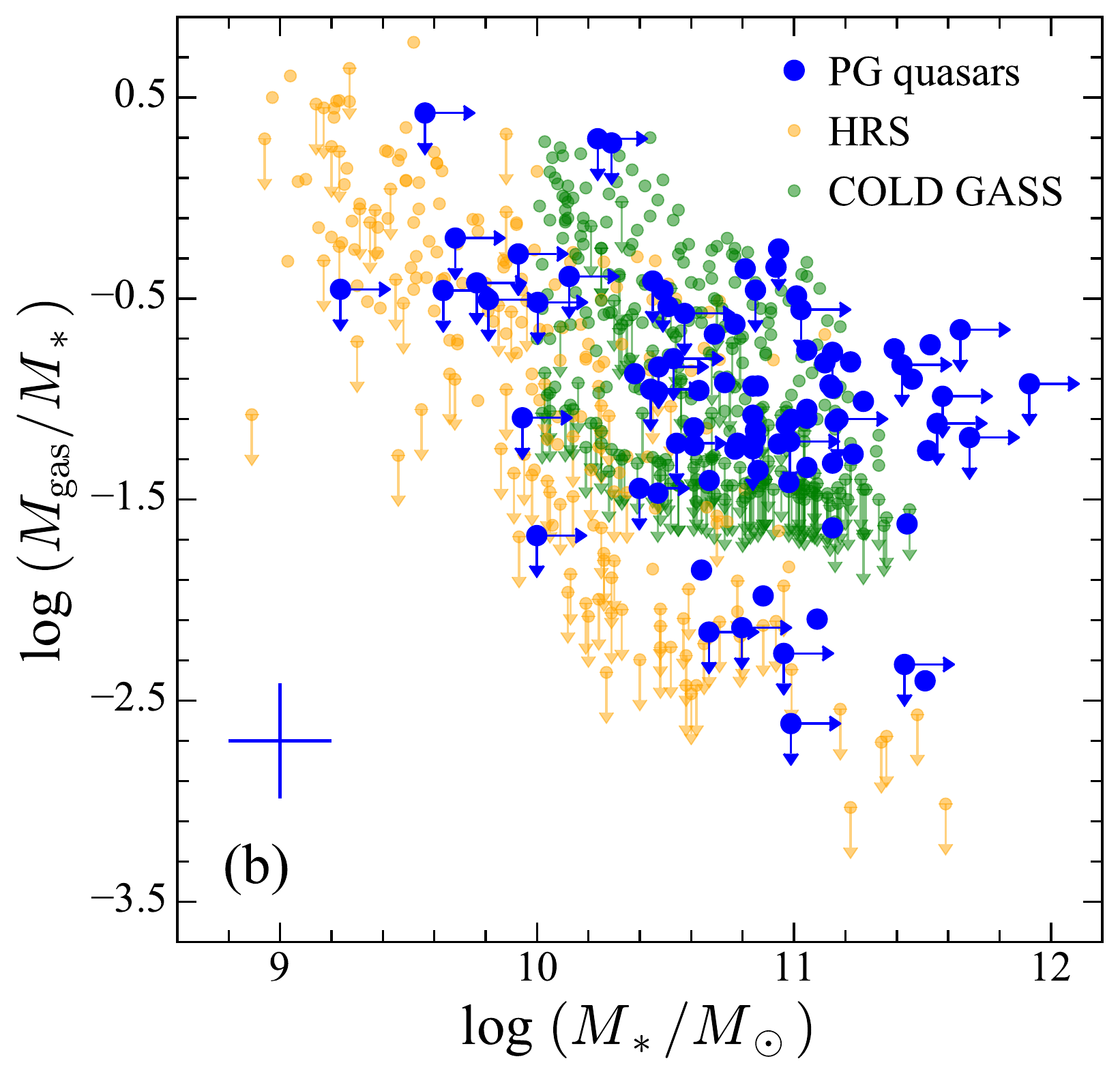}
\end{tabular}
\caption{(a) Dust-to-stellar mass ratio is plotted against the stellar mass of PG quasars 
(blue) and inactive nearby galaxies from HRS (orange; $M_*$: \citealt{Boselli2010PASP}, 
\citealt{Cortese2012AA}; $M_d$: \citealt{Ciesla2014AA}) and KINGFISH (green; $M_*$: 
\citealt{Kennicutt2011PASP}, \citealt{Skibba2011ApJ}; $M_d$: \citealt{Draine2007ApJ}).  
(b) Gas-to-stellar mass ratio is plotted against the stellar mass of PG quasars (blue) and 
inactive nearby galaxies from HRS (orange; $M_\mathrm{gas}$: \citealt{Boselli2014AA1}) 
and COLD GASS (green; \citealt{Saintonge2012ApJ}).  The typical 
uncertainty of the PG sample is shown in the lower-left corner.  Total stellar masses for 
the PG quasars come from \cite{Zhang2016ApJ}; lower limits on $M_*$ (and upper limits 
on $M_d/M_*$ and $M_{\rm gas}/M_*$; denoted with downward and rightward arrows) 
come from bulge masses estimated from the $M_\mathrm{BH}$--$M_\mathrm{bulge}$ 
relation.  Objects that only have upper limits on the dust masses are also included.
}
\label{fig:ismstar}
\end{center}
\end{figure*}

\subsubsection{Dust Mass}
\label{sssec:mdust}

One of the main goals of this study is to apply the DL07 model to our SED
fitting to measure dust masses for the PG quasars.  We derive dust masses in the
range $M_d \approx 10^{6.2} - 10^{8.7}\, M_\odot$ (Table \ref{tab:result}), with
a mean value of $10^{7.6 \pm 0.1}\, M_\odot$, properly accounting
for upper limits using the Kaplan--Meier product-limit estimator {\tt KMESTM} 
from {\tt ASURV} \citep{Feigelson1985ApJ, Lavalley1992ASPC}.

Figure \ref{fig:ismstar}(a) plots the distribution of dust-to-stellar mass
ratio as a function of stellar mass for PG quasars, comparing them with normal
galaxies from the HRS and KINGFISH samples.  As expected, the quasar hosts 
are all massive galaxies ($M_* \gtrsim 10^{10}\,M_\odot$), with the majority 
lying in a relatively narrow range of $M_* \approx 10^{11.0\pm0.5}\,M_\odot$.  
The dust-to-stellar mass ratio of PG quasars follows the general trend and
dispersion ($\sim 2$ dex) of normal galaxies.  For the objects with very low
$M_{d}/M_*$ (e.g., $\lesssim10^{-4}$), we visually check the fitting to confirm
their robustness.  Among these objects, four (PG 0804$+$761, PG 1416$-$129, 
PG 1501$+$106, and PG 1534$+$580) have highly secure dust masses because 
the detected FIR data cover the Rayleigh-Jeans tail of the SED.  In another four 
(PG 0026$+$129, PG 0049$+$171, PG 0923$+$201, and PG 2304$+$042), the 
peak of the DL07 model can be barely constrained, but, as shown in Appendix 
\ref{apd:rel}, the error on the dust mass for an individual object can hardly exceed 
0.3 dex, and hence these objects are still deficient in dust compared to the majority 
of the sample.  The dust mass for PG 1226$+$023 is very uncertain because 
the emission from its torus and synchrotron components are very strong. Another 
source of uncertainty comes from the host galaxy stellar mass or the bulge mass 
estimated from $M_\mathrm{BH}$, but it is unlikely that $M_*$ has been overestimated 
by more than 0.3 dex for these objects. The quoted uncertainty of the stellar mass 
is $\sim 0.2$ dex \citep{Zhang2016ApJ}, while the intrinsic scatter of the 
$M_\mathrm{BH}$--$M_\mathrm{bulge}$ relation is $\lesssim 0.3$ dex.

\subsubsection{Gas Mass}\label{sssec:mgas}

The dust and total gas masses are linked by

\begin{equation}\label{eq:gdm}
M_\mathrm{gas}= M_\mathrm{H\,I} + M_\mathrm{H_2} = M_{d} \, \delta_\mathrm{GDR},
\end{equation}

\noindent
where \gdr\ is the gas-to-dust ratio, which is a function of the gas-phase
metallicity \citep{Boselli2002AA,Draine2007ApJ,Leroy2011ApJ, Magdis2012ApJ}.
Assuming that the same fraction of condensable elements is locked in dust
as in the MW, and that the interstellar abundance of carbon and all of the 
heavier elements are proportional to the gas-phase oxygen abundance, 
\cite{Draine2007ApJ} suggest
$\delta_\mathrm{GDR} = 136 [\mathrm{(O/H)_{MW}}/\mathrm{(O/H)]}$,
where $\mathrm{(O/H)_{MW}}$ is the
oxygen abundance in the local MW and the factor of 136 is from MW
dust models \citep{Draine2007ApJ}, including helium and heavier elements.
\cite{Leroy2011ApJ} simultaneously constrain $\alpha_\mathrm{CO}$ and
\gdr\footnote{The heavier elements are already considered in
$\alpha_\mathrm{CO}$ and \gdr; therefore, the total gas mass derived from \gdr\
includes the contribution from heavier elements.} with spatially matched dust,
CO, and \hi\ maps of some local group galaxies.  They find a clear dependence of
\gdr\ on the gas-phase metallicity (\gzr\ relation), consistent with theoretical
expectation \citep{Draine2007ApJ}.  \cite{Magdis2012ApJ} recalibrate the \gzr\
relation of \cite{Leroy2011ApJ} to the empirical calibration of
\citet[hereafter, PP04]{Pettini2004MNRAS}, as follows,

\begin{equation}\label{eq:gdrz}
\mathrm{log}\,\delta_\mathrm{GDR} = (10.54 \pm 1.0) - (0.99 \pm 0.12)
[12 + \mathrm{log\,(O/H)}]_\mathrm{PP04},
\end{equation}

\noindent
where the scatter is 0.15 dex\footnote{It is worth mentioning that Equation
(\ref{eq:gdrz}) is, in fact, very close to the original relation of
\cite{Leroy2011ApJ},
$\mathrm{log}\,\delta_\mathrm{GDR} = (9.4 \pm 1.1) - (0.85 \pm 0.13)
[12 + \mathrm{log\,(O/H)}]$.}.
In the absence of a direct measurement of the metallicity of the galaxy, it can
be estimated from the stellar mass--metallicity (\mzr) relation (e.g., 
\citealt{Magdis2012ApJ,Santini2014AA,Berta2016AA}), if the stellar mass is
known.  Since our focus is on low-redshift objects, we adopt the \mzr\ relation
obtained for SDSS galaxies with the PP04 (N2) calibration, as given by
\cite{Kewley2008ApJ},\footnote{PP04 provide the calibration using \NII/\halpha\
(N2) and the ratio between \NII/\halpha\ and \OIII/\hbeta\ (O3N2) to obtain the
oxygen abundance.  We are not certain which one was adopted by
\cite{Magdis2012ApJ}, although N2 is preferable to match the \mzr\ relation they
adopt.  Nevertheless, the \mzr\ relation obtained with the two methods are very
similar ($\lesssim 0.05$ dex deviation; \citealt{Kewley2008ApJ}).}

\begin{equation}\label{eq:mz}
12+\mathrm{log\,(O/H)} = 23.9049 - 5.62784\,\mathrm{log}\,M_* + 0.645142\,
(\mathrm{log}\,M_*)^2 - 0.0235065\,(\mathrm{log}\,M_*)^3,
\end{equation}

\noindent
with residual scatter 0.09 dex.  As stressed by \cite{Berta2016AA}, it is
important to use the \gzr\ and \mzr\ relations self-consistently in terms of the
metallicity calibration.  The different calibrations can lead to a significant
systematic discrepancy for the \mzr\ relation in terms of both its shape and
scale (see \citealt{Kewley2008ApJ} for detailed discussions).  For example, the
\mzr\ relation obtained by \cite{Tremonti2004ApJ} from theoretical calibration
is $\gtrsim 0.4$ dex higher than the PP04 empirical calibration at
$M_* \approx 10^{11}\,M_\odot$, and the \mzr\ relation drops much steeper
toward lower $M_*$ with the former calibration than that with the latter one.

A number of recent works use the dust mass to estimate the total gas mass of
galaxies, with $\delta_\mathrm{GDR}$ estimated from combining the \mzr\ and
\gzr\ relations \citep{Santini2014AA,Magdis2012ApJ,Magdis2017AA}.  We should
bear in mind, however, that the metallicity provided by the \mzr\ relation
cannot be guaranteed to provide a proper metallicity that leads to an overall
correct \gzr, for two reasons.  First, the metallicity of a galaxy generally
decreases with increasing distance from the center \citep{Henry1999PASP}.  
When the overall metallicity is estimated from the \mzr\ relation, it is likely that
the metallicity is overestimated in the sense that the light from the inner part
of the galaxy, where the metallicity is high, dominates the observed spectrum.
Second, the \gdr\ from ``local'' relations (e.g., Equation (\ref{eq:gdrz})), can only
estimate the gas mass within the region of the galaxy that contains detectable
dust emission ($\lesssim 1.5 R_{25}$; \citealt{Ciesla2012AA,Dale2012ApJ,
Dale2017ApJ}).  However, the size of the dust disk is usually smaller than that 
of the \hi\ disk,\footnote{The molecular gas distribution is usually even less 
extended than the dust (e.g., \citealt{Bigiel2008AJ,Pappalardo2012AA}), so 
it mainly contributes to the total gas mass in the inner region of the galaxy.} 
unless the \hi\ distribution is truncated by environmental effects
\citep{Thomas2004MNRAS,Cortese2010AA,Cortese2012AA}.  \cite{Smith2016MNRAS}, in
fact, reveal extended dust emission out to $2R_{25}$, combining \herschel\ maps
of 110 HRS galaxies and reaching 10 times higher sensitivity than the map
tracing dust to $\sim 1.2R_{25}$.  \cite{Munoz2009ApJ} also find that the
dust-to-gas ratio drops faster than the metallicity gradient toward the
outskirts of the galaxy, perhaps a consequence of the detailed physics of the
evolution of dust grains \citep{Mattsson2012MNRAS}.  Therefore, the
aforementioned method tends to underestimate the total gas mass, mainly by
excluding the extended \hi\ gas.  This problem becomes critical when one
compares the gas mass estimated from dust with that directly measured from CO
and \hi\ observations.  In order to understand how serious the problem is and
provide a correction for \gdr, we study the HRS sample, for which 176 galaxies
have measurements of dust, CO, and \hi.

\begin{figure}
\centering
\includegraphics[width=0.9\textwidth]{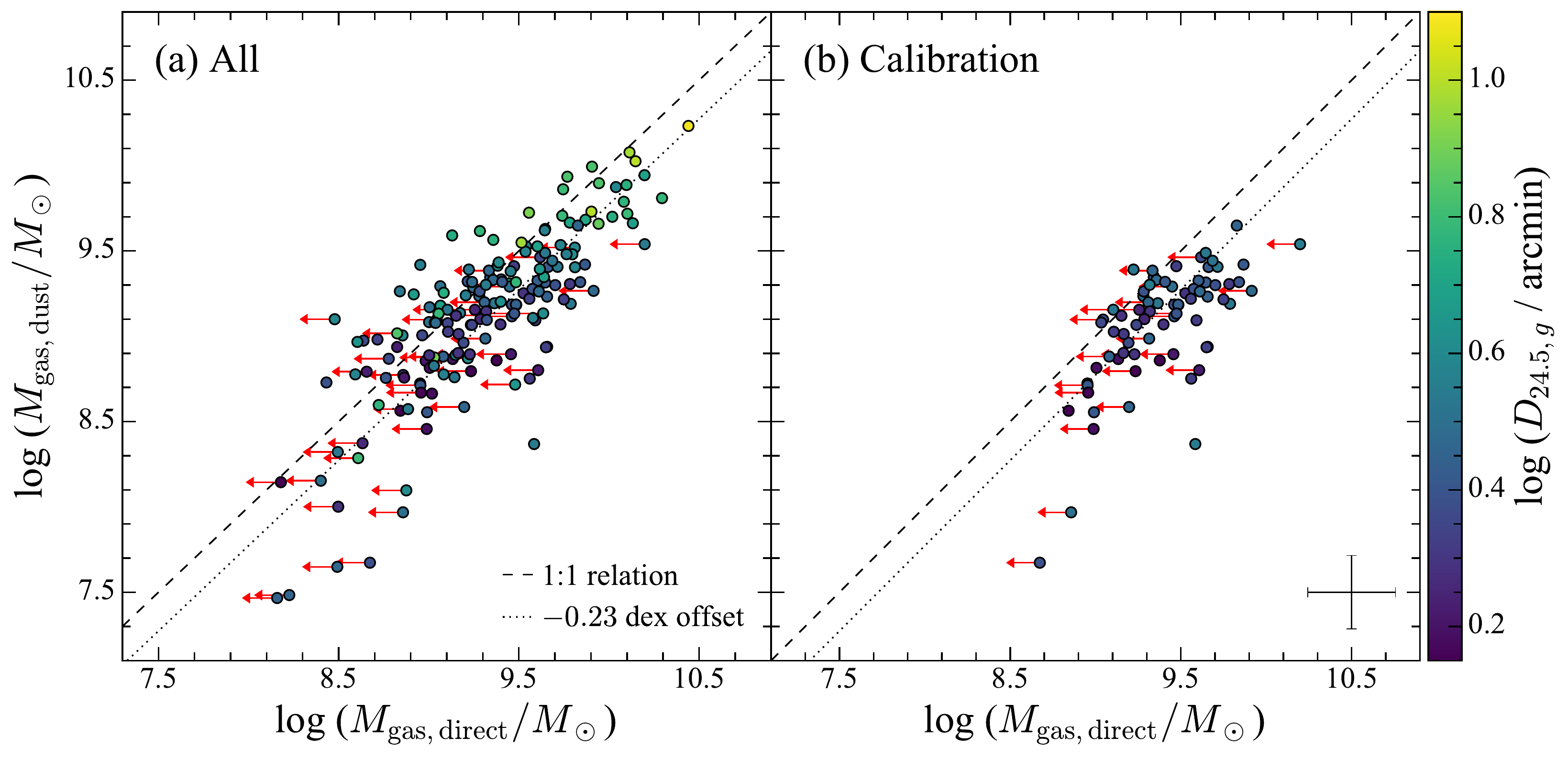}
\caption{Comparison of gas masses estimated from dust and direct measurements
for 176 HRS galaxies having dust, CO, and \hi\ observations.  The color code is
$D_{24.5, g}$, the diameter of the galaxy at a surface brightness of 24.5 $g$
$\mathrm{mag\,arcsec^{-2}}$.  (a) Comparison for the entire sample.  The mean 
deviation (y$-$x) is $-0.09 \pm 0.02$ dex; the $50^{+25}_{-25}$th percentile of 
the deviation is $-0.08^{+0.17}_{-0.19}$ dex.  (b) Comparison for the subsample 
of calibration galaxies with diameters $D_{24.5,g}< 3\farcm5$ and \hi\ deficiency 
def$_\mathrm{H\,I}<0.5$ (see the text for details). The mean deviation (y$-$x) is 
$-0.23 \pm 0.03$ dex; the $50^{+25}_{-25}$th percentile of the deviation is 
$-0.22^{+0.18}_{-0.11}$ dex.  The statistics are obtained using {\tt KMESTM}, 
which accounts for the upper limits.  The dashed line is the 1:1 correlation; the 
dotted line is 0.23 dex below the dashed line, showing the necessary correction 
to obtain the correct total gas mass.  The typical uncertainties of the gas masses 
are shown in the bottom-right corner.}
\label{fig:gasoff}
\end{figure}

\begin{figure*}
\begin{center}
\includegraphics[width=0.85\textwidth]{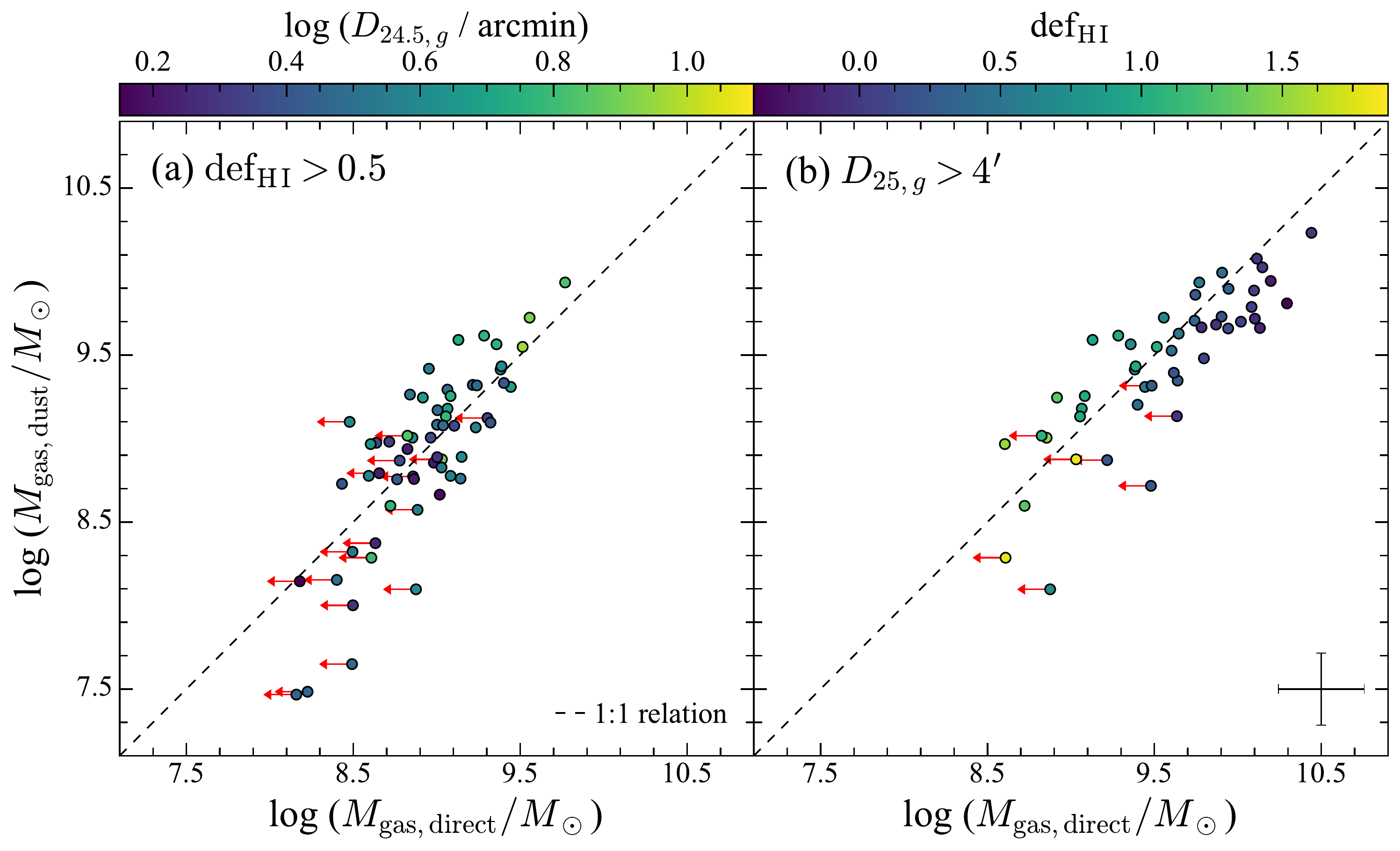}
\caption{(a) Comparison of total gas masses estimated from dust and
direct measurements for \hi-deficient HRS galaxies.  The color code is
$D_{24.5, g}$, the diameter of the galaxy at a surface brightness of 24.5 $g$
$\mathrm{mag\,arcsec^{-2}}$.  The dashed line is the 1:1 relation.  The mean
deviation (y$-$x) is 0.10$\pm$0.03 dex; the $50^{+25}_{-25}$th percentile of the
deviation is $0.11^{+0.17}_{-0.15}$ dex.  (b) Same comparison, but for the
subset of galaxies with $D_{24.5,g}>4\farcm0$.  The color code is the \hi\
deficiency of the galaxy.  The mean deviation (y$-$x) is $-0.04 \pm 0.04$ dex;
the $50^{+25}_{-25}$th percentile of the deviation is $-0.04^{+0.19}_{-0.17}$
dex.  The statistics are obtained using {\tt KMESTM}, which accounts for the
upper limits.  The typical uncertainties of the gas masses are show in the
bottom-right corner.}
\label{fig:hrsmatch}
\end{center}
\end{figure*}

We find that the total gas mass estimated by dust is on average 0.09$\pm$0.02
dex lower than that obtained from direct measurement, with the
$50^{+25}_{-25}$th percentile of the deviation $-0.08^{+0.17}_{-0.19}$ dex
(Figure \ref{fig:gasoff}(a)).  The deviation is apparent but not very
significant, because there is a considerable fraction of highly \hi-deficient
galaxies, whose \hi\ disks are truncated.  If we select only the subsample of
HRS galaxies with small angular size
($D_{24.5, g} < 3\farcm5$)\footnote{$D_{24.5, g}$ is the isophotal
diameter of the galaxy at a surface brightness of 24.5 $g$
$\mathrm{mag\,arcsec^{-2}}$.  $D_{24.5,g}$ is close to $D_{25} \equiv 2R_{25}$.
The typical beam size of the Arecibo telescope is $3\farcm5$.  We require the
extended \hi\ gas not to be missed due to the beam size.} and low
\hi\ deficiency\footnote{The \hi\ deficiency (def$_\mathrm{\hi}$) is defined as
the difference, on a logarithmic scale, between the observed \hi\ mass and the
value expected from an isolated galaxy given the same morphological type and
optical diameter \citep{Haynes1984AJ}.} (def$_\mathrm{\hi}<0.5$), the mean
deviation is $-0.23 \pm 0.03$ dex (dotted line in Figure \ref{fig:gasoff}(b)) with 
the $50^{+25}_{-25}$th percentile $-0.22^{+0.18}_{-0.11}$ dex, which is much 
more prominent than that for the entire sample.

To ascertain whether \gdr\ is accurately estimated for HRS galaxies within
$\sim R_{25}$ (comparable to the detectable dust emission region), we need to
compare dust-derived gas masses to the directly measured gas masses within
$\sim R_{25}$.  Gas-stripping processes, which truncate the large-scale \hi\
distribution to produce \hi-deficient galaxies, provide a natural tool in this
regard.  \cite{Cortese2010AA} find that the \hi\ disks of \hi-deficient galaxies
(def$_\mathrm{\hi}>0.5$) match well the sizes of their dust disks.  Figure
\ref{fig:hrsmatch}(a) demonstrates that the systematic deviation between the
gas masses obtained with the two methods is small for the subset of HRS galaxies
classified as \hi\ deficient (def$_\mathrm{\hi}>0.5$).  The $50^{+25}_{-25}$th
percentile of the gas mass deviation is $0.11^{+0.17}_{-0.15}$ dex.  Similarly,
when a galaxy is large enough to be well-resolved, the dust and \hi\ gas should
also be spatially better matched than those in unresolved galaxies.  In view of 
the $3\farcm5$ beam of Arecibo, Figure \ref{fig:hrsmatch}(b) isolates the subset
of HRS galaxies with $D_{24.5,g}> 4\farcm0$: the deviation,
$-0.04^{+0.19}_{-0.17}$ dex, essentially vanishes.  Therefore, we conclude that
dust masses can estimate total gas masses inside $\sim R_{25}$ with reasonably
good accuracy.  However, in order to consistently compare the dust-inferred gas
masses with the directly measured gas masses, including the extended \hi\ gas
(e.g., Figures \ref{fig:ismstar}(b) and \ref{fig:direct}(a)), we provide
an empirical correction to \gdr\ obtained from the stellar mass:

\begin{equation}\label{eq:gdrcorr}
\mathrm{log}\,\delta_\mathrm{GDR, total} = \mathrm{log}\,\delta_\mathrm{GDR}
+ (0.23 \pm 0.03),
\end{equation}

\noindent
where the 0.23 dex correction is determined from the mean offset of the
subsample of calibration galaxies with presumably intact \hi\ disk completely
measured (Figure \ref{fig:gasoff}(b)).  \cite{RemyRuyer2014AA} found a
correction factor of 1.55 (or 0.19 dex) for the \hi\ gas in dwarf galaxies, very
close to our value.  This supports the critical underlying assumption that the
radial profiles of the \hi\ gas are the same for spiral and dwarf galaxies
\citep{Wang2014MNRAS,Wang2016MNRAS}.  The \hi\ gas may be more extended in
early-type galaxies than in spirals, reflecting their possible accretion origin
\citep{Wang2016MNRAS}.  Therefore, it is possible, but by no means certain, that
our corrected \gdr\ may underestimate the total gas mass of PG quasars residing
in early-type galaxies.

We use the corrected \gdr\ (Equation (\ref{eq:gdrcorr})) to estimate the total gas
mass (Col. 14 of Table \ref{tab:result}) from the dust mass.  The uncertainty of
the \gdr\ is assumed to be 0.2 dex, dominated by the scatter of the scaling relations.
The gas masses of PG quasars span $\sim 10^{8.3}$--$10^{10.8}\,M_\odot$, 
with a mean value of $10^{9.7 \pm 0.1}\,M_\odot$, accounting for the upper 
limits.  \hst\ images reveal that many of the hosts of PG quasars are not early-type
galaxies (\citealt{Kim2008ApJ,Kim2017ApJS}; Y. Zhao et al. 2018, in preparation).  
As the host galaxies are very massive (Cols. 5 and 8 of Table \ref{tab:result}), 
the extended \hi\ gas is likely retained.  We find a median $\delta_\mathrm{GDR} \approx 124\pm6$, 
which, given the uncertainties of $\alpha_\mathrm{CO}$ and the dependence 
of \gdr\ on $M_*$, is generally consistent with the values reported by \cite{Draine2007ApJ}.  
The distribution of \gdr\ is very narrow because the \mzr\ relation (Equation
(\ref{eq:mz})) flattens at the high-$M_*$ end.  Therefore, we adopt the median
\gdr\ for objects without a stellar mass measurement.  Comparison of the gas
masses estimated from direct and indirect methods (Section \ref{sssec:cmpgas})
shows that our method is unbiased.

Figure \ref{fig:ismstar}(b) compares the gas-to-stellar mass ratio of PG
quasars with normal galaxies from HRS and star-forming galaxies from COLD GASS.
The total gas masses (including heavier elements) of the HRS and COLD GASS
galaxies are measured from direct CO and \hi\ observation.  It is clear that
most quasar host galaxies have as much gas as the gas-rich COLD GASS galaxies.
The typical gas mass fraction\footnote{We define the gas mass fraction as
$M_\mathrm{gas}/M_*$, in accordance with \cite{Saintonge2016MNRAS}.} of the
gas-rich quasar host galaxies is $M_\mathrm{gas}/M_* \approx 0.1$.  At the same
time, we note that nine ($\sim$ 10\%) of the quasar host galaxies show
$M_\mathrm{gas}/M_* \lesssim 0.01$, equivalent to the gas fraction of quenched galaxies,
$\sim1$ dex below the star-forming galaxy main sequence \citep{Saintonge2016MNRAS}.  
All of these objects have $M_d/M_* \lesssim 10^{-4}$.  These
objects genuinely lack cold dust and hence are truly deficient in cold ISM.
As discussed in Section \ref{ssec:feedback}, \hst\ images reveal that the hosts
of PG 0026$+$129, PG 0804$+$761, PG 0923$+$201, PG 1226$+$023, and PG 1416$-$129
are likely elliptical galaxies \citep{Kim2008ApJ}.  Although we caution that the
gas masses of these galaxies may be underestimated, it is unlikely that this can
be as large as 0.5 dex.  These galaxies should be gas poor, anyway.  According
to the evolutionary scenario \citep{Sanders1988ApJ}, the IR-luminous galaxies
triggered by gas-rich major mergers are presumably the progenitors of quasars.
\cite{Larson2016ApJ} report $M_\mathrm{H2}/M_* \approx 0.3-0.5$ for the
intermediate- to late-stage mergers with $M_* \approx 10^{10.8}\,M_\odot$.
Taken at face value, the molecular gas mass fraction of these starburst galaxies
is 3--5 times the total gas mass fraction of PG quasars.  Unfortunately, the
huge uncertainty of $\alpha_\mathrm{CO}$ makes the comparison insecure.
\cite{Larson2016ApJ}, following \cite{Scoville2016ApJ}, adopt
$\alpha_\mathrm{CO}$ = 6.5 \acounit, but \cite{Downes1998ApJ} advocate
$\alpha_\mathrm{CO}$ = 0.8 \acounit for starburst systems.  In view of the
nearly 1 dex uncertainty in $\alpha_\mathrm{CO}$, future comparisons using
dust-based gas masses may be more robust.

\subsubsection{Comparison with Other Methods}
\label{sssec:cmpgas}
\begin{figure}
\begin{center}
\begin{tabular}{c c c}
\includegraphics[height=0.28\textheight]{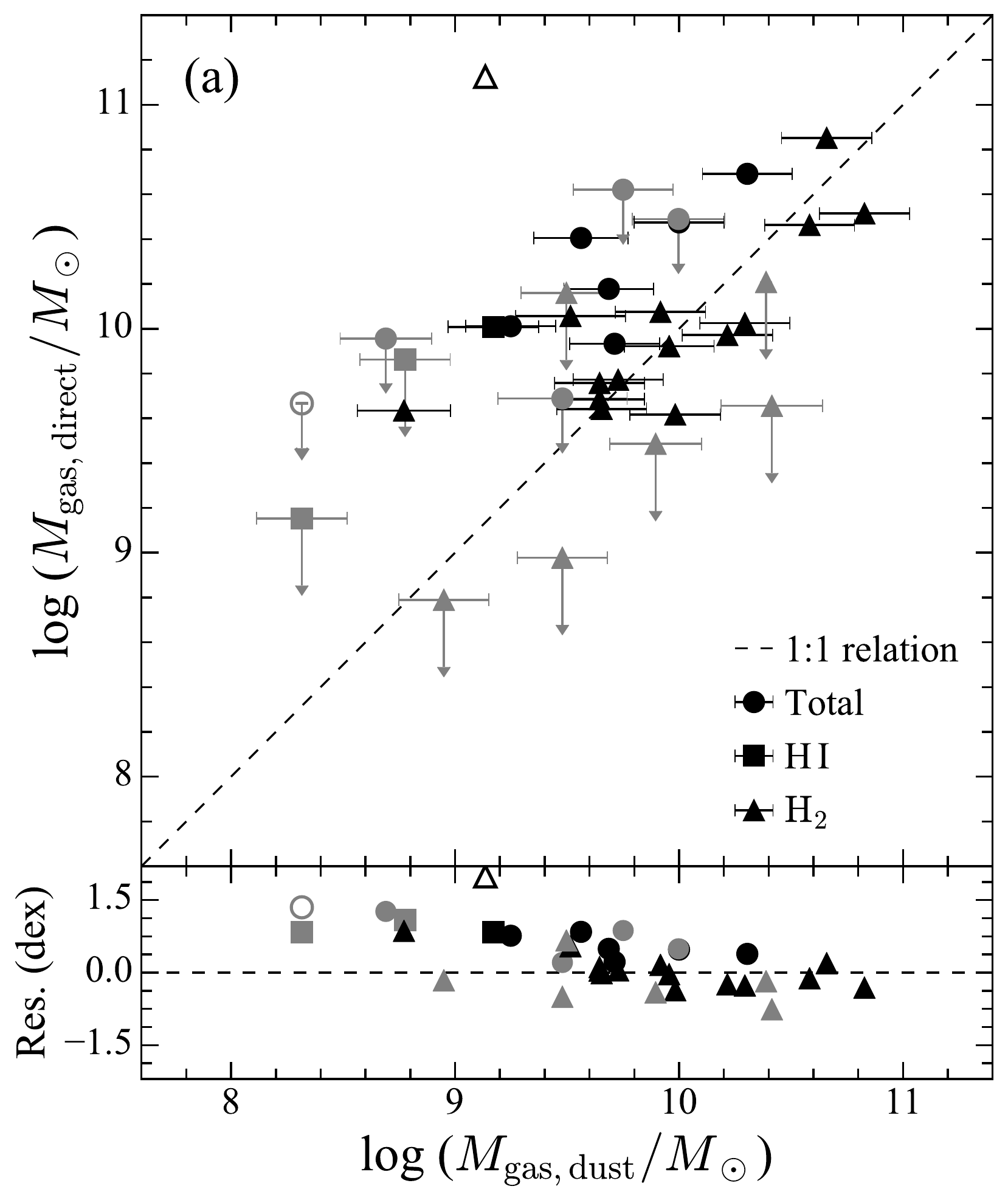} &
\includegraphics[height=0.28\textheight]{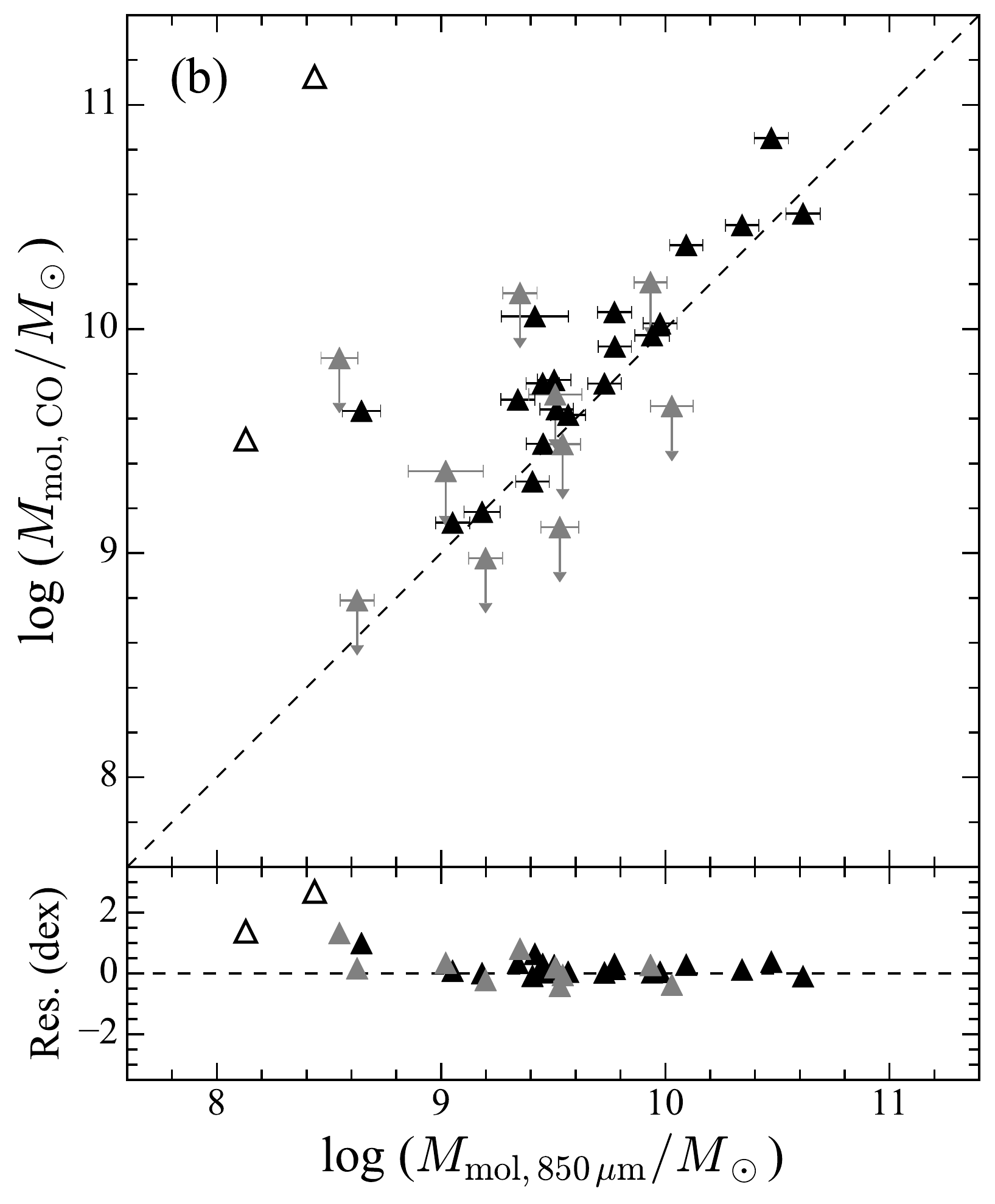} &
\includegraphics[height=0.28\textheight]{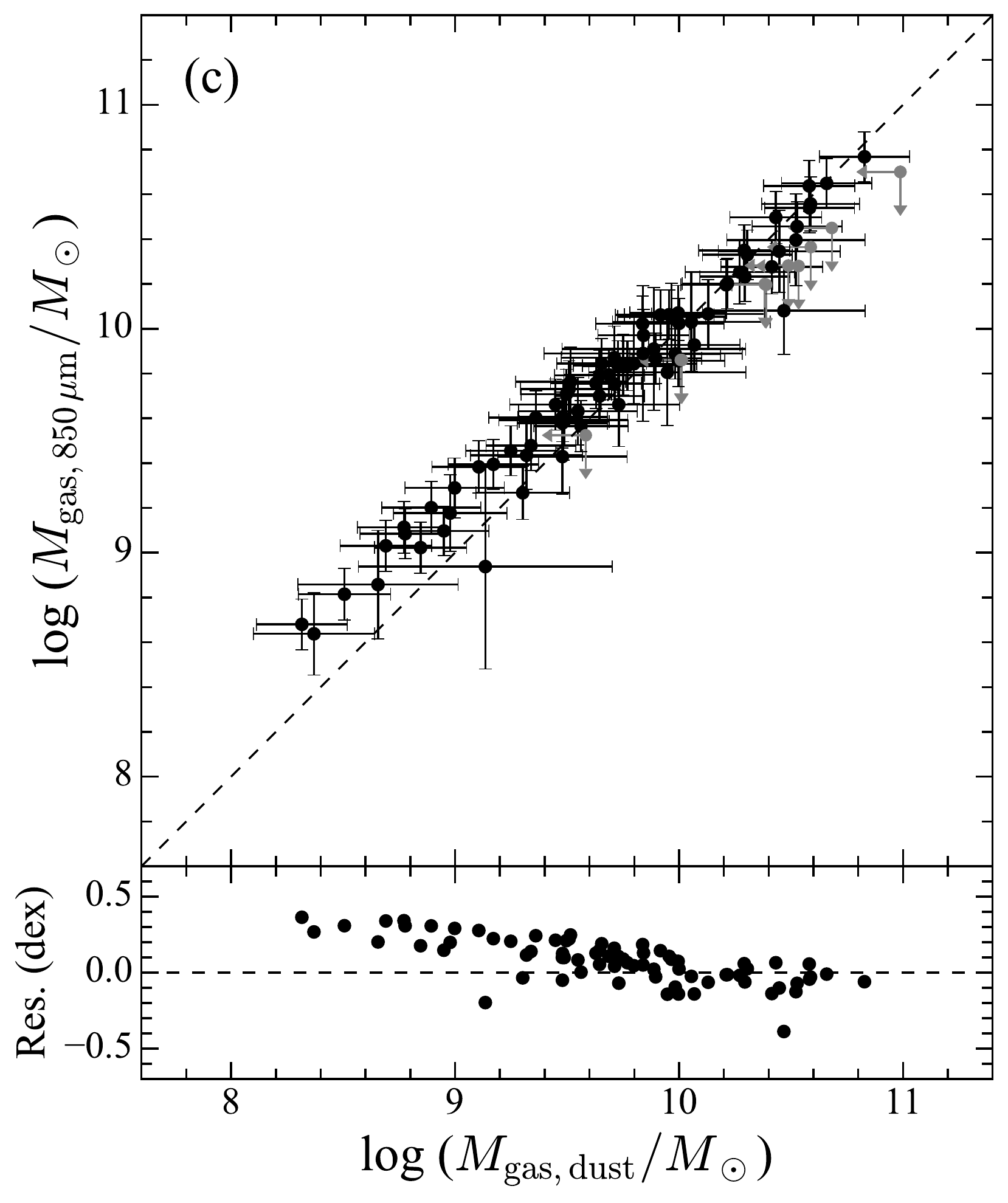}
\end{tabular}
\caption{(a) Comparison of the directly measured gas masses with the
dust-converted gas masses of PG quasars with available measurements of 
both CO and \hi\ (circles), only \hi\ (squares), or only CO (triangles).  Comparing 
to the dust-derived total gas mass, the deviations (y$-$x) of the directly measured
total (\hi+\MH), \hi, and \MH\ gas masses are on average 0.47$\pm$0.08 dex, 
0.43$\pm$0.10 dex, and $-0.22\pm$0.08 dex, respectively, with the $50^{+25}_{-25}$th
percentile being $0.42^{+0.08}_{-0.20}$ dex, $0.39^{+0.33}_{-0.29}$ dex, and
$-0.24^{+0.28}_{-0.14}$.  (b) Comparison of molecular gas masses measured
from CO observations with those derived from $L_{\nu_\mathrm{850\,\mu m}}$.
The gray symbols are the objects for which only CO upper limits are available.  
The mean deviation (y$-$x) is 0.08$\pm$0.06; the $50^{+25}_{-25}$th percentile 
of the deviation is $0.05^{+0.22}_{-0.14}$ dex. The statistics above properly
consider upper limits.  PG 0003$+$199  and PG 1226$+$023 (empty symbols), 
which show large deviations from the one-to-one relation line in (a) and (b), have 
only marginal CO detections, and hence their CO-based molecular gas masses 
may have large errors; we exclude these two objects from the statistics.  (c) Total 
gas masses estimated from the specific luminosity at $850\,\mu m$ are, on average, 
consistent with our total gas masses converted from dust masses: the median 
deviation (y$-$x) is 0.09 dex with an RMS of 0.14 dex.  Since the upper limits are 
for both axes, we do not consider the upper limits here.
}
\end{center}
\label{fig:direct}
\end{figure}

A subset of the PG quasars have published CO(1--0) (32 objects) and
\hi\ (13 objects) observations (Table \ref{tab:gas}).  We
calculate molecular and atomic gas masses following

\begin{equation}\label{eq:h2}
M_\mathrm{H2}\,(M_\odot) = (2.45 \times 10^3)\,\alpha_\mathrm{CO}\,
S_\mathrm{CO}\Delta v\,D_{L}^2\,(1+z)^{-1},
\end{equation}
\begin{equation}\label{eq:hi}
M_\mathrm{\hi}\,(M_\odot) = (3.204 \times 10^5)\,D_{L}^2\,
S_\mathrm{\hi}\Delta v,
\end{equation}

\noindent
where $S_\mathrm{CO}\Delta v$ and $S_\mathrm{\hi}\Delta v$ are the integrated
line fluxes (Jy km s$^{-1}$), $D_{L}$ is the luminosity distance (Mpc), and $z$
is the redshift.  The conversion factor $\alpha_\mathrm{CO}$ is still uncertain
for quasar host galaxies.  For consistency with the literature
\citep{Solomon1997ApJ,Evans2001AJ}, we adopt
$\alpha_\mathrm{CO}=4.3\,M_\odot\,\mathrm{(K\,km\,s^{-1}\,pc^2)^{-1}}$, the
typical value of the MW disk \citep{Bolatto2013ARAA}.  Helium and heavier
elements are included in the gas masses\footnote{For molecular gas, the heavy
element fraction is included in $\alpha_\mathrm{CO}$, while for atomic gas we
multiply a factor 1.36 to the atomic hydrogen mass in Equation (\ref{eq:hi}).}.

Figure \ref{fig:direct}(a) compares the gas masses estimated from the dust
mass ($M_\mathrm{gas,dust}$) with those derived from direct observations
($M_\mathrm{gas,direct}$).  Since the number of objects with both CO and \hi\
measurements is limited, we also plot separately objects with either CO or \hi\
data only.  For the objects with both CO and \hi\ measurements, the
$50^{+25}_{-25}$th percentile of the deviation between measured  and
dust-inferred total gas masses is $0.42^{+0.08}_{-0.20}$ dex.  The excursion,
however, is driven almost entirely by the \hi\ gas. Replacing the directly
measured total gas masses with \hi\ gas masses alone, the deviation distribution
becomes $0.39^{+0.33}_{-0.29}$ dex.  By contrast, the deviation of CO-based 
\MH\ gas masses is $-0.24^{+0.28}_{-0.14}$ dex.

What is responsible for the discrepancy with the \hi\ masses?  We inspect 
the host galaxy morphology and environment of the 13 PG quasars with \hi\
measurements using optical images from \hst\ (\citealt{Kim2008ApJ,Kim2017ApJS}; 
Y. Zhao et al. 2018, in preparation) and SDSS.  As documented in the notes 
of Table \ref{tab:gas}, all the eight objects with directly measured gas masses 
$>0.3$ dex higher than the dust-derived gas masses appear to be disturbed 
systems in various stages of merging.  The \hi\ line profiles of most of these 
objects show broad and/or asymmetric features, indicating that the target \hi\ 
gas suffers confusion and/or dynamical disturbance.  The remaining five objects 
are all likely isolated.  We speculate, but cannot prove, that the \hi\ in these 
merger systems may be exceptionally extended (e.g., \citealt{vanGorkom1996AJ,Gereb2016MNRAS}).
For objects residing in dense environments (e.g., PG 0007$+$106 and
PG 1119$+$120), the reported \hi\ detections may be significantly overestimated
by contamination from neighboring companions.  More detailed \hi\ observations
with higher angular resolution with interferometers will be crucial to reveal
the true \hi\ gas masses of quasar host galaxies.

\cite{Scoville2014ApJ, Scoville2016ApJ} proposed a method to estimate
molecular gas masses ($M_\mathrm{mol}$) from the specific luminosity at
rest-frame $850\,\mu m$, $L_\mathrm{\nu,850\,\mu m}$.  The conversion factor,

\begin{equation}\label{eq:scov}
\alpha_\mathrm{850\,\mu m} =
\frac{L_\mathrm{\nu, 850\,\mu m}}{M_\mathrm{mol}}
\approx 1.1 \times 10^{20} \, \mathrm{erg \, s^{-1}\, Hz^{-1}} \, M_\odot^{-1},
\end{equation}

\noindent
is empirically calibrated using low-redshift star-forming and ultraluminous IR
galaxies, as well as $z \approx 2$ submillimeter galaxies.
\cite{Scoville2016ApJ} used
$\alpha_\mathrm{CO} = 6.5\,M_\odot\,\mathrm{(K\,km\,s^{-1}\,pc^2)^{-1}}$
to calculate the molecular gas masses of the calibration galaxies.  For
consistency with our convention, we multiply the original value of
$\alpha_\mathrm{850\,\mu m}$ by a factor 1.5.  We obtain
$L_\mathrm{\nu,850\,\mu m}$ from extrapolation of the best-fit SED model. Figure
\ref{fig:direct}(b) shows that the molecular gas masses estimated from
$L_\mathrm{\nu,850\,\mu m}$ agree quite well with those derived directly from
CO observations.  The mean deviation is 0.08$\pm$0.06 dex with the
$50^{+25}_{-25}$th percentile of the deviation $0.05^{+0.22}_{-0.14}$ dex. The
deviation can easily be explained by the fact that we may have overestimated the
molecular gas masses using the $\alpha_\mathrm{CO}$ value of the MW, which is
much less massive than the quasar host galaxies. \cite{Leroy2011ApJ} find that
$\alpha_\mathrm{CO}$ is inversely correlated with galaxy stellar mass.  In
addition, the incompleteness of the currently compiled sample may also bias the
statistics.  Note that the two most extreme outliers, PG 0003$+$199 and
PG 1226$+$023, were only marginally detected.  We omit these two objects in all
statistics.

\cite{Hughes2017MNRAS} recently provide another relation between
$L_\mathrm{\nu,850\,\mu m}$ and the total gas mass,

\begin{equation}\label{eq:hug}
\mathrm{log}\,M_\mathrm{gas,850\,\mu m} = (0.84 \pm 0.02)\,\mathrm{log}\,
L_{\nu,850\,\mu m} - (14.95 \pm 0.54).
\end{equation}

\noindent
They assume
$\alpha_\mathrm{CO} = 4.6\,M_\odot\,\mathrm{(K\,km\,s^{-1}\,pc^2)^{-1}}$, close
enough to our choice.  We use Equation (\ref{eq:hug}) to estimate the total gas
masses of PG quasars and compare them with the dust-derived gas masses (Figure
\ref{fig:direct}(c)).  The two methods are also closely consistent with each other, 
with median deviation 0.09$\pm$0.14 dex. The slight, systematic trend seen in 
Figure \ref{fig:direct}(c) stems from the sublinear slope of Equation (\ref{eq:hug}).

To summarize: the gas masses of PG quasars estimated indirectly from dust
masses and from the 850 $\mu$m specific luminosity are consistent with each
other, as well as with the \MH\ masses directly measured from CO observations.
Dust-inferred gas masses systematically underestimate directly measured \hi\
gas masses, which may suffer from confusion from the low angular resolution of 
the existing \hi\ observations.

\input{tab6.tex}

\section{Discussion}\label{sec:discuss}

\subsection{Dust Masses from the Photometric SED Alone}

\begin{figure}
\centering
\includegraphics[width=0.45\textwidth]{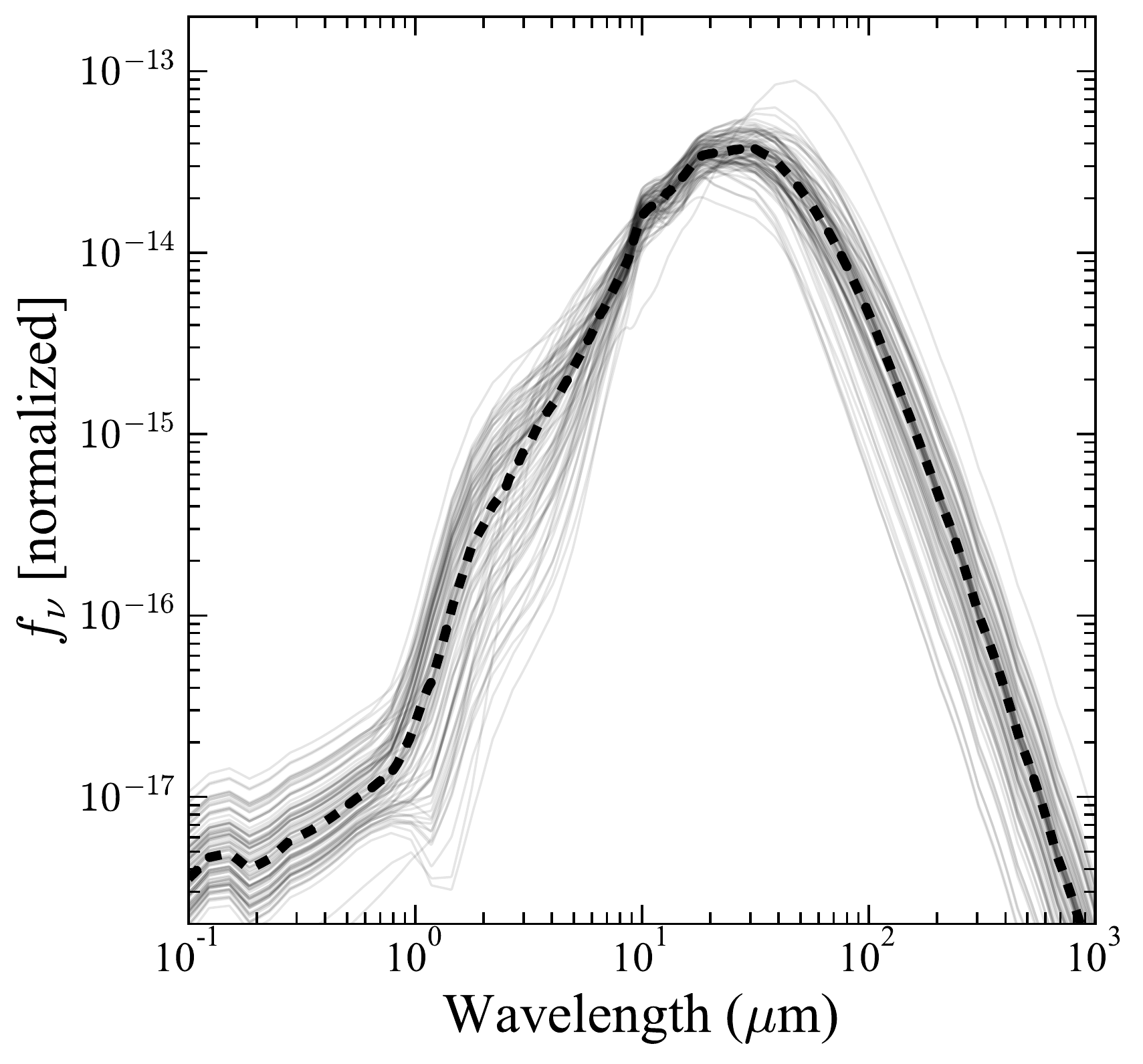}
\caption{Best-fit CLUMPY models for the PG quasars (light gray lines) and
their median value (thick dashed line), plotted on a scale with their integrated
fluxes normalized to 1.  Objects whose dust mass was fixed to estimate its upper
limit are omitted.  We use the median CLUMPY template to fit the photometric
SEDs.}
\label{fig:clumed}
\end{figure}

The \irs\ spectra play a significant role in constraining the AGN dust torus
component in our SED fits.  However, not all AGNs have  \spitzer\ \irs\
observations.  Thus, it is crucial to understand how well one can measure the
dust mass using only photometric data.  Over the wavelength range of \irs, the
\wise\ \w3\ and \w4\ bands\footnote{The entire PG sample also has \mips\ 24
\micron\ data; however, the 24 $\mu$m band is very close to the \wise\ \w4\ band
and thus does not provide additional constraints.} alone cannot fully constrain
the CLUMPY model.  We reduce the flexibility of the model by fixing it to the
``median'' CLUMPY template, which we calculate from the entire set of best-fit
torus models for the PG sample obtained from the full SED fits (Figure
\ref{fig:clumed}).

Using the median CLUMPY template (only varying its amplitude) along with the
rest of the components (BC03, BB, and DL07), we fit the photometric SEDs
(typically 13 bands) with six physical parameters for radio-quiet objects and with 
eight physical parameters for radio-loud objects.  We fix $\gamma = 0.03$ and
$q_\mathrm{PAH} = 0.47$. Since $\gamma$ is highly degenerate with the torus
component, if allowed to be free, it will always try to fit the mismatch between
the torus model and the data.  As long as $\gamma$ is small (e.g., $<0.1$), it
will not bias the fitting results.  Since the photometric data do not have
sufficient coverage to be particularly sensitive to PAH features,
$q_\mathrm{PAH}$ does not affect the fitting results with any fixed value.

\begin{figure}[h]
\centering
\begin{tabular}{c c c}
\includegraphics[height=0.23\textheight]{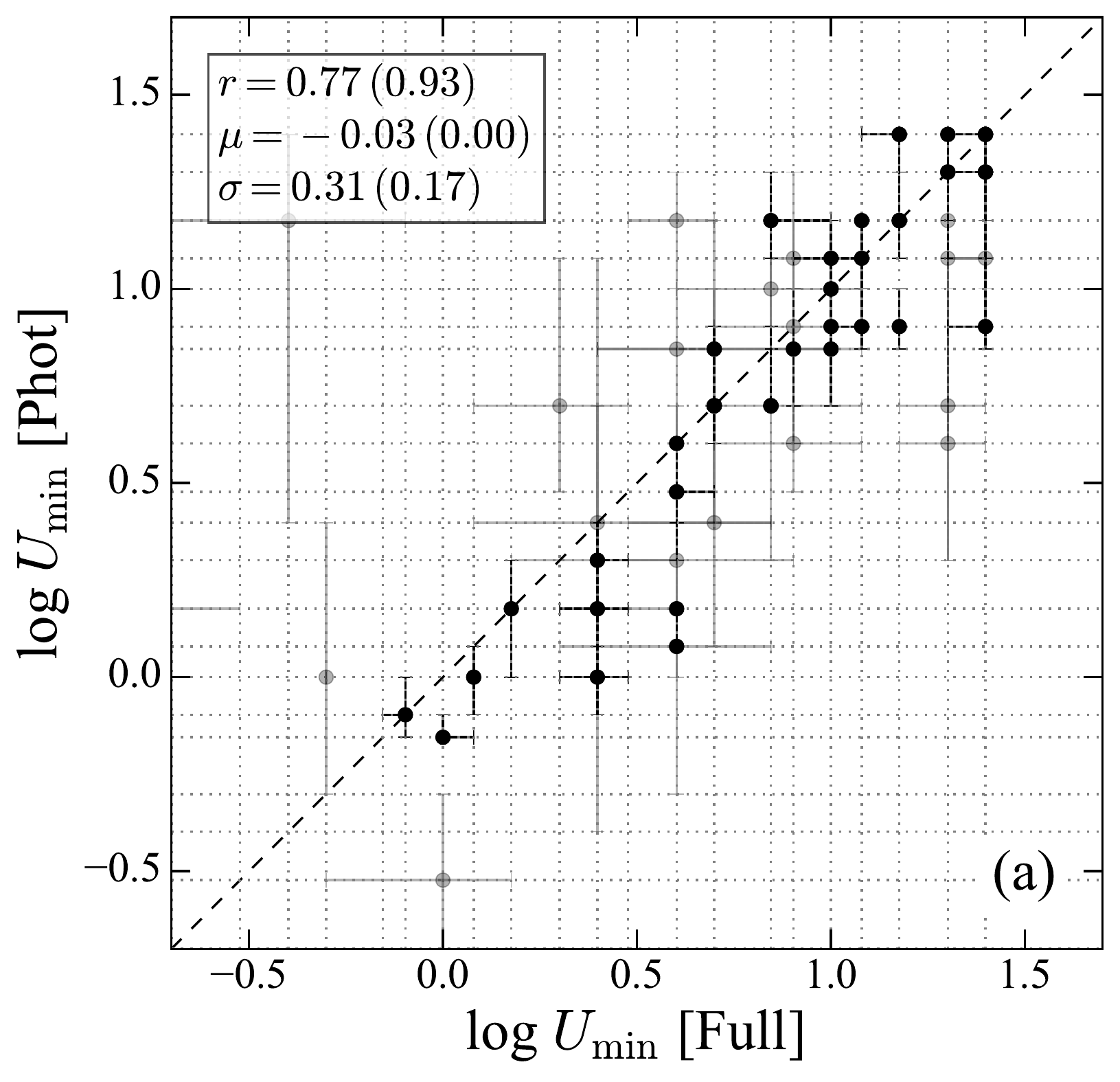} &
\includegraphics[height=0.23\textheight]{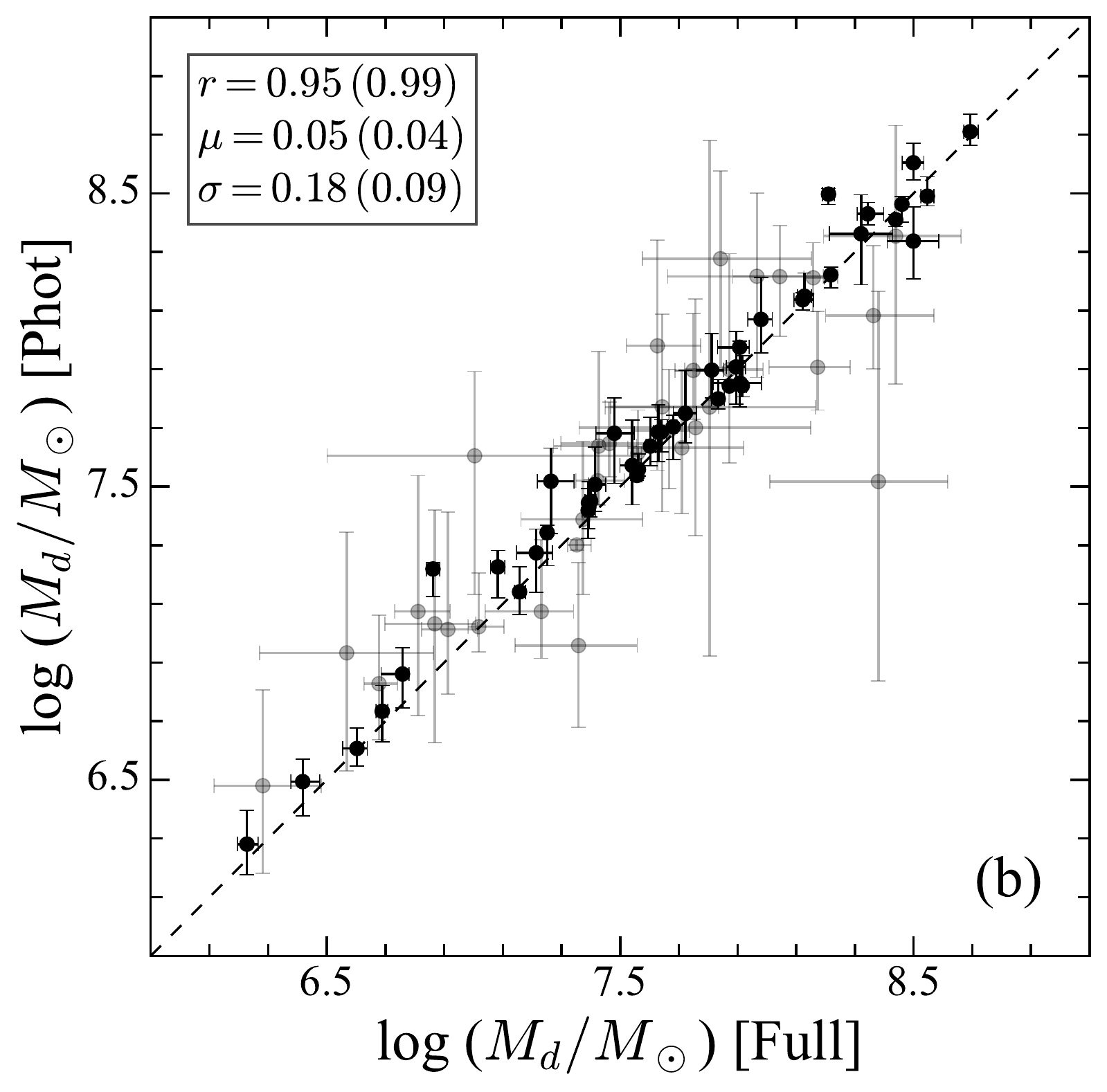} &
\includegraphics[height=0.23\textheight]{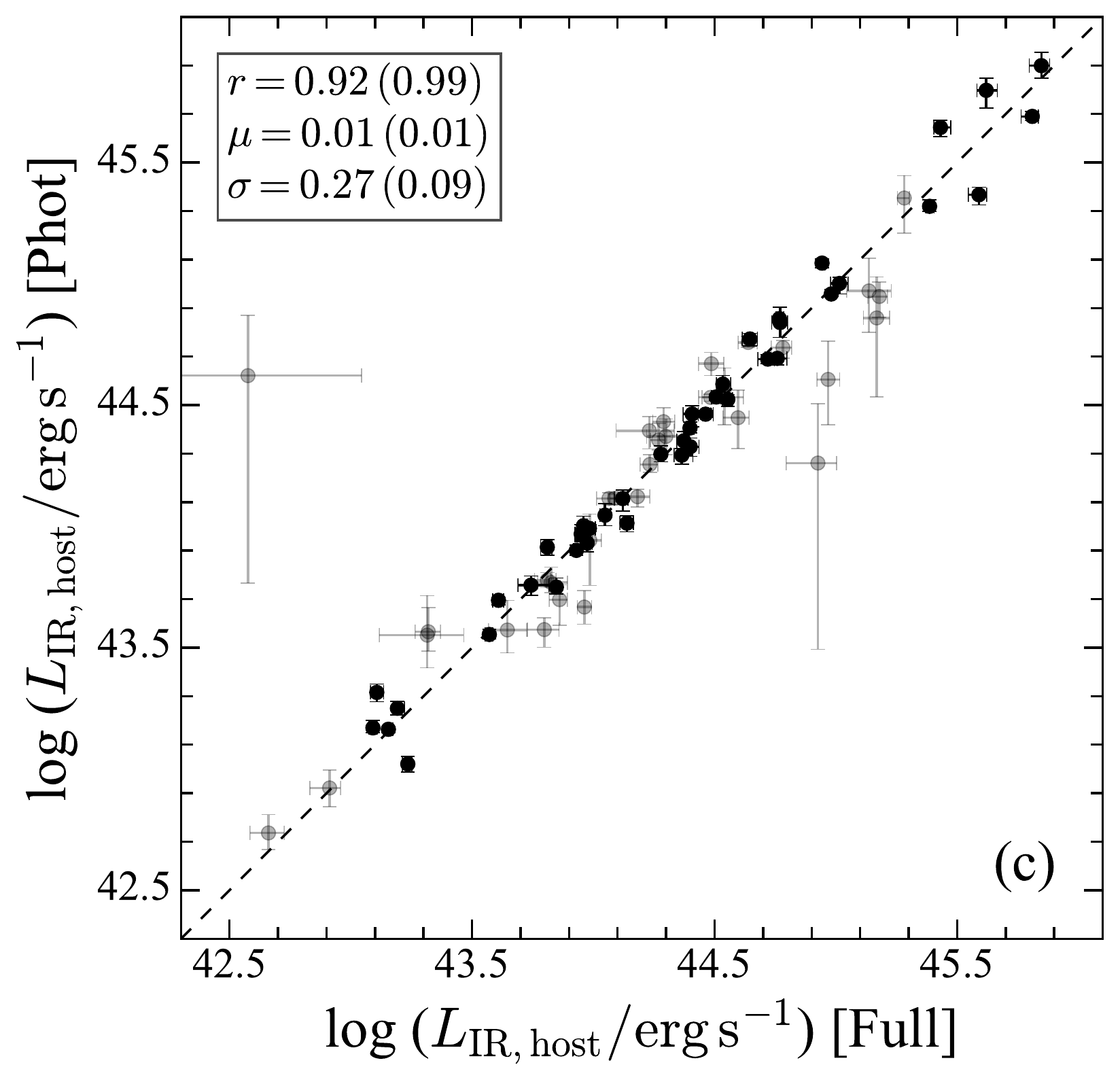}
\end{tabular}
\caption{Comparison of fitting results for the full SED, including the IRS
spectra, versus fitting only the photometric data, for the derived DL07
parameters (a) $U_\mathrm{min}$, (b) $M_d$, and (c) IR (8--1000 \micron) 
luminosity $L_{\rm IR, host}$.  Since $U_\mathrm{min}$ is a discrete parameter, 
the results are located on the dashed grids and sometimes overlap with each 
other; the errors are not resolvable if they are smaller than the grid size.  The 
black points are objects with FIR data good enough to reliably constrain the 
DL07 model, while the gray points are the remaining objects with less robust 
fits.  Objects with only upper limits on dust mass are omitted.  The dashed 
line is the one-to-one relation.  The upper-left corner of each panel shows 
the Pearson correlation coefficient ($r$) and the median ($\mu$) and standard 
deviation ($\sigma$) of the deviation from the linear relation.  The first set 
of values is for the entire sample; the values for the most robust subsample 
(black points) are given in parentheses.}
\label{fig:phot}
\end{figure}

We compare the best-fit values of $U_\mathrm{min}$, $M_{d}$, and
$L_\mathrm{IR, host}$ (the integrated IR luminosity over 8--1000 \micron)
obtained from the full SED and those by using the photometric SED alone 
(Figure \ref{fig:phot}).  The three quantities measured using the two methods 
all follow the one-to-one relation reasonably well and do not show systematic
deviations.  In particular, the objects with the most robust FIR measurements
(black points) have a Pearson's correlation coefficient  $r>0.9$.  The relatively 
large scatter mainly comes from objects whose FIR SEDs barely constrain 
the peak of the cold dust emission.  These objects, however, have correspondingly 
large and properly assigned uncertainties.  The scatter in $U_\mathrm{min}$ 
is relatively large.  The fits of the photometric SEDs seem to systematically 
underestimate $U_\mathrm{min}$,  especially for low values of $U_\mathrm{min}$, 
compared to fits of the full SEDs, mainly due to the mismatch between the simple 
torus template and the MIR data.  The dust masses themselves are not impacted.  
The DL07 model is insensitive to the mismatch of the MIR model as long as the 
FIR data constrain the cold dust emission well.

\subsection{Comparison with the MBB Model}
\label{ssec:mbb}

\begin{figure*}
\begin{center}
\begin{tabular}{c c}
\includegraphics[height=0.3\textheight]{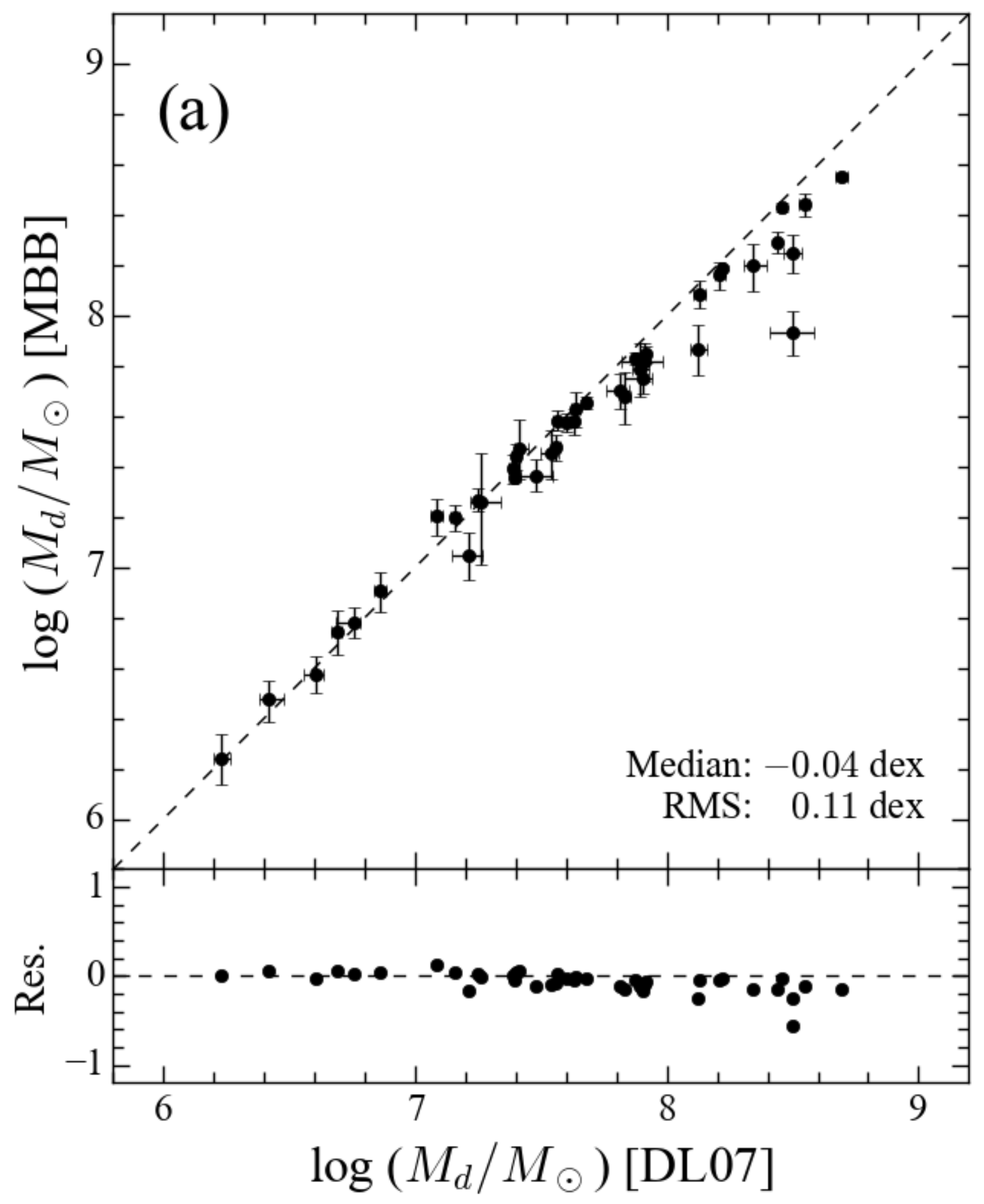} &
\includegraphics[height=0.3\textheight]{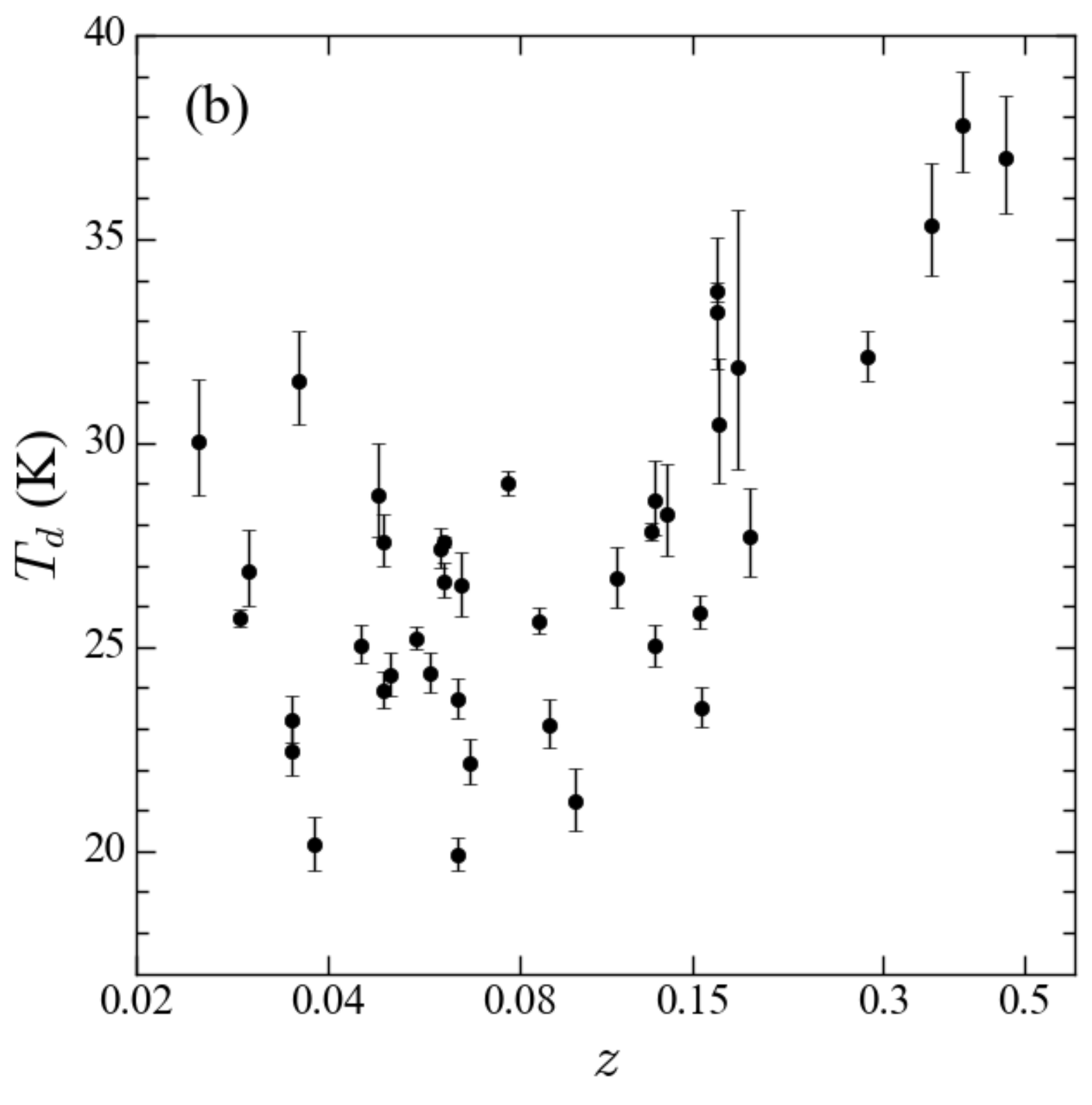}
\end{tabular}
\caption{(a) Comparison of dust masses derived from the MBB model with
those from full SED fitting with the DL07 model.  We only compare the 41
radio-quiet objects with at least four \herschel\ bands detected.  The MBB model
is only applied to fit the \herschel\ data, excluding the 70 \micron\ band to
avoid strong contamination from warm dust.  For the MBB model, we adopt the dust
absorption cross-section recommended by \cite{Bianchi2013AA}.  $M_d$[MBB] is
offset by $-$0.04$\pm$0.11 dex relative to $M_d$[DL07]; the deviation is
systematically larger toward higher dust masses.  (b) Distribution of dust
temperature versus redshift for the objects fitted with the MBB model.  The
apparent trend of $T_d$ rising for $z \approx 0.1-0.5$ may be due to the bias of
warm dust (see text).}
\end{center}
\label{fig:pgmbb1}
\end{figure*}

Many works use the MBB model to fit the FIR SED to estimate the dust mass and
temperature (e.g., \citealt{U2012ApJS,Magdis2013AA,Magnelli2014AA}), assuming
that the FIR emission comes from dust grains with the same size and temperature:

\begin{equation}
f_\mathrm{\nu, MBB} = \frac{(1+z)^2\,M_{d}\,\kappa_\mathrm{abs}\,
B_\nu(T_d)}{D_{L}^2},
\end{equation}

\noindent
where $f_\mathrm{\nu, MBB}$ is the rest-frame flux density, $D_L$ is the
luminosity distance, $z$ is the redshift, and $B_\nu(T_d)$ is the Planck function
with dust temperature $T_d$. The grain absorption cross-section per unit mass
$\kappa_\mathrm{abs}$ is usually assumed to be 

\begin{equation}
\kappa_\mathrm{abs} = \kappa_\mathrm{abs}(\lambda_0)\,
\left(\frac{\lambda_0}{\lambda}\right)^\beta,
\end{equation}

\noindent
where $\kappa_\mathrm{abs}(\lambda_0)$ is the absorption cross-section at the
given wavelength $\lambda_0$, calculated from MW dust models.
\cite{Bianchi2013AA} uses a sample of local star-forming galaxies to argue that
MBB fits can provide dust masses within $\lesssim10$\% of those derived from
the DL07 model, as long as $\kappa_\mathrm{abs}$ and $\beta$ are consistently
chosen.

In order to see whether this much simpler method also works for AGN host
galaxies, we use the MBB model to fit the \herschel\ data of PG quasars and
compare the dust masses measured from the DL07 model.  We limit this test 
to the subset of 41 objects with detections in four or more \herschel\ bands, 
with at least one detected on the Rayleigh-Jeans tail (see Appendix \ref{apd:mbb}). 
Following \cite{Bianchi2013AA}, we adopt
$\kappa_\mathrm{abs}(250\,\mathrm{\mu m}) = 4.0~\mathrm{cm^2\,g^{-1}}$ and fix
$\beta=2.08$.  The median deviation between the dust masses derived from the MBB
model and those from the DL07 model is $-0.04\pm0.11$ dex (Figure
\ref{fig:pgmbb1}(a)).  The dust temperatures cluster around $\sim 25$ K,
although $T_{d}$ seems to rise with increasing redshift for $z \approx 0.1-0.5$
(Figure \ref{fig:pgmbb1}(b)).  As discussed in Appendix \ref{apd:mbb}, when
redshift increases (especially $z>0.1$), warm dust emission increasingly
affects the \herschel\ bands, raising $T_d$ and hence lowering $M_d$.  Since
more distant, more luminous objects tend to have more dust, the deviation
increases systematically toward higher $M_d$ (Figure \ref{fig:pgmbb1}(a)).
We conclude that the simple MBB method can provide robust dust masses for quasar
host galaxies, as long as the SED covers the peak and Rayleigh-Jeans tail of the
cold dust emission.  However, one should be wary about contamination by warm
dust emission and data quality.  With the typical wavelength coverage and noise
level of \herschel\ data, the MBB model can easily underestimate the dust mass
by more than a factor of 2, even for moderately high-$z$ objects (e.g., $z>0.1$).

\subsection{Implications for AGN Feedback}
\label{ssec:feedback}

Our study shows that PG quasar host galaxies have a wide dispersion in ISM
content.  Most (90\%) of the sample have gas mass fractions indistinguishable
from those of massive star-forming galaxies.  Only nine objects ($\sim$ 10\% of 
the sample) are notably gas poor; they have $M_\mathrm{gas}/M_* \lesssim 0.01$, 
which is in the regime of quenched early-type galaxies (Figure \ref{fig:ismstar}(b)).
PG 1226$+$023 (3C~273) is a flat-spectrum radio-loud quasar, while the rest are
radio quiet.  Six objects (PG 0026$+$129, PG 0804$+$761, PG 0923$+$201, 
PG 1226$+$023, PG 1416$-$129, and PG 1534$+$580) have available high-resolution
\hst\ optical images, the analysis of which indicates that the host galaxies have
stellar light distributions consistent with elliptical galaxies 
(\citealt{Kim2008ApJ,Kim2017ApJS}; Y. Zhao et al. 2017, in preparation).

\begin{figure*}
\centering
\includegraphics[width=0.8\textwidth]{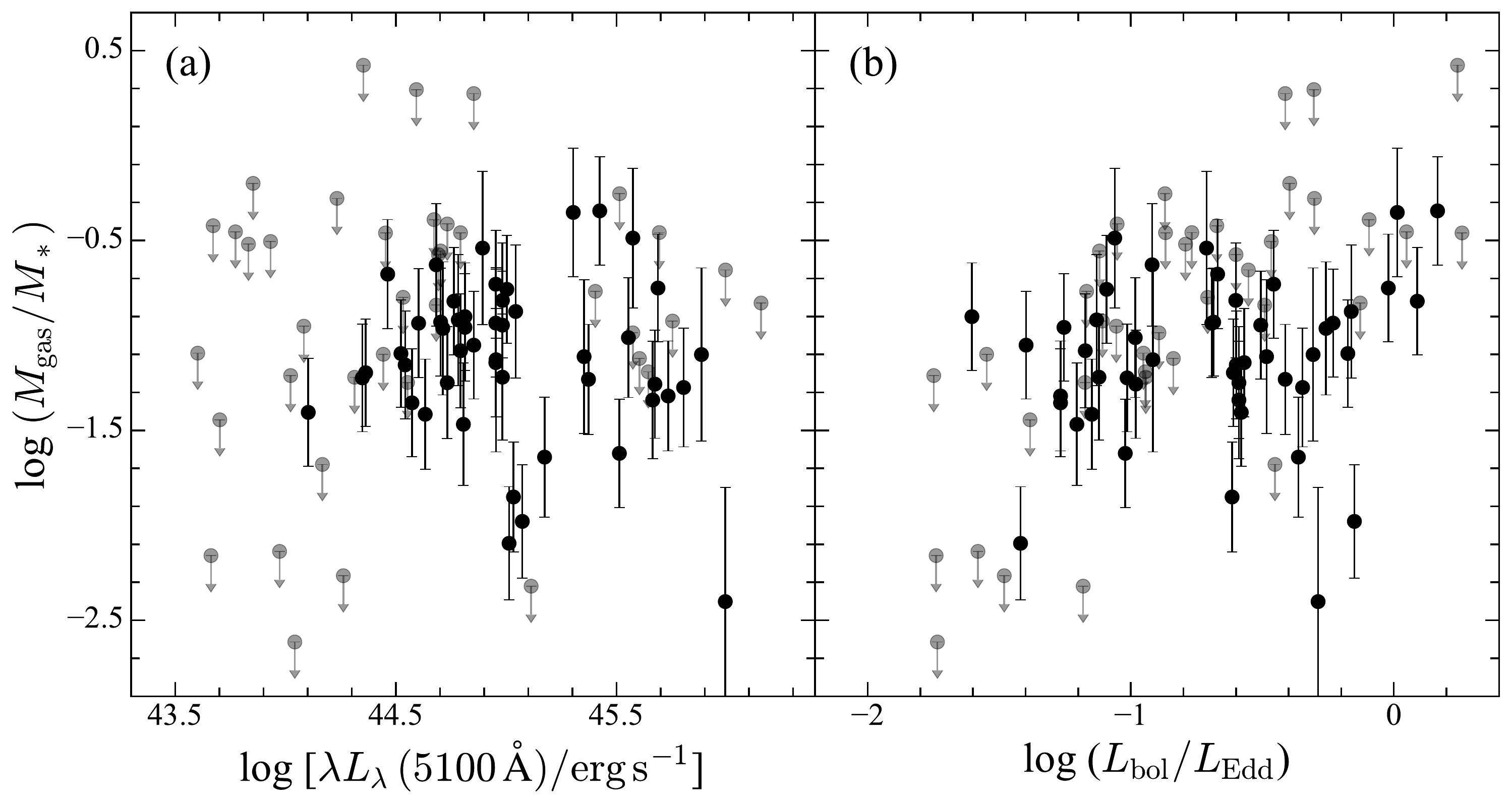}
\caption{Dependence of the gas-to-stellar mass ratio on (a) optical (5100 \AA) AGN 
luminosity and (b) Eddington ratio.  Upper limits include sources whose dust masses 
are upper limits or whose host galaxy stellar masses come from the bulge masses 
estimated from the $M_\mathrm{BH}$--$M_\mathrm{bulge}$ relation.  For clarity, the 
upper limits are displayed in gray.}
\label{fig:gasdef}
\end{figure*}

There are no obvious connections between gas content and AGN properties.  Figure
\ref{fig:gasdef} shows no relationship at all between gas mass fraction and
optical AGN luminosity [$\lambda L_{\lambda}$(5100 \AA)] or Eddington ratio
($L_\mathrm{bol}/L_\mathrm{Edd}$).  The bolometric luminosity is estimated from
$L_\mathrm{bol} = 10\,\lambda L_{\lambda}$(5100 \AA) \citep{McLure2004MNRAS,
Richards2006ApJS}, and the Eddington luminosity is
$L_\mathrm{Edd}=1.26\times 10^{38}(M_\mathrm{BH}/M_\odot)$.  In fact, the
gas-deficient quasars are not the most luminous members, all having
$\lambda L_{\lambda}$(5100 \AA)$\lesssim 10^{45}\,\mathrm{erg\,s^{-1}}$, 
except PG 1226+023.  The Eddington ratio of the gas-deficient quasars 
span a wide range and tend to lie below $L_\mathrm{bol}/L_\mathrm{Edd} \approx 0.1$.  
This suggests that the wide dispersion of gas fractions likely reflects the evolutionary stage 
of the host galaxy rather than any direct influence of AGN feedback.

Our results challenge the popular merger-driven evolutionary scenario for AGNs,
wherein the cold gas content of unobscured quasars should be depleted, or at the
very least diminished, toward the late stages of the merger process as a
consequence of AGN feedback.  If gas clumps are accelerated by the AGN above the
escape velocity of the galaxy, the gas depletion time scale should be less than
a few hundred Myr (e.g., \citealt{Cicone2014AA}).  It is thus very surprising
that we see little evidence that the quasar properties have any connection to
the ISM content of the host galaxies.

\section{Summary}
\label{sec:sum}

The cold gas content of quasar host galaxies provides a powerful probe of the
efficiency of AGN feedback.  We describe our approach of inferring total gas
masses from dust masses for quasars, making use of gas-to-dust ratios estimated
from the stellar mass, and hence metallicity, of the host galaxies.

We construct full (1--500 \micron) IR SEDs of the entire sample of 87
low-redshift ($z < 0.5$) PG quasars, using photometric data from \tmass, \wise,
and \herschel, complemented by \spitzer\ \irs\ spectra. We develop a Bayesian
MCMC method to simultaneously fit the photometric and spectroscopic data, using
physically motivated emission components for the starlight, AGN dust torus, and
large-scale galactic dust.  The templates for the dust component on galactic
scales are based on the dust emission models of \cite{DraineLi2007ApJ}. The
reliability of our fitting method is thoroughly investigated using mock SEDs and
controlled experiments.  We demonstrate that we can derive robust dust-inferred
gas masses, which are unbiased with respect to gas masses obtained through other
direct and indirect methods.

Our main conclusions are as follows:

\begin{enumerate}

\item The vast majority of quasar host galaxies are gas-rich systems, having
total dust masses of $M_d \approx 10^{6.2} - 10^{8.7}\, M_\odot$, with a 
mean value of $10^{7.6 \pm 0.1}\, M_\odot$.  These translate to 
total gas masses of $M_{\rm gas} \approx 10^{8.3} - 10^{10.8}\, M_\odot$, 
with a mean value of $10^{9.7 \pm 0.1}\, M_\odot$.

\item Most (90\%) quasar host galaxies have similar dust and gas content to 
normal star-forming galaxies of similar stellar mass.  Only a minority (10\%) of
the quasar hosts are gas-poor systems.  The gas mass fraction of quasar host
galaxies depends on neither the AGN luminosity nor the Eddington ratio.

\item The rich ISM content of quasars and its insensitivity to AGN properties
indicate that AGN feedback is ineffective in low-redshift quasars.

\item The dust grains in quasar host galaxies appear to be exposed to a
systematically stronger interstellar radiation field than normal, star-forming
galaxies, suggesting that the AGN radiation field contributes to dust heating
on galactic scales.  We caution against the common practice of inferring
star formation rates from the integrated FIR luminosity.

\item Quasar host galaxies exhibit systematically weaker PAH emission than
normal galaxies.  This suggests that either PAH molecules are destroyed in
AGN environments or quasar hosts experience lower levels of ongoing
star formation.

\item The common practice of fitting IR SEDs using modified blackbody models
tends to systematically overestimate the dust temperature and underestimate the
dust mass when applied to SEDs with inadequate coverage of the Rayleigh-Jeans
tail of the spectrum.

\end{enumerate}

\acknowledgments

We are very grateful to an anonymous referee for providing helpful, 
expert suggestions. 
J.S. thanks the \herschel\ help desk and Yali Shao for advice on the reduction 
of \herschel\ data; Michael Gully-Santiago, Yanrong Li, Seth Johnson, Brendon
Brewer, Feng Long, and the PKU pulsar group for help on the SED fitting;
Minjin Kim and Yulin Zhao for access to \hst\ images of some of the quasars; 
and Robert Nikutta and Ralf Siebenmorgen for information on dust torus models.
He is also grateful to Aigen Li, Ran Wang, Linhua Jiang, Jing Wang, and David
Sanders for much useful scientific and technical advice.  The work of L.C.H. 
was supported by the National Key Program for Science and Technology Research
and Development (2016YFA0400702) and the National Science Foundation of
China (11303008, 11473002, 11721303).  Y.X. is supported by China Postdoctoral 
Science Foundation Grant 2016 M591007.  This publication makes use of data 
products from the Two Micron All-Sky Survey, which is a joint project of the University 
of Massachusetts and the Infrared Processing and Analysis Center/California Institute 
of Technology, funded by the National Aeronautics and Space Administration and the 
National Science Foundation.  It also makes use of Astropy, a community-developed 
core Python package for astronomy \citep{Astropy2013AA}.

\bibliography{qsed}

\appendix
\section{Data Systematics}
\label{apd:sys}
In order to understand how well the \wise, \spitzer, and \herschel\ data match
each other, we study potential systematic deviations among the three data
sets.  \spitzer\ IRS spectra cover the wavelength range of \wise\ \w3 and \w4,
so we can compare the actually observed \w3 and \w4 flux densities with
synthetic values generated from IRS spectra.  To estimate the systematics of
\herschel\ and \spitzer\ data, we can compare the 70 $\mu$m and 160 $\mu$m
measurements of \spitzer\ \mips\ and \herschel\ \pacs.  As the \irs\ spectra
have been scaled to match \mips\ 24 $\mu$m photometry \citep{Shi2014ApJS},
comparing \mips\ and \pacs\ at a mutual wavelength is a reasonable approach.

\begin{figure*}
\begin{center}
\begin{tabular}{c c}
\includegraphics[height=0.35\textheight]{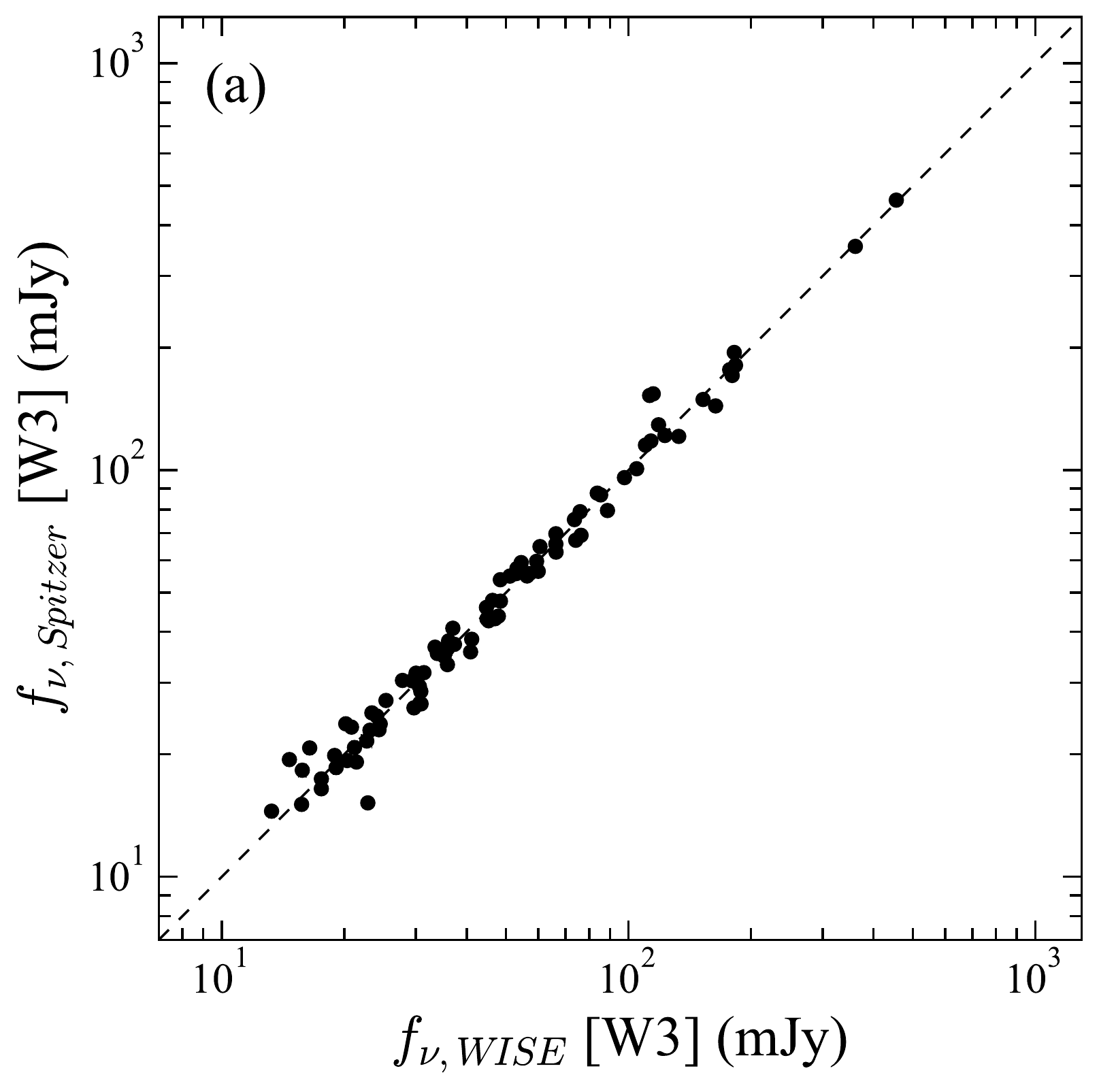} &
\includegraphics[height=0.35\textheight]{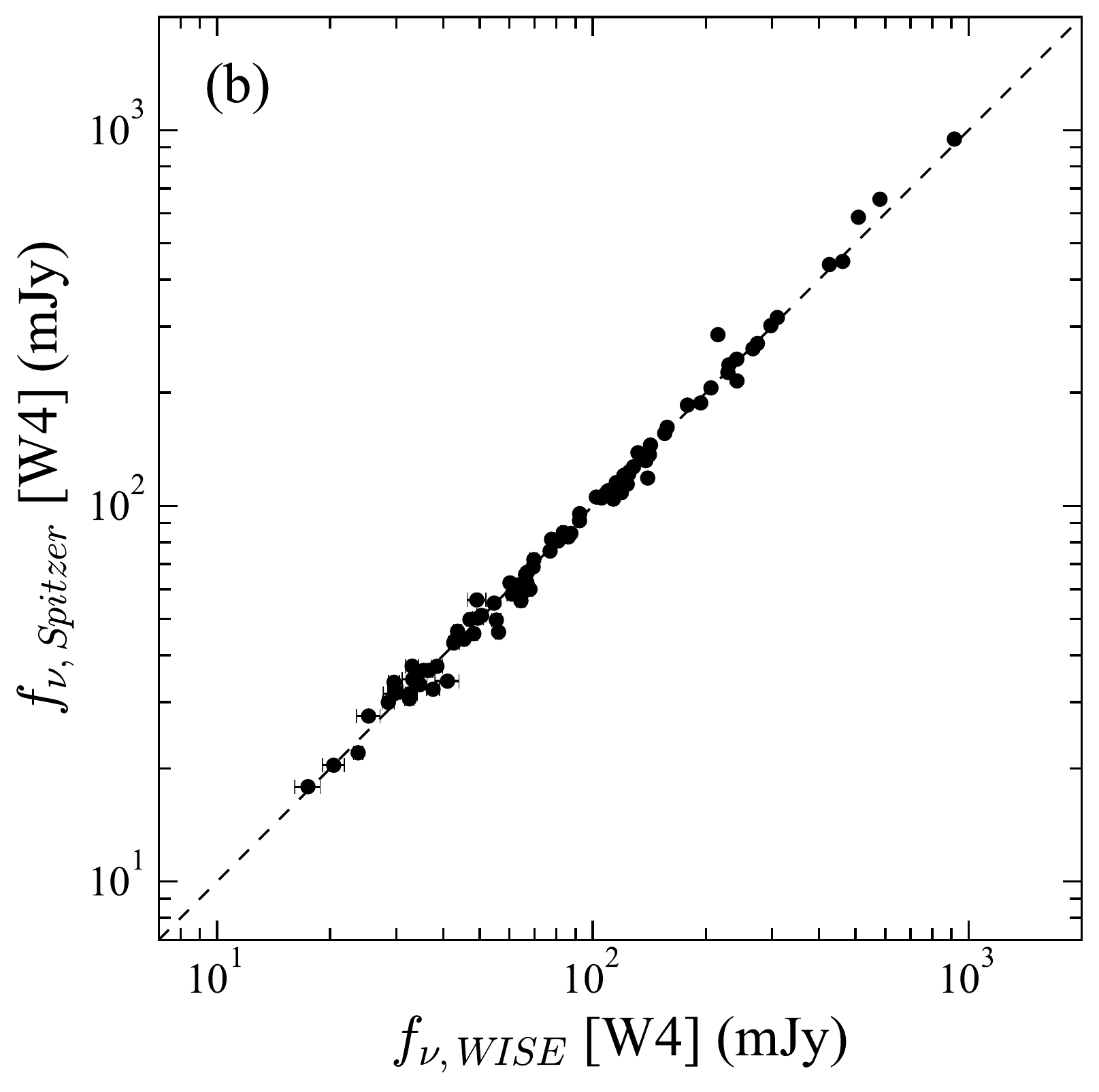}
\end{tabular}
\caption{Comparison between flux densities from \wise\ and synthetic
measurements of \wise\ bandpasses applied to \spitzer\ IRS spectra, 
for the (a) \w3\ and (b) \w4\ bands.  The median deviation is 0.0$\pm$0.04
dex for \w3\ and 0.0$\pm$0.03 dex for \w4.}
\end{center}
\label{fig:wise_sys}
\end{figure*}

The \wise\ magnitudes are converted into monochromatic flux densities using
the isophotal flux densities,\footnote{For the \w1--\w4 bands, the isophotal flux
densities, $f_\nu\mathrm{(iso)}$, are 309.540$\pm$4.582, 171.787$\pm$2.516,
31.674$\pm$0.450, and 8.363$\pm$0.293 Jy, corresponding to wavelengths
3.3526, 4.6028, 11.5608, and 22.0883 $\mu$m \citep{Wright2010AJ}.}
$f_\nu\mathrm{(iso)}=f_\nu\mathrm{(Vega)}$ \citep{Jarrett2011ApJ}.
The \w3 and \w4 bands are known to have a systematic, color-dependent bias:
red sources (typically $f_\nu\propto\nu^{-2}$) are 17\% fainter and 9\%
brighter than blue sources (typically $f_\nu\propto\nu^{2}$), which are used
for calibration \citep{Wright2010AJ}.  Since quasars usually have
$f_{\nu} \propto \nu^{-\alpha}$, with $\alpha \gtrsim 1$, we need to apply 
a correction to the \wise\ \w4 flux densities (see also WISE Data 
Processing\footnote{\url{http://wise2.ipac.caltech.edu/docs/release/allsky/expsup/sec4_4h.html}}):
$f'_{\nu}(W4) \approx 0.90 f_{\nu} (W4)$.  For \w3, we simply increase the
$f_\nu (W3)$ of all objects by 17\%.  The corrected \wise\ \w3 and \w4 bands
agree remarkably well with the \spitzer\ synthetic flux densities (Figure
\ref{fig:wise_sys}).  Thus, we conclude that the \spitzer\ IRS spectra match
the \wise\ bands very well, considering the 1.5\% calibration uncertainty of
\wise.

\begin{figure*}
\begin{center}
\begin{tabular}{c c}
\includegraphics[height=0.35\textheight]{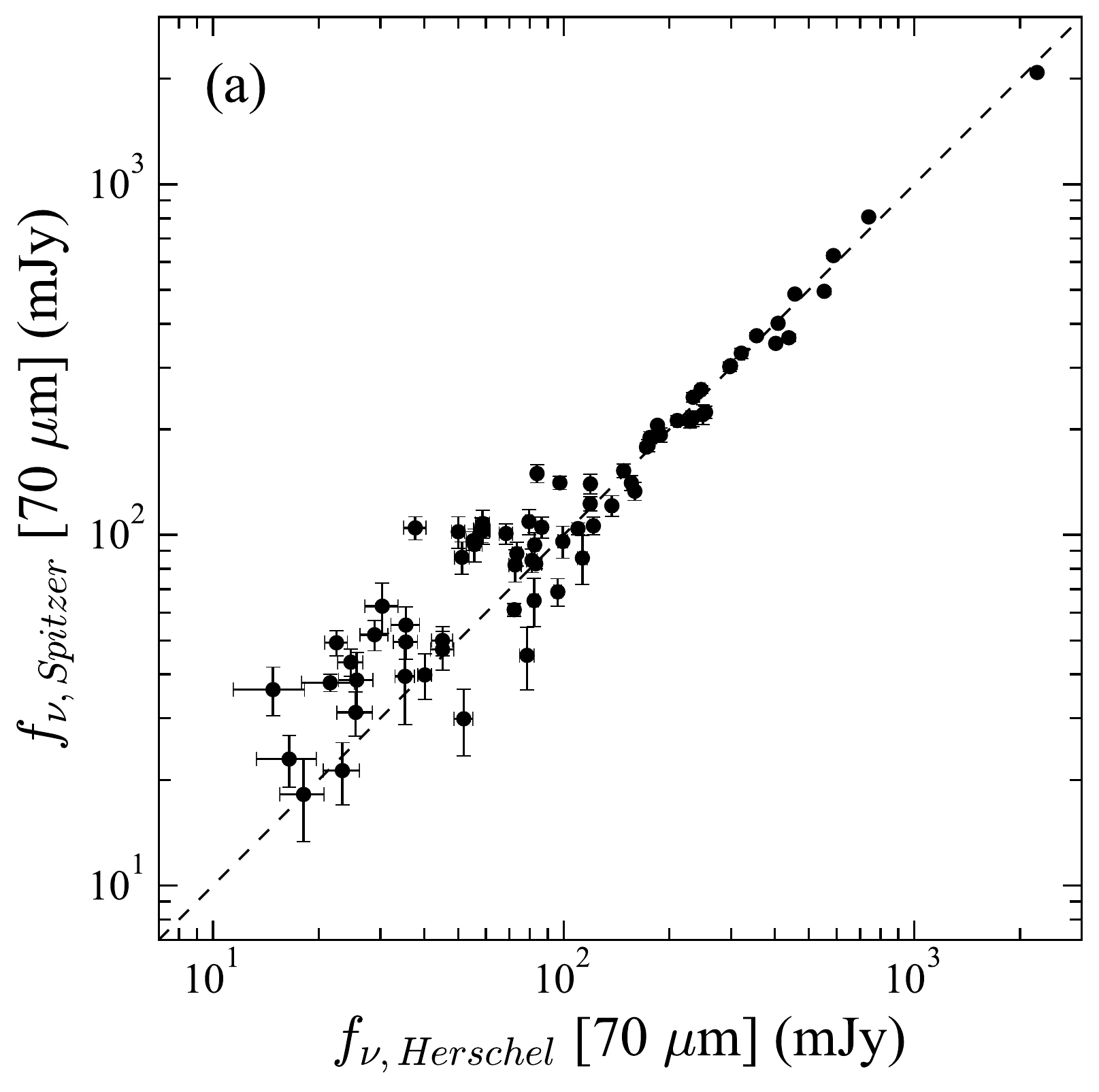} &
\includegraphics[height=0.35\textheight]{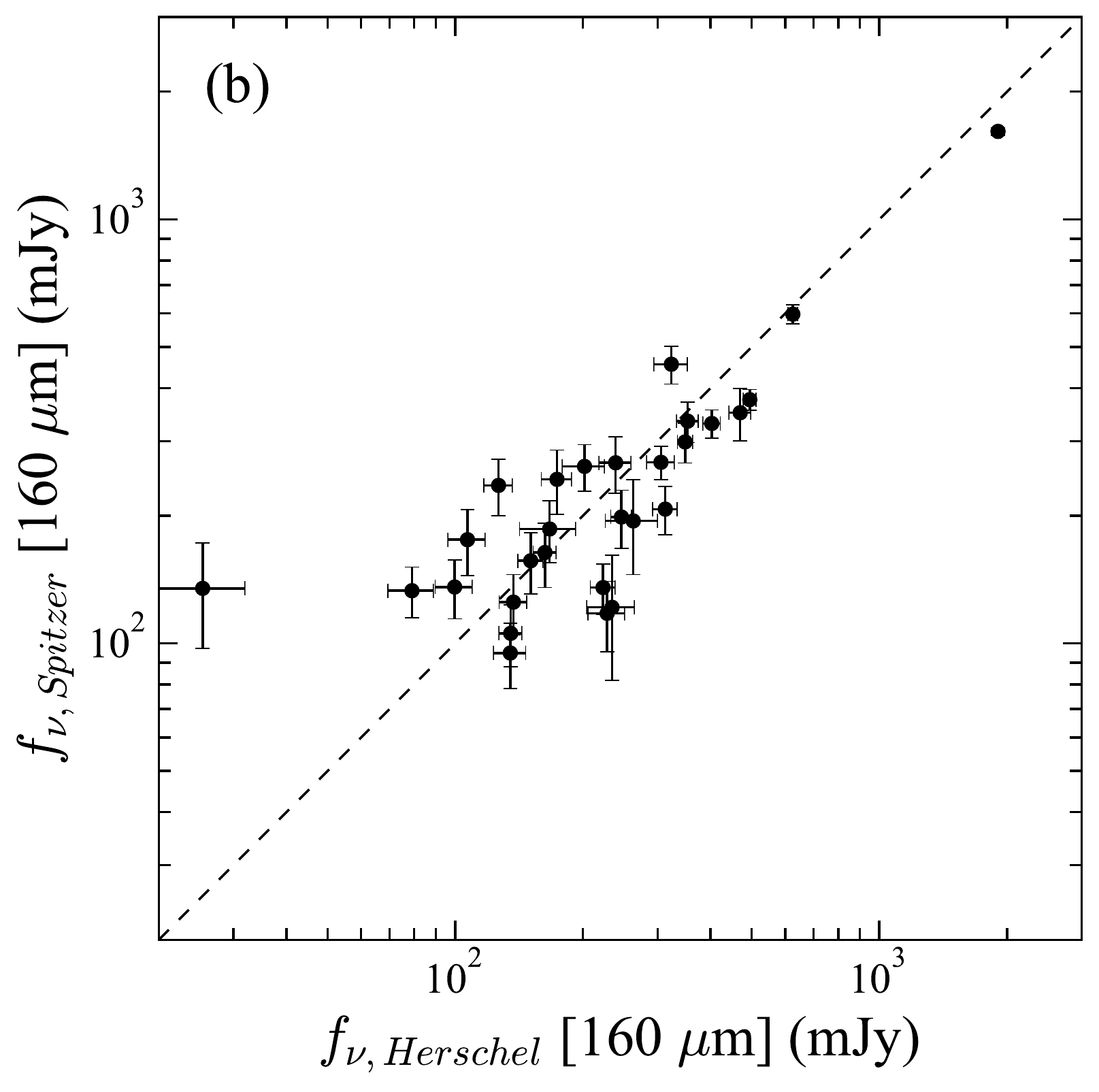}
\end{tabular}
\caption{Comparison between flux densities from \herschel\ and \spitzer\ \mips\ 
for the (a) 70 \micron\ and (b) 160 \micron\ bands.  The median deviation (y$-$x) 
is 5\%$\pm$43\% for the 70 \micron\ bands and $-$9\%$\pm$86\% for the 160 
\micron\ bands.}
\end{center}
\label{fig:hers_sys}
\end{figure*}

We obtain the \spitzer\ \mips\ 70 and 160 $\mu$m measurements from
\cite{Shi2014ApJS} to compare with the \herschel\ PACS measurements (Figure
\ref{fig:hers_sys}).  The \spitzer\ 70 $\mu$m data tend to be systematically
higher than our \herschel\ measurements for objects $\lesssim 100$ mJy.  This
is likely due to confusion within the \mips\ beam, which is most serious for
fainter objects, because the \mips\ 70 \micron\ PSF (FWHM $\approx 18\arcsec$)
is much broader than that of \pacs\ (FWHM $\approx 6\arcsec$).  The median
deviation at 160 \micron\ is $-$9\%$\pm$86\%.  We conclude that the flux scales 
of \spitzer\ and \herschel\ data are well-matched to $\lesssim 10$\%.

\section{Comparison of Herschel Data Reduction}
\label{apd:herdata}
The \herschel\ data for the PG sample have previously been analyzed by
\cite{Petric2015ApJS}.  Our work is based on a completely new reduction and
analysis of the same data set (Section \ref{sec:data}), and here we present a
comparison between these two independent efforts.

\begin{figure*}
\begin{center}
\begin{tabular}{c c c}
\includegraphics[height=0.28\textheight]{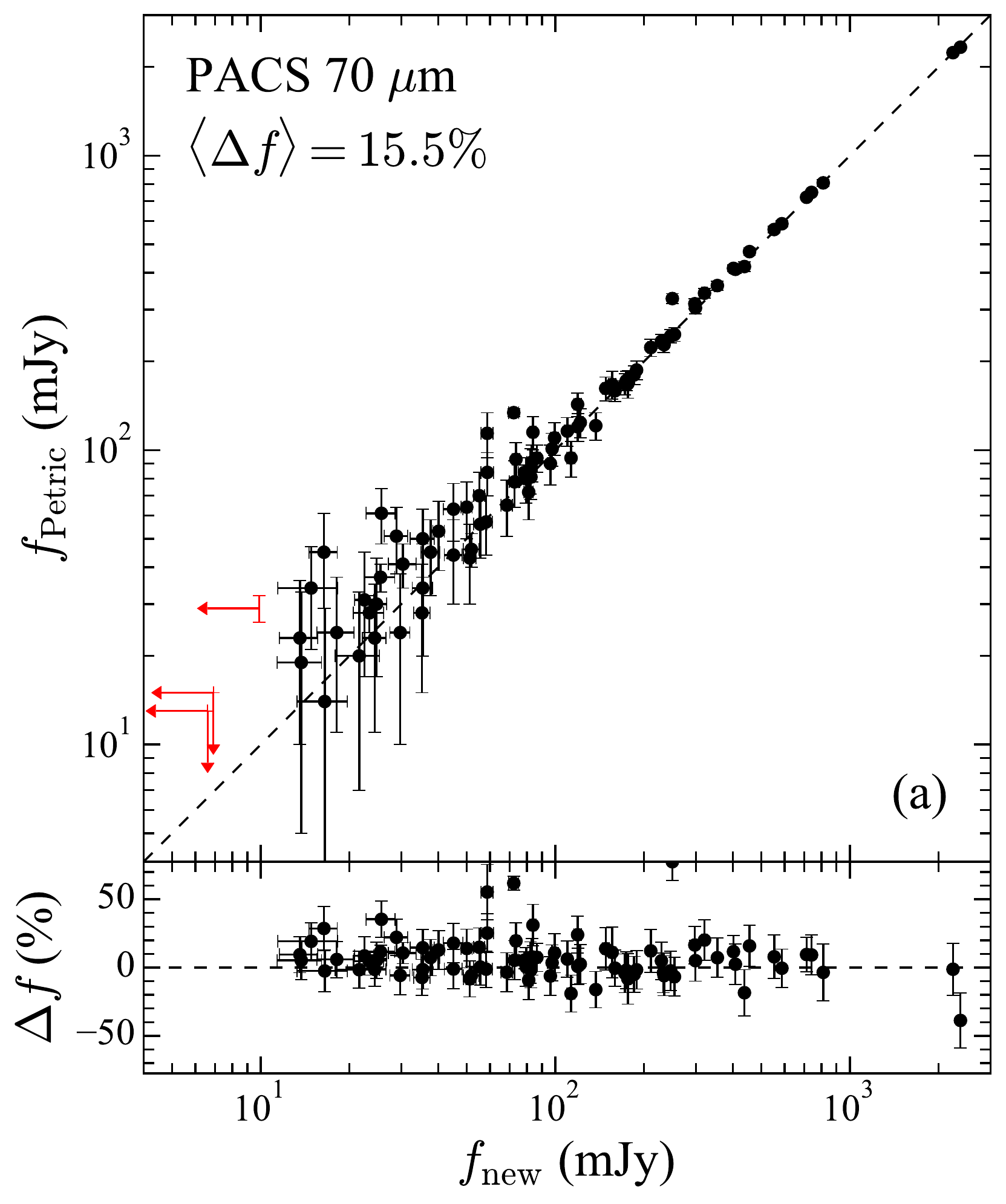} &
\includegraphics[height=0.28\textheight]{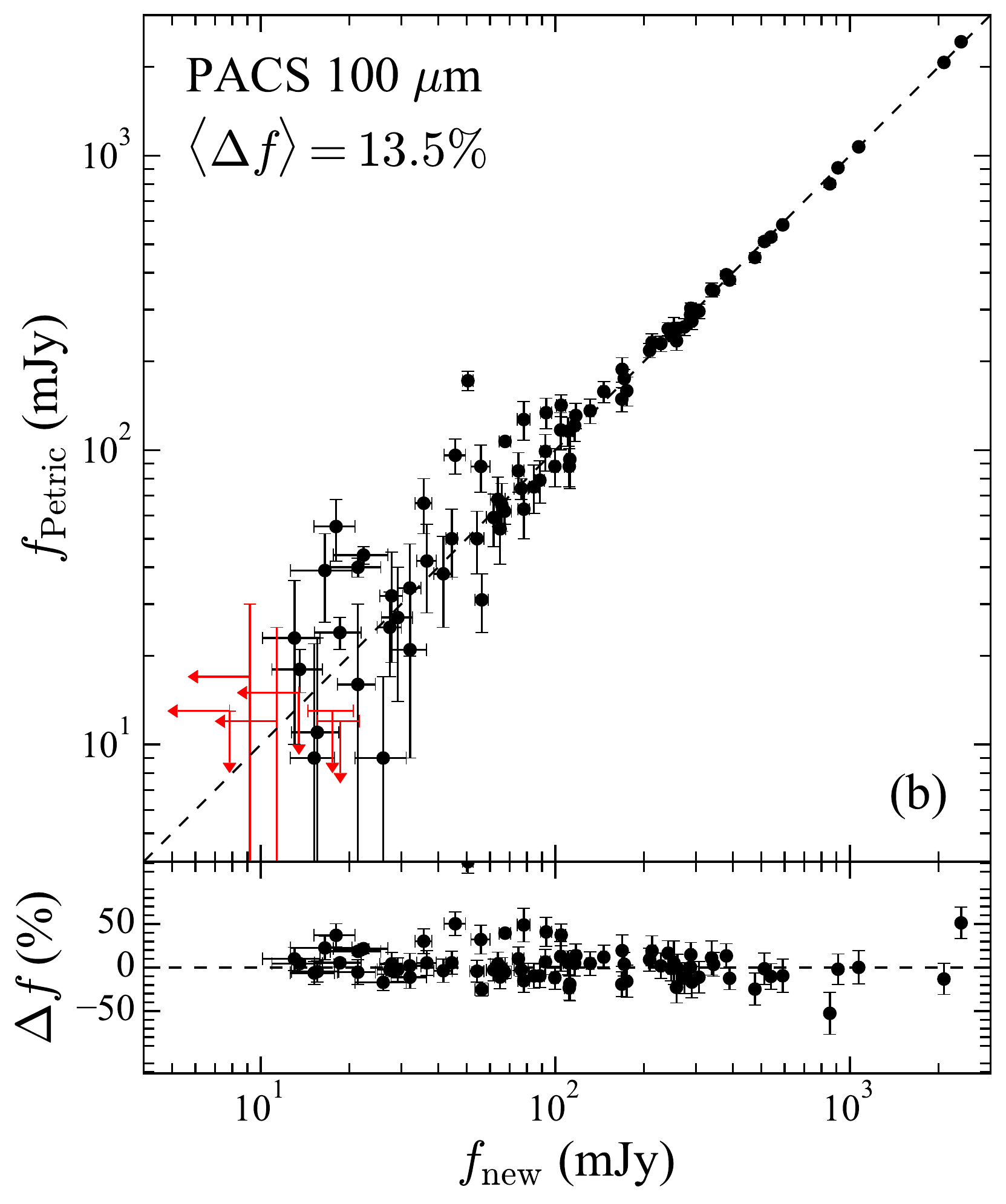} &
\includegraphics[height=0.28\textheight]{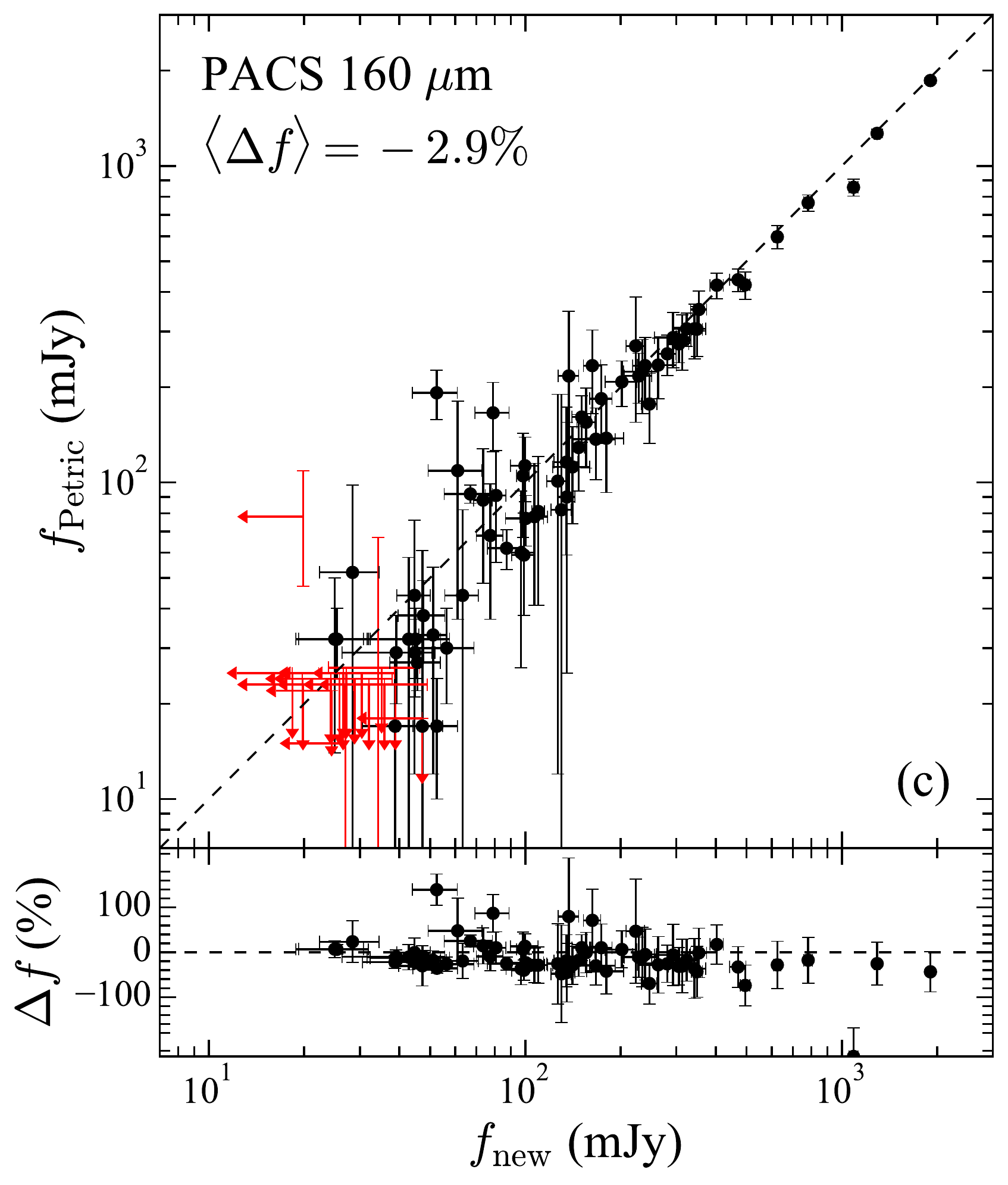}
\end{tabular}
\caption{Comparison between our new flux densities from \herschel\ with those
measured by \cite{Petric2015ApJS}, for the \pacs\ (a) 70 \micron, (b) 100 \micron, 
and (c) 160 \micron\ bands.  Detected sources are plotted in black, and those that 
are undetected by either Petric et al. or us, or both are plotted in red.  The dashed 
line is the one-to-one relation.  $\Delta f$ is the relative fractional deviation.  The 
measurements of Petric et al. are systematically higher than ours, by $15.5\%$ at 
70 \micron\ and by $13.5\%$ at 100 \micron; their uncertainties are also larger than 
ours.  For the 160 \micron\ band, the deviation is $-2.9\%$.  This is mainly due to the
aperture size effect (see the text for details).}
\end{center}
\label{fig:comppacs}
\end{figure*}

\cite{Petric2015ApJS} perform aperture photometry on the \pacs\ data by summing
all the pixels within a circular aperture radius of $\sim 20\arcsec$ for all three bands.  
They estimate uncertainties by randomly measuring the sky with the same aperture 
size and calculate the standard deviation of all the measurements.  The large aperture 
size is likely to include more contaminating sources and introduce higher noise.  
Indeed, we find that our measurements at 70 and 100 \micron\ are systematically 
lower than those of Petric et al. by 15.5\% and 13.5\%, respectively (Figure \ref{fig:comppacs}).  
The uncertainties of Petric et al.'s measurements are also larger than ours.  The two 
sets of measurements at 160 \micron\ are more consistent because the aperture sizes 
are comparable.

\begin{figure*}
\begin{center}
\begin{tabular}{c c c}
\includegraphics[height=0.28\textheight]{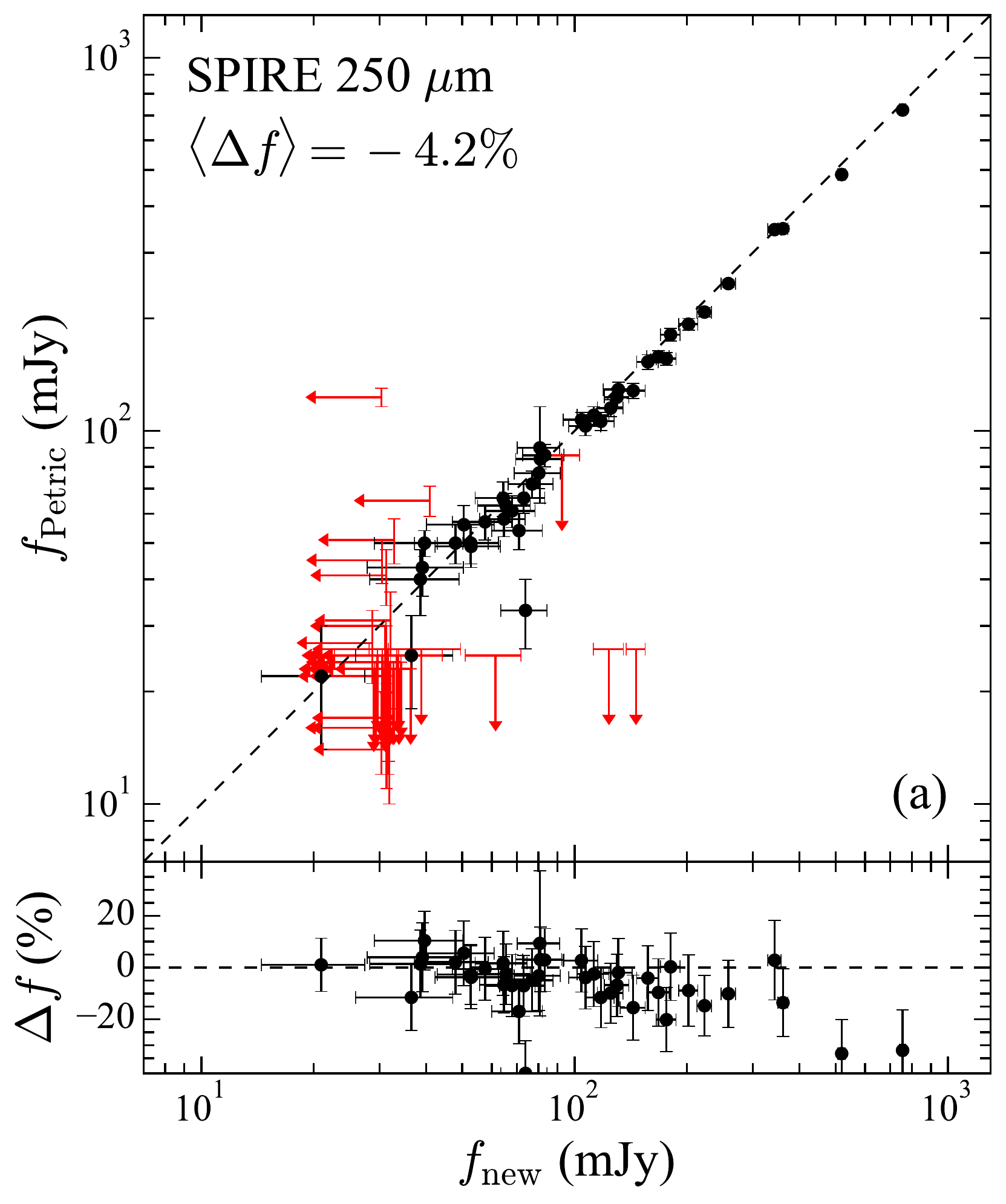} &
\includegraphics[height=0.28\textheight]{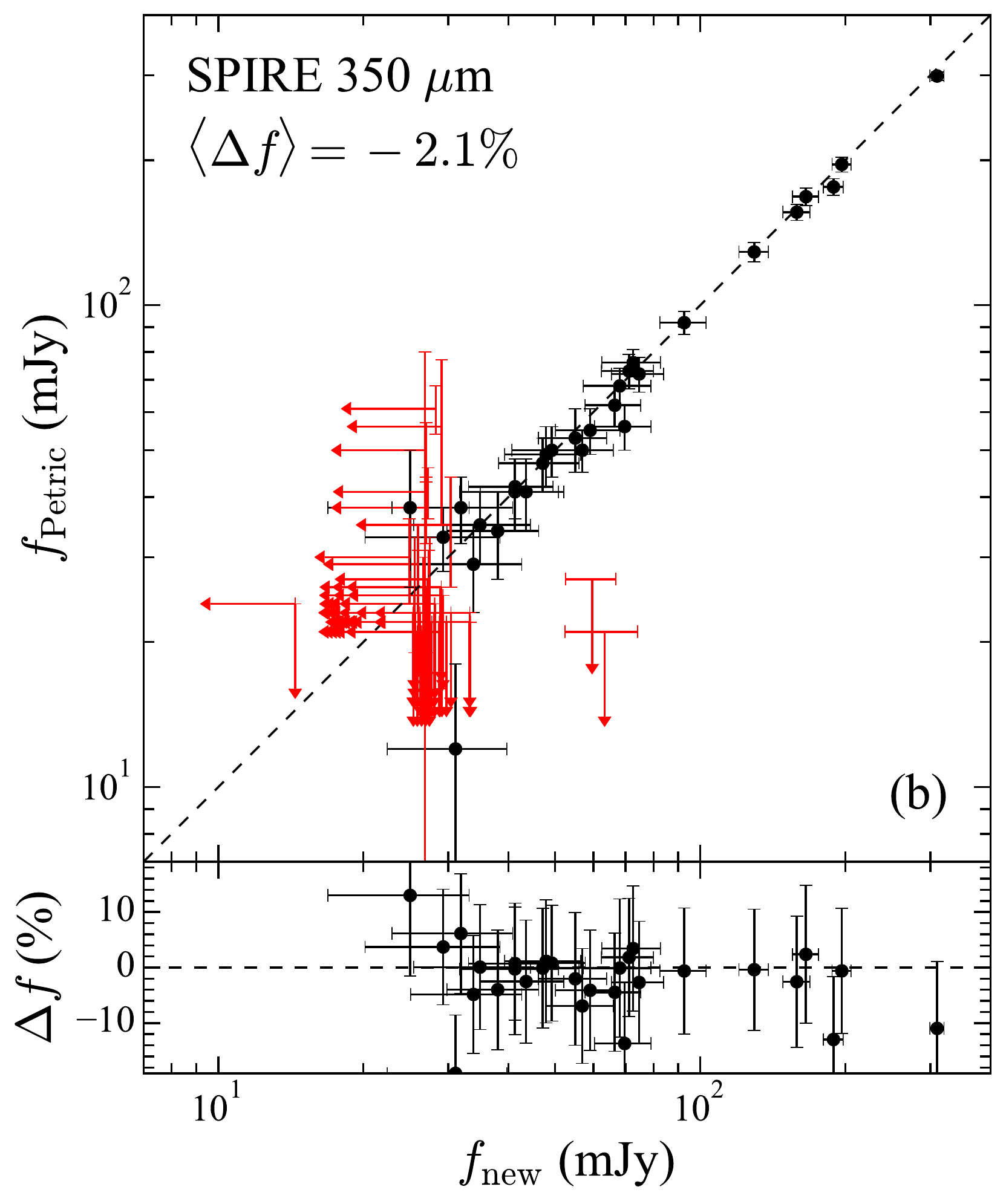} &
\includegraphics[height=0.28\textheight]{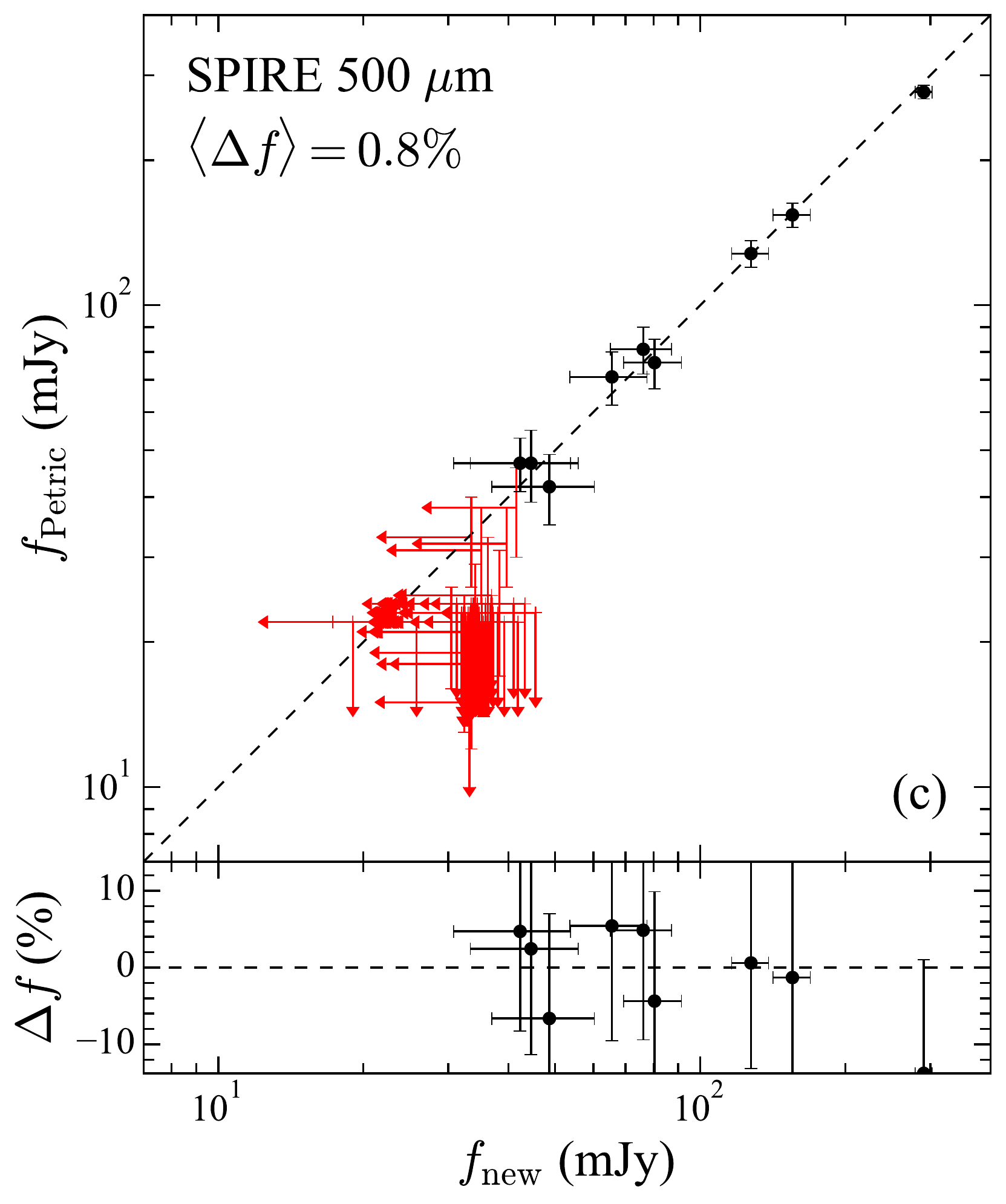}
\end{tabular}
\caption{Comparison between our new flux densities from \herschel\ with those
measured by \cite{Petric2015ApJS} for the \spire\ (a) 250 \micron, (b) 350 \micron, 
and (c) 500 \micron\ bands.  Detected sources are plotted in black, and those that 
are undetected by either Petric et al. or us, or both are plotted in red.  The dashed 
line is the one-to-one relation.  $\Delta f$ is the relative fractional deviation.  The 
measurements of Petric et al. are consistent with ours within $\sim 5\%$ for the 
detected sources.  However, there are also a number of notable discrepancies. 
Some objects are detected by our method but not by Petric et al., and vice versa.}
\end{center}
\label{fig:compspire}
\end{figure*}

For the reduction of the \spire\ data, Petric et al. use the {\tt Timeline
Fitter} to measure the sources by directly fitting the timeline data
\citep{Pearson2014ExA}.  They estimate uncertainties by randomly choosing the
location on the map to run the {\tt Timeline Fitter} and then calculate the
standard deviation.  As Figure \ref{fig:compspire} shows, our measurements and
those of Petric et al. are consistent within 5\% for the detected sources.
However, our upper limits for the undetected sources are generally higher. On
the one hand, we regard a source as undetected whenever its measured flux is
below $3\,\sigma$, regardless of whether the source extractor deems it to be
real.  On the other hand, the method used by Petric et al. to estimate the
uncertainty may not be proper.  {\tt Timeline Fitter} is not suitable for
measuring faint sources ($< 30$ mJy; \citealt{Pearson2014ExA}), and to estimate
the uncertainty one needs to inject fake sources into the timeline (e.g.,
\citealt{Ciesla2012AA}) instead of fitting the timeline randomly.

\section{Fitting the SEDs with Different Dust Torus Models}
\label{apd:xl17}

We evaluate the impact of the choice of torus model on the derived parameters
for the galactic dust properties.  Different from the radiative transfer CLUMPY
model, \cite{Xie2017ApJS} demonstrate that the \irs\ spectra of PG quasars can 
be successfully fitted by MBB models with theoretical dust absorption coefficients 
calculated from Mie theory \citep{Bohren1983Natur}.  We briefly summarize the 
XLH17 model and refer to \cite{Xie2015ApJ,Xie2017ApJS} for detailed discussions.

The silicate emission in the \irs\ spectra indicates that the dust torus is
optically thin in the MIR, so that it is possible to model the torus emission
without considering radiative transfer.  The XLH17 model assumes that the torus
consists of, on average, two chemical compositions, silicate and carbonaceous
dust.  We mainly use astronomical silicate \citep{Draine1984ApJ} and sometimes
amorphous olivine or pyroxene \citep{Dorschner1995AA} for the silicate dust and
graphite for the carbon dust, following the compositions suggested by XLH17.  For
each of the two compositions, we assume that the dust has two representative
temperatures, warm and cold.  The rest-frame model flux density is

\begin{equation}
f_\mathrm{\nu, XLH17} = \frac{(1+z)^2}{D_{L}^2} \sum_i \{B_\nu(T_{w, i}) \,
\kappa_{\mathrm{abs}, i}(a_d, \nu) \, M_{w, i} + B_\nu(T_{c, i}) \,
\kappa_\mathrm{abs, i}(a_d, \nu) \, M_{c, i} \},
\end{equation}

\noindent
where the summation is over two dust compositions denoted with the subscript 
$i$, $D_{L}$ is the luminosity distance, $\kappa_\mathrm{abs, i}(\nu)$ is the 
mass absorption coefficient for the dust with characteristic size $a_d$ and at
frequency $\nu$, $M$ is the dust mass, and $B_\nu(T)$ is the Planck function
with temperature $T$.  The subscripts $w$ and $c$ correspond to the ``warm'' 
and ``cold'' dust components, respectively.  The naming convention is consistent 
with that used in \cite{Xie2015ApJ}, even though our galactic dust (DL07
component) is even colder than the ``cold'' torus component here.  The mass
ratios between carbon and silicate dust are between 0.2 and 2.0 for the warm
($r_{w, \mathrm{G/S}}$) and cold ($r_{c, \mathrm{G/S}}$) components,
respectively.  There are altogether nine free parameters in the XLH17 model
(Table \ref{tab:xl17}).  The grain size $a_d$ is discrete, from 0.1 to 1.5
\micron, while the rest of the parameters are continuous.

\begin{table}
  \begin{center}
  \caption{The parameters and priors of the XLH17 dust torus model}
  \label{tab:xl17}
  \begin{tabular}{clcll}
    \hline
    \hline
    Parameter               & Unit    & Discreteness & Prior                 \\\hline
    $a_d$                   & \micron & \ding{52}    & [0.1, 1.5]            \\
    $T_{w, \mathrm{sil}}$ & K         & \ding{56}    & [250, 1500]           \\
    $M_{w, \mathrm{sil}}$ & $M_\odot$ & \ding{56}    & [$10^{-5}$, $10^{8}$] \\
    $T_{c, \mathrm{sil}}$ & K         & \ding{56}    & [40, 200]             \\
    $M_{c, \mathrm{sil}}$ & $M_\odot$ & \ding{56}    & [$10^{-5}$, $10^{8}$] \\
    $T_{w, \mathrm{gra}}$ & K         & \ding{56}    & [250, 1500]           \\
    $r_{w, \mathrm{G/S}}$ & --        & \ding{56}    & [0.2, 2.0]            \\
    $T_{c, \mathrm{gra}}$ & K         & \ding{56}    & [40, 200]             \\
    $r_{c, \mathrm{G/S}}$ & --        & \ding{56}    & [0.2, 2.0]            \\\hline
    \hline
  \end{tabular}
  \end{center}
\end{table}

\begin{figure*}
\begin{center}
\begin{tabular}{c c}
\includegraphics[height=0.35\textheight]{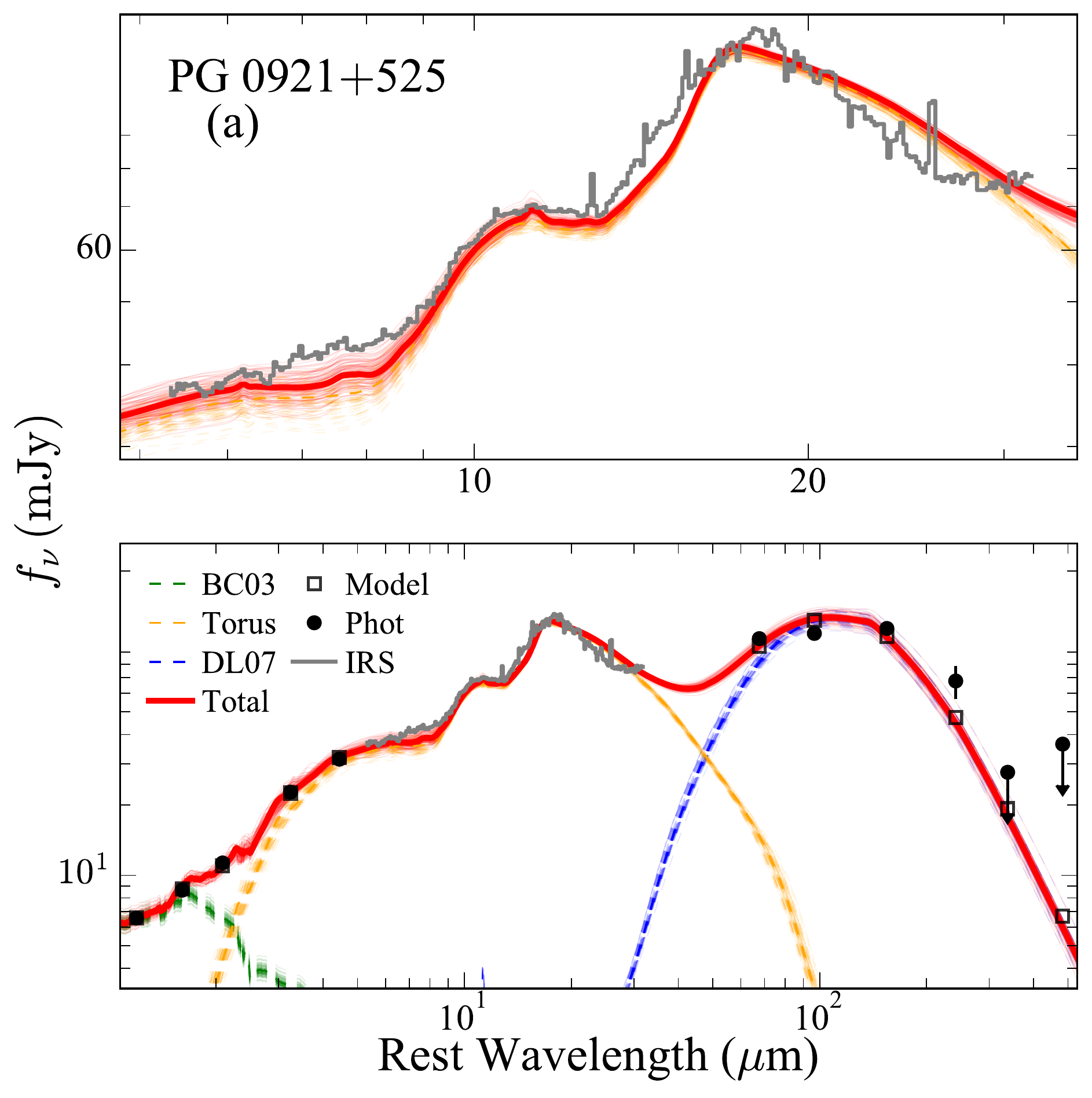} &
\includegraphics[height=0.35\textheight]{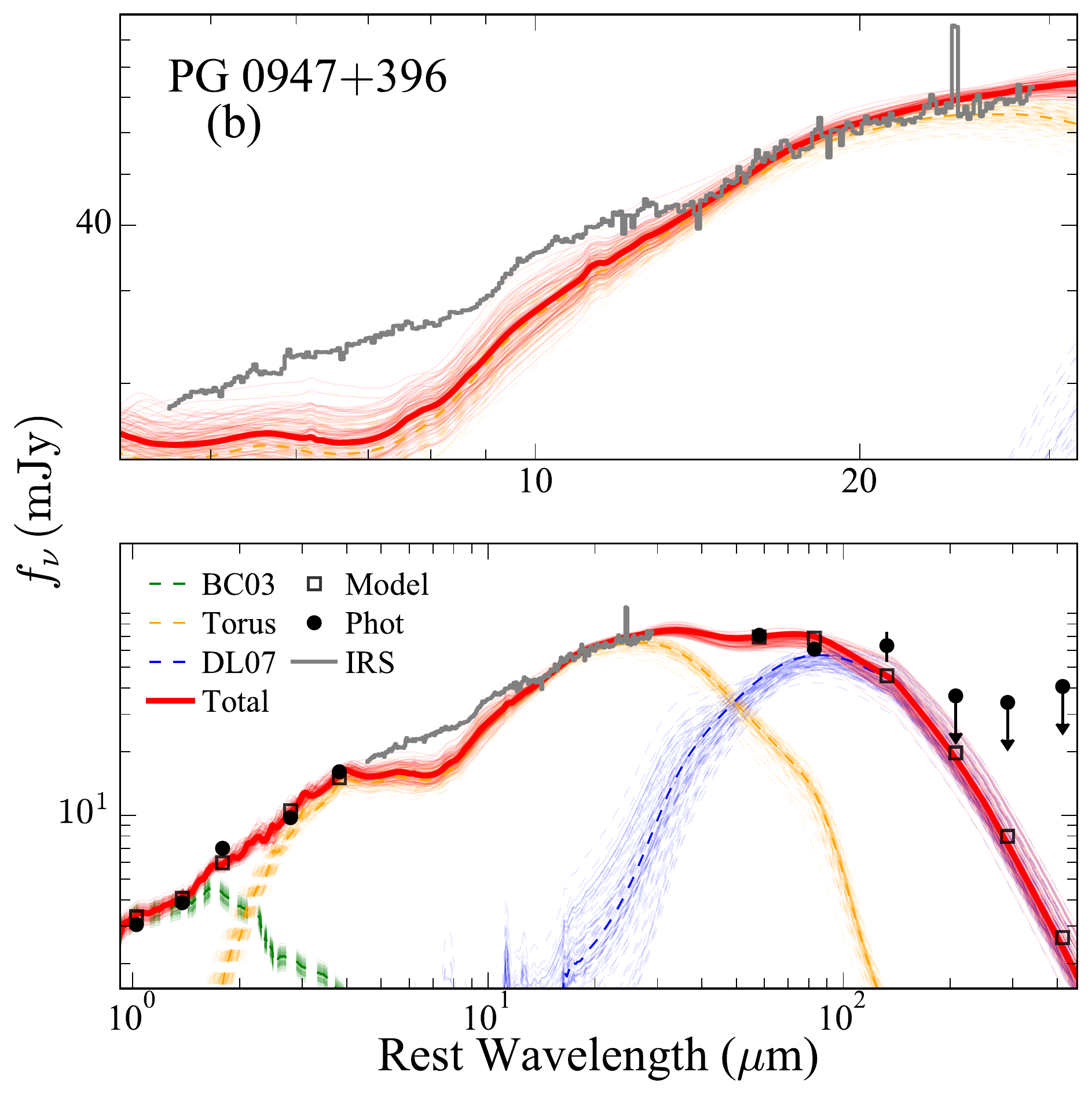}
\end{tabular}
\caption{Examples of SED fits using the XLH17 torus model.  (a) PG 0921+525 
is better fit with the XLH17 model than with the CLUMPY model because
the former uses an amorphous olivine model for the silicate dust.  The XLH17
model is plotted as a red line, and the rest of the conventions are the same as
in Figure \ref{fig:fitgood}.  (b) By contrast, the XLH17 model does not give a 
good fit to the short end of the \irs\ spectrum for PG 0947+396. {\it The best-fit 
results of the entire sample (76 objects for which the FIR data are good enough 
to constrain the DL07 model) can be found in the online version.}
}
\end{center}
\label{fig:fitxl17}
\end{figure*}

For many objects, the XLH17 model performs as well, if not better, than the CLUMPY 
model (Figure \ref{fig:fitxl17}(a)).  However, for $\sim 1/3$ of the sample, the best-fit 
models show a light deficit at the short end ($\sim 7$ \micron) of the \irs\ spectra 
(Figure \ref{fig:fitxl17}(b)). This is likely due to the simplicity of the XLH17 model, which 
contains only four discrete temperature components, two for each composition of dust, 
whereas in reality the complex systems under consideration have a continuum of dust 
temperatures.  When the temperature gaps are too large, there are light deficits in the 
best-fit models.  This problem is not obvious if only the \irs\ spectrum is fitted (XLH17), 
but it becomes apparent when we incorporate the \tmass\ and \wise\ bands.  An extra 
hot component is necessary to account for the emission in NIR.

\begin{figure*}[htbp]
\begin{center}
\begin{tabular}{c c}
\includegraphics[height=0.35\textheight]{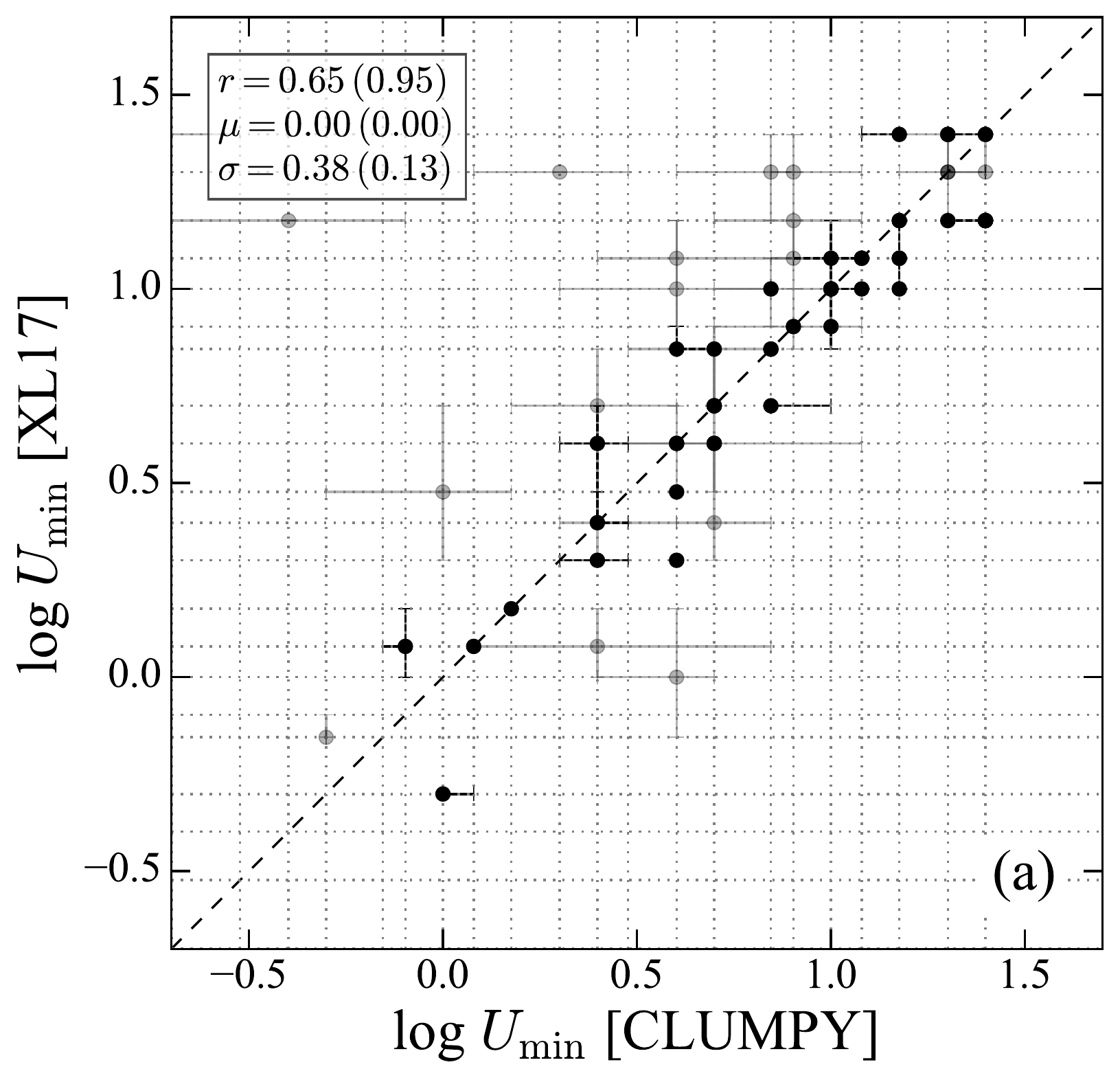} &
\includegraphics[height=0.35\textheight]{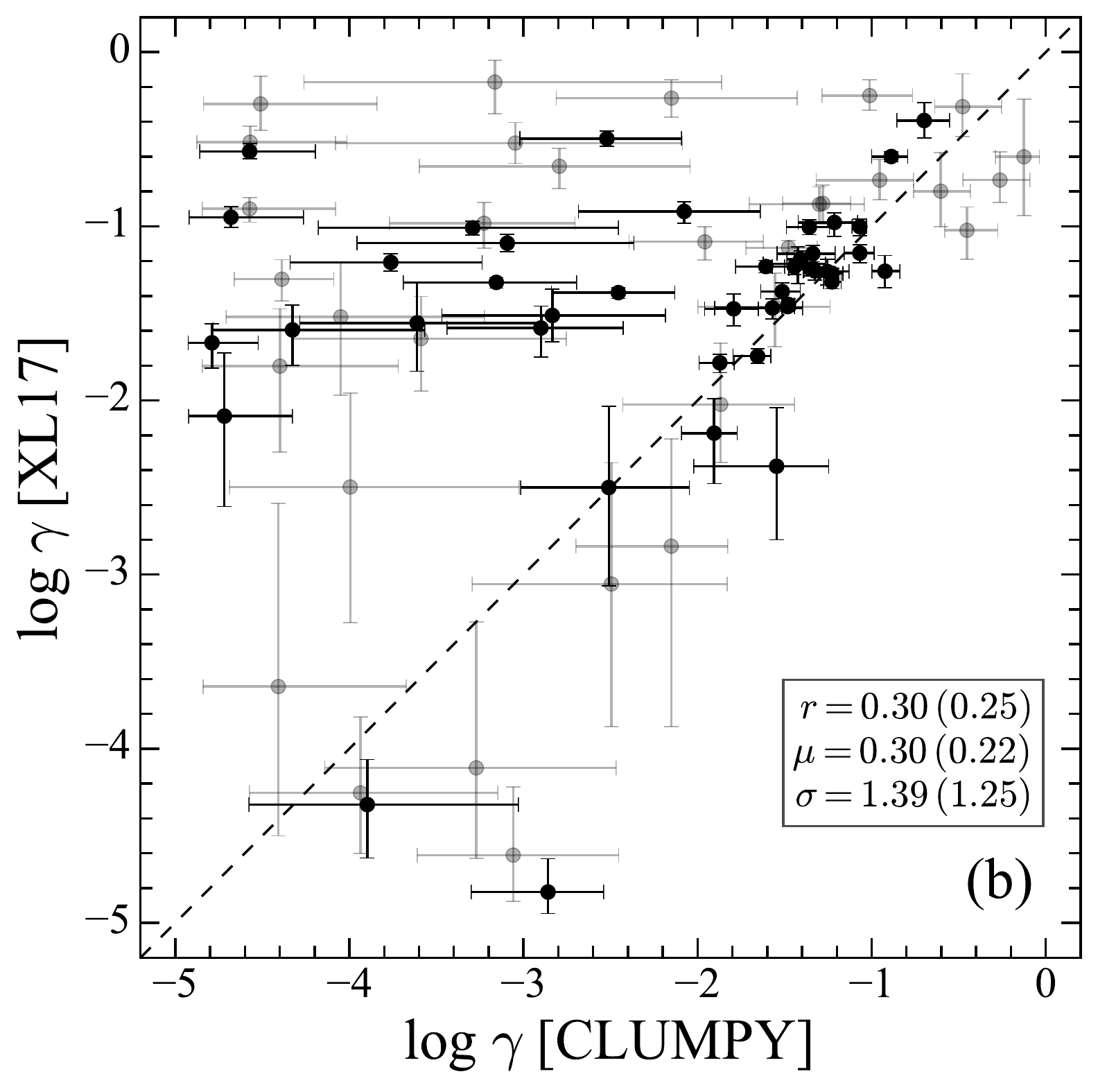} \\
\includegraphics[height=0.35\textheight]{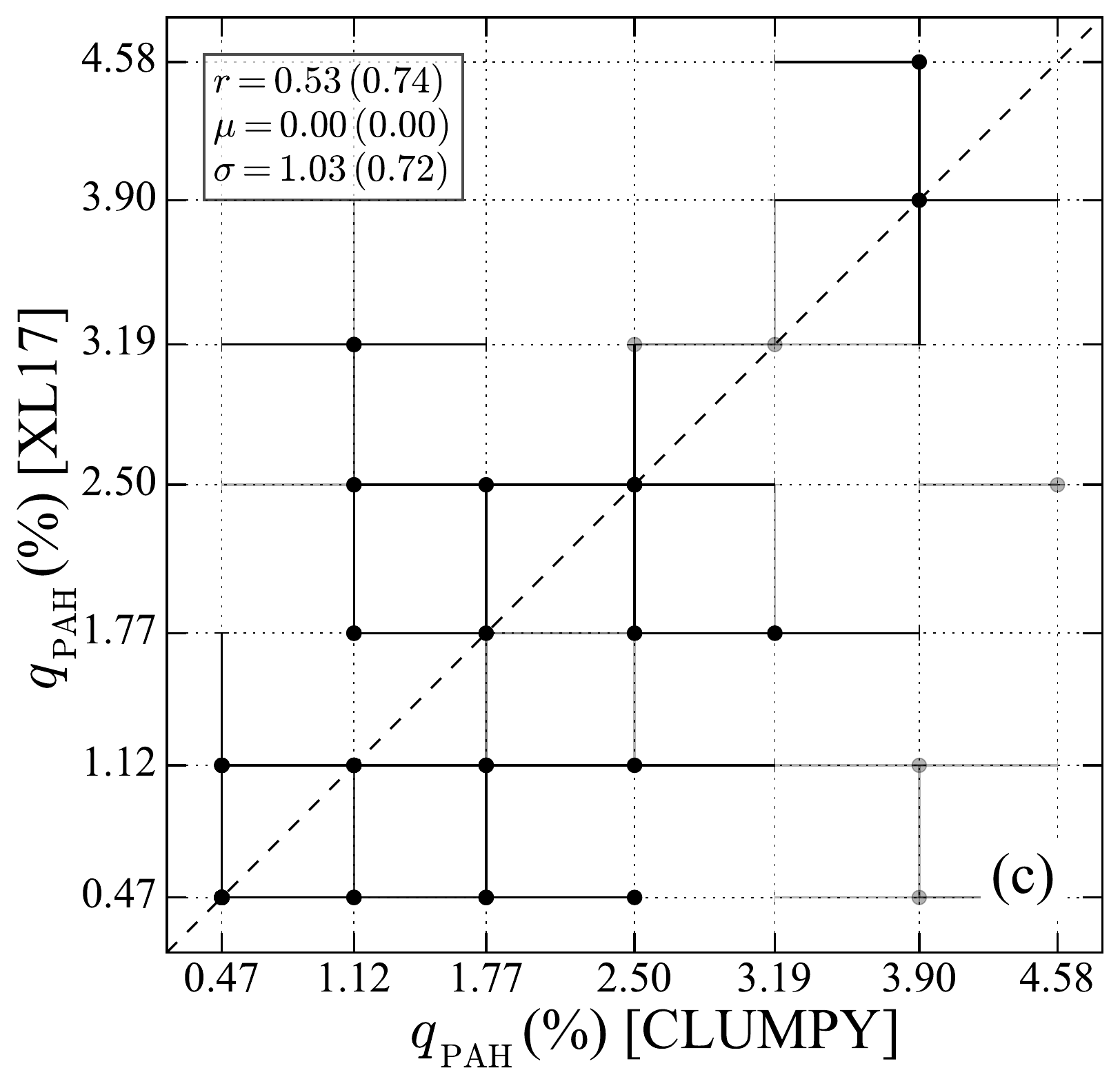} &
\includegraphics[height=0.35\textheight]{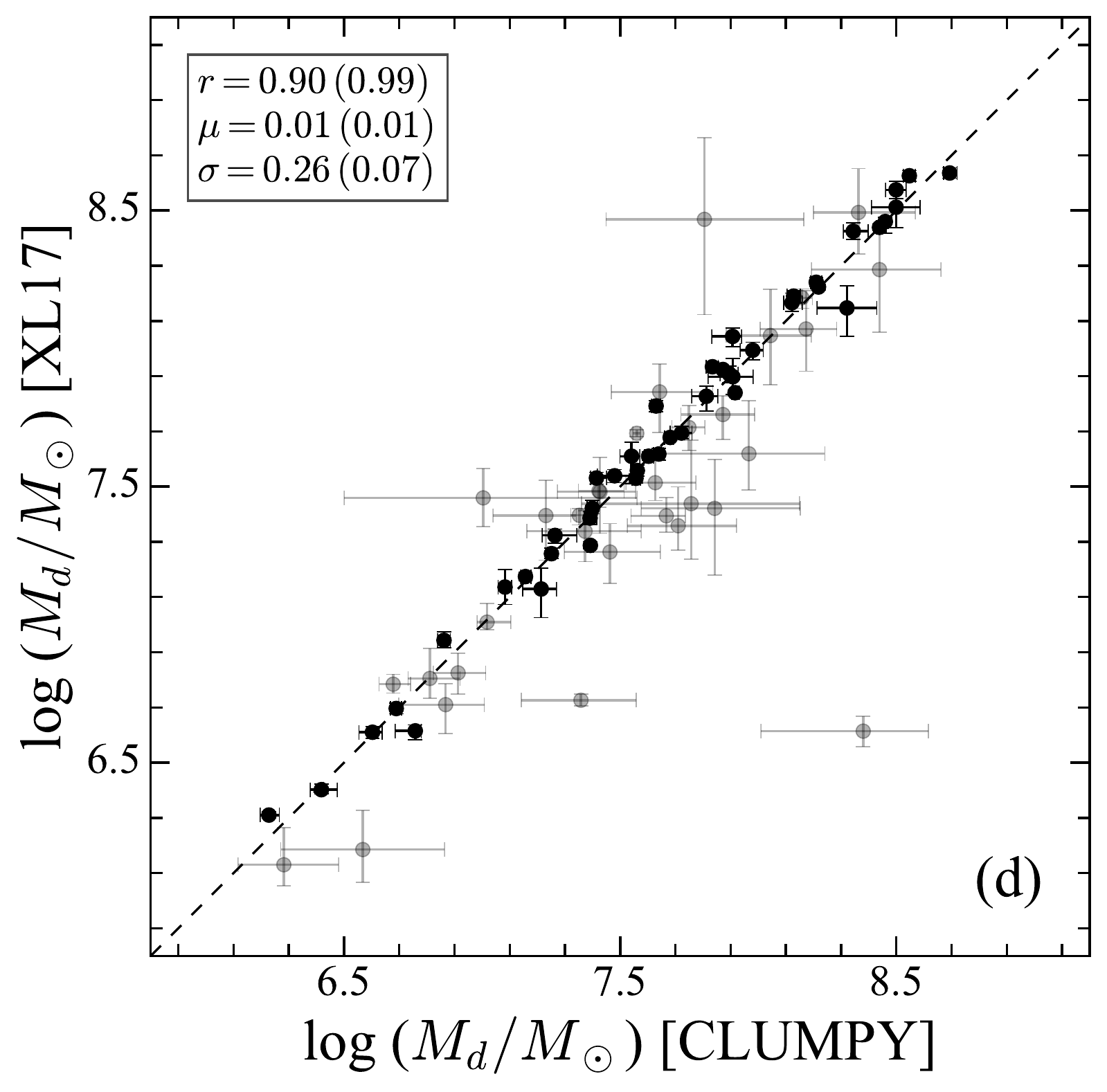}
\end{tabular}
\caption{Comparison of the impact of the choice of torus model (XLH17 or CLUMPY)
on the best-fit DL07 parameters (a) $U_\mathrm{min}$, (b) $\gamma$, (c) $q_\mathrm{PAH}$, 
and (d) $M_{d}$.  The black points are the 44 PG quasars whose FIR data well 
constrain the peak and the Rayleigh-Jeans tail of the dust emission.  The gray points 
are the 32 objects whose FIR data can still constrain the DL07 model.  The dashed 
line is the one-to-one relation.  The legend in each panel shows the Pearson correlation 
coefficient ($r$) and the median ($\mu$) and standard deviation ($\sigma$) of the 
deviation from the linear relation (y$-$x).  The first set of values is for the entire 
sample; the values for the most robust subsample (black points) are given in parentheses.  
The errors of the discrete parameters ($U_\mathrm{min}$ and $q_\mathrm{PAH}$) 
are sometimes not resolvable if they are smaller than the grid size. The error bars 
for $M_d$ are sometimes smaller than the symbols (especially for the black points).}
\end{center}
\label{fig:compxl17}
\end{figure*}

Figure \ref{fig:compxl17} compares the best-fit DL07 parameters using the two
torus emission models. Both yield consistent measurements of $U_\mathrm{min}$
and $M_{d}$, especially for objects with sufficient \herschel\ detections that
cover well the peak and the Rayleigh-Jeans tail of the FIR SED (black points).
For the rest of the objects (gray points), the scatter is considerably larger,
reflecting the fact that the DL07 model is more sensitive to the torus models
when the SED coverage does not provide sufficient constraints.  The best-fit
values of $\gamma$, as expected, show large and systematic deviations, since
$\gamma$, which mainly controls the MIR emission of the DL07 model, is strongly
degenerate with the torus component.  The parameter $q_\mathrm{PAH}$ exhibits
the worst performance with the XLH17 model, most likely due to its inability to
properly handle the SED at wavelengths $\lesssim 7$ \micron, where many PAH
features lie.

\section{Fitting and reliability}
\subsection{SED Fitting with emcee}
\label{apd:mcmc}

In order to fit a model with up to 19 free parameters,\footnote{There are 14
parameters for the IR SED models, two parameters for the synchrotron emission 
of radio-loud objects, and three parameters to describe the uncertainties.} the
MCMC method is preferred to sample the posterior probability distribution to
find a physically optimal solution.  We use the Python package \mc\
\citep{ForemanMackey2013}, a widely used tool, to perform the MCMC sampling.
This package uses the Affine Invariant MCMC Ensemble sampler to probe the
parameter space with a number of ``walkers.''  A walker randomly proposes the
next step based on the position of the other walkers.  The likelihood of the new
position is calculated based on the model and data.  The chance for the walker
to move to the new position depends on the ratio of the likelihood of the
current position to that of the new position.  The sequence of the positions
visited by a walker forms a ``chain'' following the Markov process.  When the
chains become long enough, they can be used to construct  the posterior PDF of
the parameters.

Since our likelihood function is very complicated, we need to use many walkers
and long enough burn-in runs in order to find the global optimum. We first use
128 walkers, initialized randomly in the parameter space allowed by the prior.
Then we run the sampling three times with chain length [8000, 5000, 3000] 
steps as the burn-in rounds.  After each burn-in round, we find the ``maximum 
a posteriori'' (MAP) and randomly initialize the walkers within a hyper-ball with
a radius of 10\% of the parameter prior ranges, centering at the MAP and resetting
the sampler for the next round.  After the first three burn-in rounds, we initialize 
the walkers within a hyper-ball with a radius of 1\% of the prior ranges, still 
centered at the MAP of the previous sampling, and run another burn-in sampling 
with a length of 800 steps. We repeat the initialization step and run a 600 step 
final sampling.  We drop the first 300 steps of all the walkers and use the rest 
of the chains ($128 \times 300$ points) to define the full posterior PDF.  
To determine the best fit of one parameter, we marginalize the posterior
PDF of all the other parameters and calculate the median.  All quoted
uncertainties represent the 68 percent confidence interval determined from 
the 16th and 84th percentiles of the marginalized posterior PDF.  The strategy 
to run the MCMC is found effective to obtain reliable fits, although the fitting
results are not sensitive to the detailed choices of the number of burn-in
rounds and the length of the chains, as long as the burn-in rounds are long
enough.

In the current work, we simply use uniform priors for all the parameters.
Namely, for a given parameter $X$, $\mathrm{ln}\, prior=0$ for
$X\in [X_\mathrm{min}, X_\mathrm{max}]$, while $\mathrm{ln}\,prior=-\infty$ for
$X$ out of the range.  The prior ranges are chosen to be wide enough to include
physically meaningful parameter ranges.  It is worth mentioning that the prior
ranges of the covariance model, $a$ and $\tau$, are crucial for a successful fit.
In order to achieve a reasonable fitting result, the prior of $a$ should be, at
most, comparable to the typical flux of the spectrum, and the prior of $\tau$
should be comparable to the length of the typical structure in the spectrum.
Much wider prior ranges may lead the fit to be trapped into some unphysical
solutions.  Taking these considerations into account, we choose
$-10<\mathrm{ln}\,a<5$ and $-5<\mathrm{ln}\,\tau<2.5$ as the fiducial priors.
The lower boundaries are not important, as long as they are small enough.  For
some of the bright objects, we need to enlarge the prior to
$-10<\mathrm{ln}\,a<10$, so that the posterior probability distribution of
$\mathrm{ln}\,a$ can be a regular, Gaussian-like profile. Meanwhile, for some
objects whose \irs\ spectra are usually not well described by the torus model,
we need to constrain $a$ with a hard boundary (e.g., $\mathrm{ln}\,a<1$);
otherwise, the torus model cannot match the spectrum well. For these objects
(17 objects with CLUMPY model), the model-dependent uncertainty may be larger
than the rest of the objects.

The ensemble sampler of \mc\ requires the parameters of the model to be a
continuous variable.  Therefore, we need to interpolate the discrete parameters
that determine the precalculated templates.  We use the k-Nearest Neighbor
(kNN) method to overcome the discreteness of the DL07 model and the grain size
of the XLH17 model.  We build the K-D Tree \citep{Bentley1975} with the
{\tt scikit-learn} package \citep{scikit-learn} and find the nearest parameters
of the templates for the input parameters.  Since CLUMPY consists of more than
$10^6$ templates densely sampling the relevant parameter space, we use a
dedicated code\footnote{\url{https://github.com/rnikutta/ndiminterpolation}}
(R. Nikutta 2017, private communication) to interpolate the templates with
multilinear interpolation.  The discreteness, especially the coarse grid, may
influence the posterior probability distribution of the parameters.  For example,
the uncertainty of the dust mass may be underestimated because the grids of
$U_\mathrm{min}$ are too coarse.  We know that $U_\mathrm{min}$ and $M_d$ 
are degenerate: a smaller $U_\mathrm{min}$ leads to a larger $M_d$.  Therefore, 
the distribution of $M_d$ is likely limited, since usually all of the walkers are
trapped between two grid points of $U_\mathrm{min}$.  Fully addressing this
problem is beyond the scope of the current work; nevertheless, we test the
reliability of our fitting code and the uncertainty estimation, as described
below.

\subsection{Reliability of Fitting}
\label{apd:rel}
We generate mock SEDs based on the real data of the PG quasars\footnote{We
exclude objects whose \herschel\ data cannot constrain the DL07 model.}. We use
the best-fit parameters of each quasar to generate the SED model.  We calculate
the synthetic spectral and photometric data at the same rest-frame wavelengths
as the real data.  The uncertainties of the real SED are also used as the
uncertainties of the mock SED.  The mock data values are then perturbed around
themselves assuming a Gaussian probability distribution with the standard
deviation as the uncertainties.  Sometimes, the uncertainties of some spectral
points and/or {\it Herschel}\ photometric points are larger than one-third of the
synthetic mock values.  In that case, we use one-third of the mock value as the
standard deviation so that the perturbation will not be too large.  Some targets
have upper limits in some of the {\it Herschel}\ bands.  Their synthetic
photometric data are then replaced by the real upper limits.  Furthermore, we
perturb the \wise, \spitzer, \pacs, and \spire\ data by 3\%
\citep{Jarrett2011ApJ}, 5\% (\mips\
handbook\footnote{\url{http://irsa.ipac.caltech.edu/data/SPITZER/docs/irac/iracinstrumenthandbook/}}),
5\% \citep{Balog2014ExA}, and 5\% \citep{Pearson2014ExA}, respectively, to model
the calibration systematic uncertainties.

\begin{figure*}
\begin{center}
\begin{tabular}{c c}
\includegraphics[height=0.35\textheight]{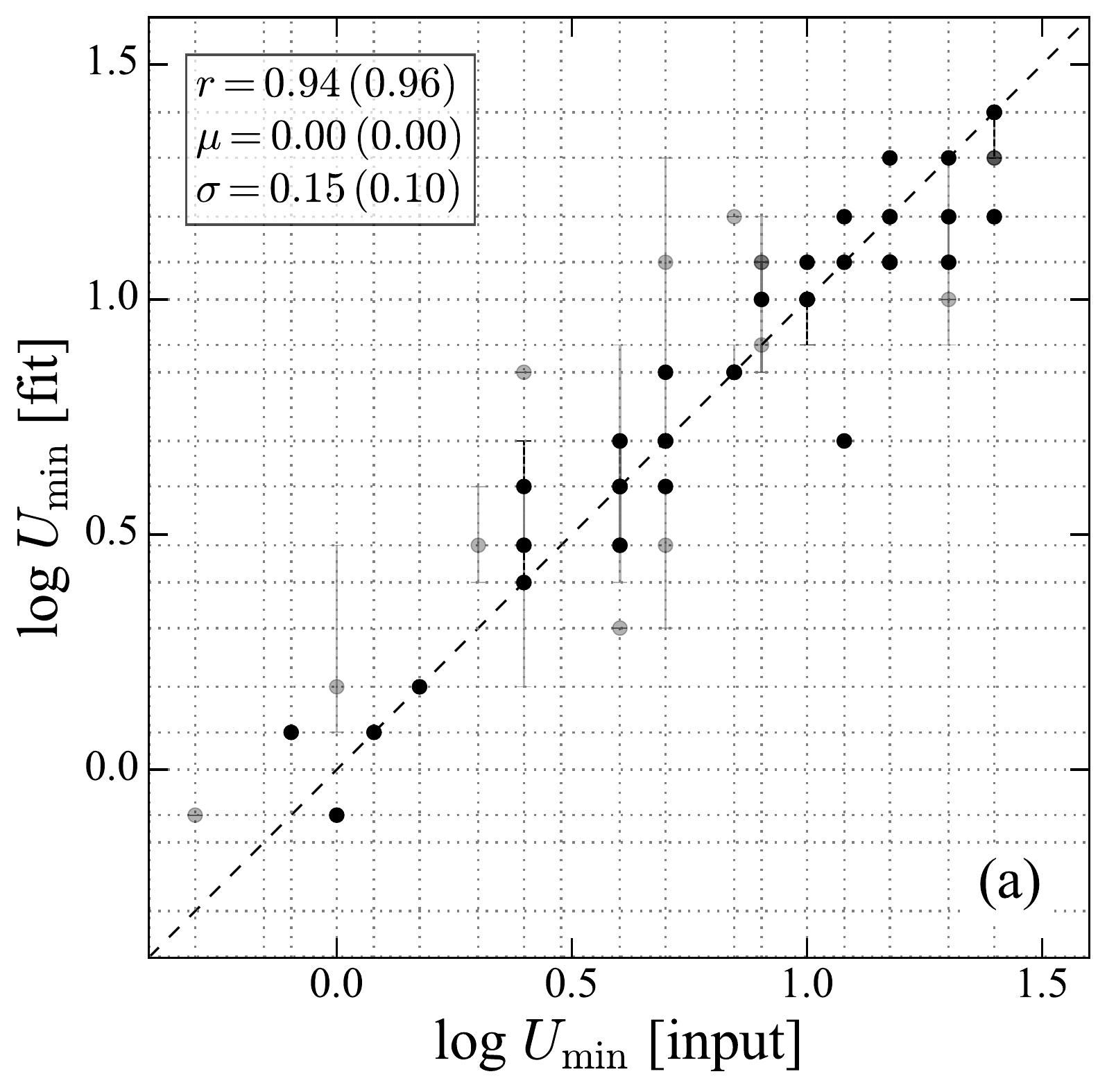} &
\includegraphics[height=0.35\textheight]{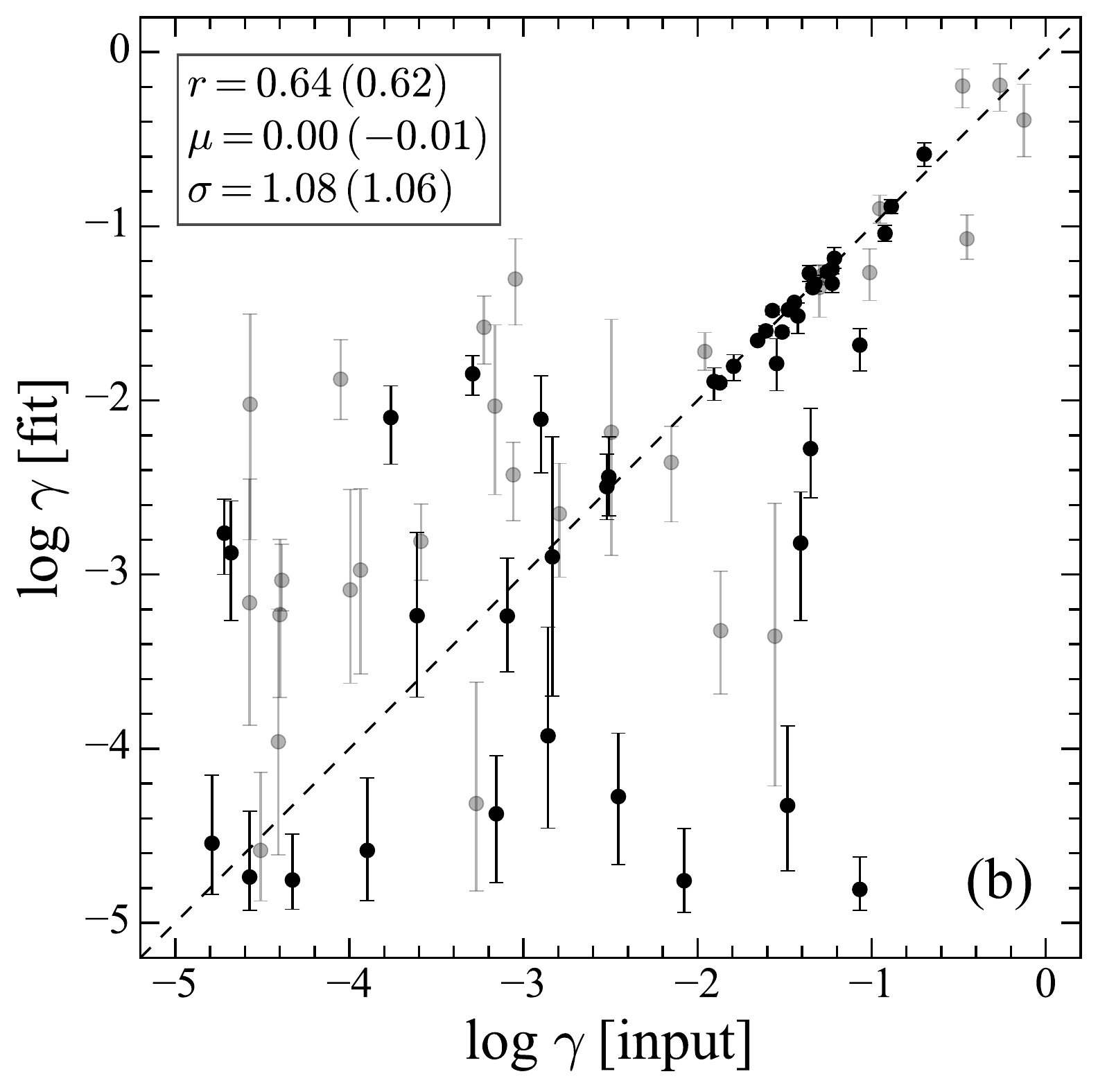} \\
\includegraphics[height=0.35\textheight]{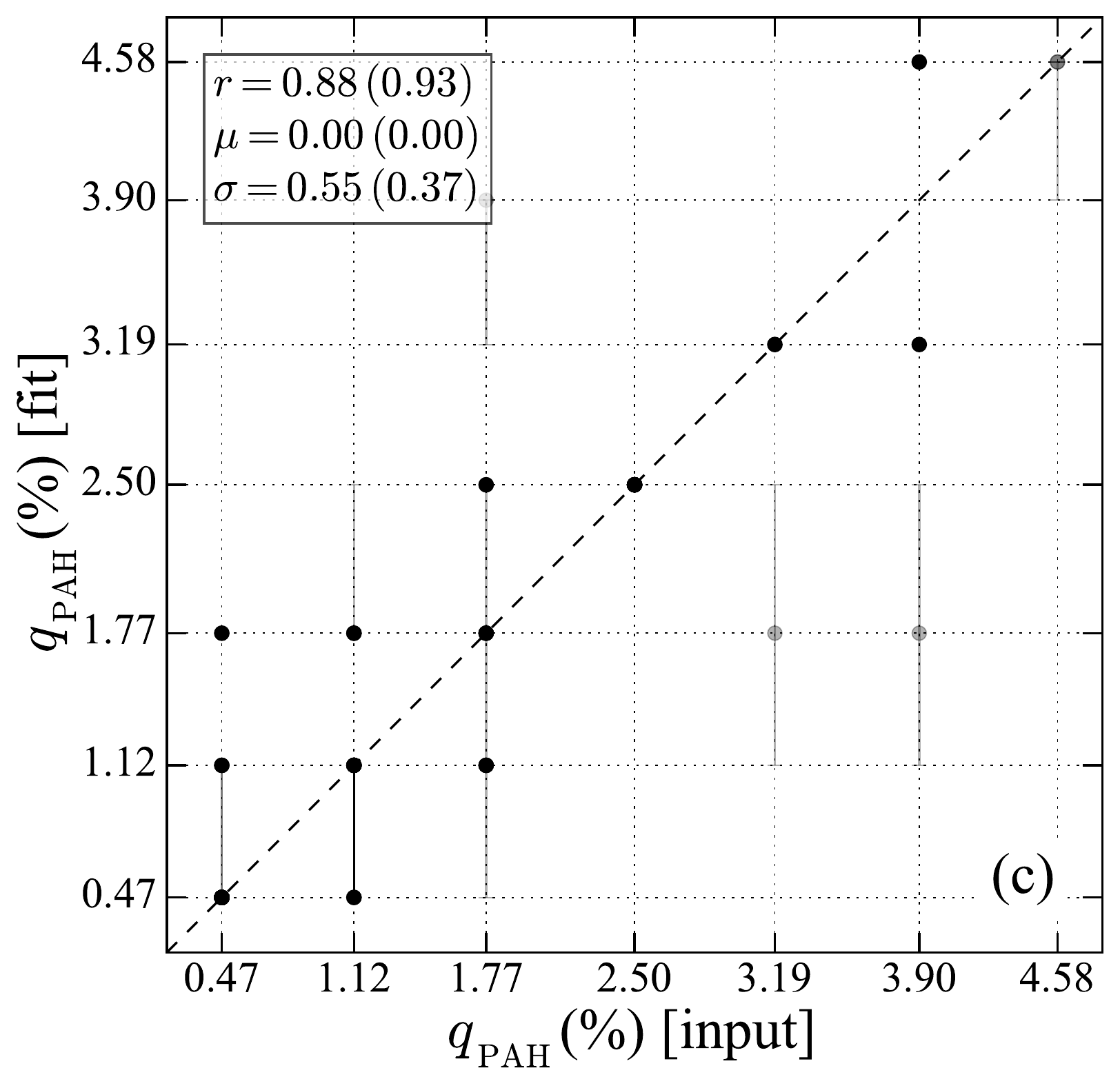} &
\includegraphics[height=0.35\textheight]{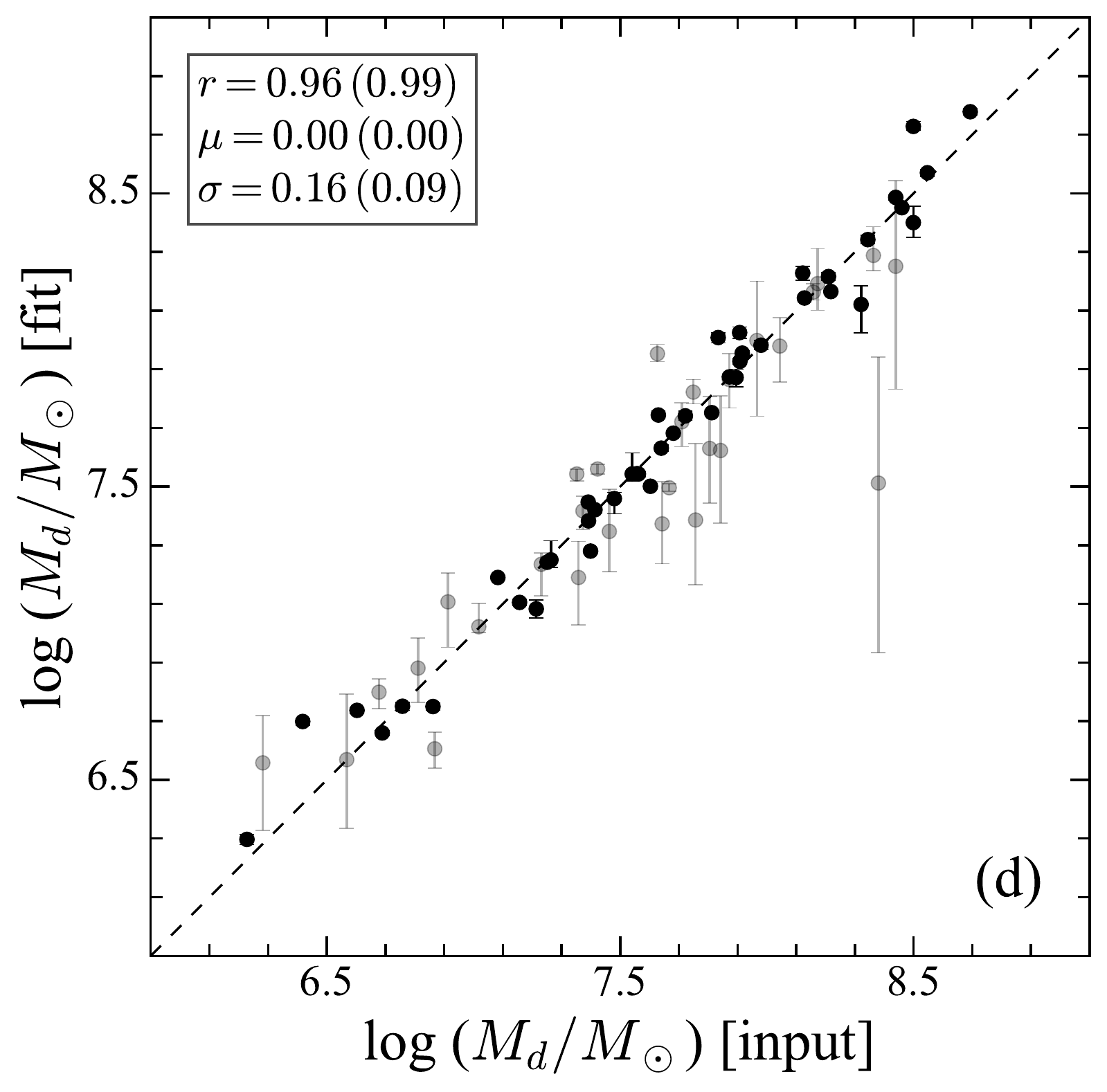}
\end{tabular}
\caption{Fits of mock SEDs to evaluate the degree to which the input parameters
of the DL07 model parameters, (a) $U_\mathrm{min}$, (b) $\gamma$, (c) $q_\mathrm{PAH}$, 
and (d) $M_d$, can be recovered.  The black points are the 44 PG quasars whose FIR 
data well constrain the peak and the Rayleigh-Jeans tail of the dust emission.  The gray 
points are the 29 objects whose \herschel\ data can still constrain the DL07 model; three 
objects with FIR data from the archive are not included.  The dashed line is the one-to-one 
relation.  The legend of each panel shows the Pearson correlation coefficient ($r$) and 
the median ($\mu$) and standard deviation ($\sigma$) of the deviation from the linear 
relation (y$-$x).  The first set of values is for the entire sample; the values for the most 
robust subsample (black points) are given in parentheses.  The errors of the discrete 
parameters ($U_\mathrm{min}$ and $q_\mathrm{PAH}$) are sometimes not resolvable 
if they are smaller than the grid size.  The error bars for $M_d$ are sometimes smaller 
than the symbols (especially for the black points).}
\end{center}
\label{fig:relDL07}
\end{figure*}

We fit the mock SEDs using the same method as that used for the real SEDs.  
The best-fit and input parameters of the DL07 model are compared in Figure
\ref{fig:relDL07}.  The input and best-fit parameters of $U_\mathrm{min}$,
$q_\mathrm{PAH}$, and $M_{d}$ are tightly correlated without systematic
deviation.  As $U_\mathrm{min}$ and $q_\mathrm{PAH}$ are discrete parameters, 
we plot their grid as dotted lines.  The typical scatter in $U_\mathrm{min}$ and
$q_\mathrm{PAH}$ is $\lesssim 2$ grid points, especially for objects whose FIR
data well constrain the peak and Rayleigh-Jeans tail of the dust emission.  For
$\gamma$, the correlation is reasonably good for $\gamma \gtrsim 0.01$, below
which the scatter becomes large, albeit showing no systematic deviation.  The
reasons are as follows: (1) $\gamma$ mainly controls the MIR emission of the
DL07 model, while the MIR emission of the quasar is dominated by the AGN 
torus; (2) $\gamma$ is usually small, rendering it more sensitive to mismatch 
between the torus model and the spectra; (3) systematic uncertainties in the 
\irs\ spectra with respect to the FIR data may lead to a large error on $\gamma$, 
especially when it is small.  In view of these complications, we should exercise 
caution in interpreting the results of $\gamma$.

As just mentioned, the uncertainties of the fitting results may not reflect
their true errors for parameters that are discrete.  For example, in the case 
of $U_\mathrm{min}$, even some of the robust fits with small error bars 
(black points in Figure \ref{fig:relDL07}) deviate from the one-to-one line 
by two grid points.  The same holds for $M_d$.  The situation is not as 
serious for the less robust fits (gray points).  Note that the primary goal 
of this study is to derive gas masses, whose final uncertainty is likely 
dominated by the uncertainty in \gdr, especially the unknown systematic 
uncertainty due to the \hi\ gas distribution (Section \ref{sssec:mgas}).  As 
the exact dust mass uncertainty will not affect our main conclusions, we 
will directly quote its value provided by our MCMC code.  Future refinement 
of the parameter grids of the DL07 templates would be valuable to obtain 
more accurate dust masses. 

\begin{figure*}
\begin{center}
\begin{tabular}{c c}
\includegraphics[height=0.35\textheight]{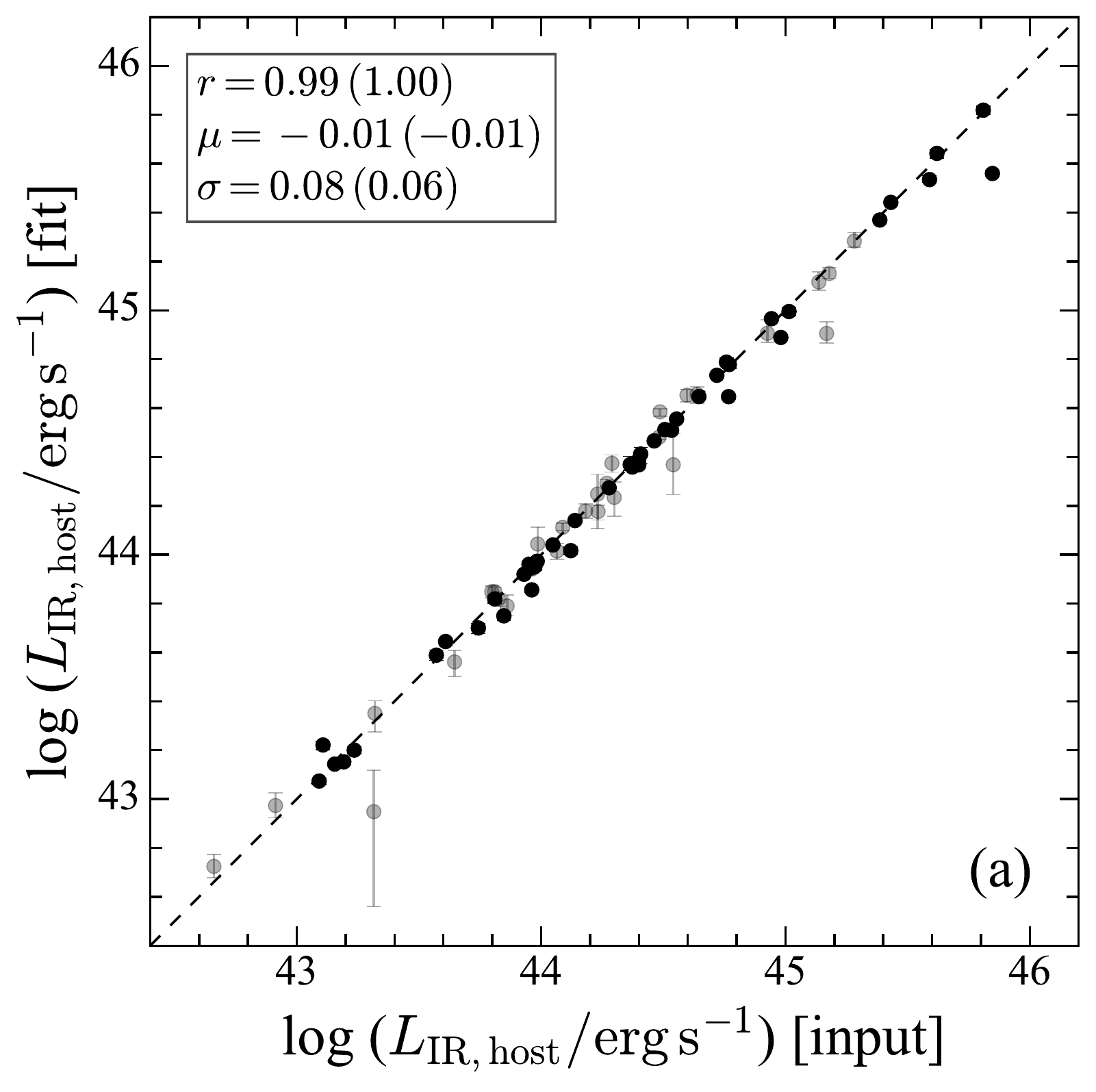} &
\includegraphics[height=0.35\textheight]{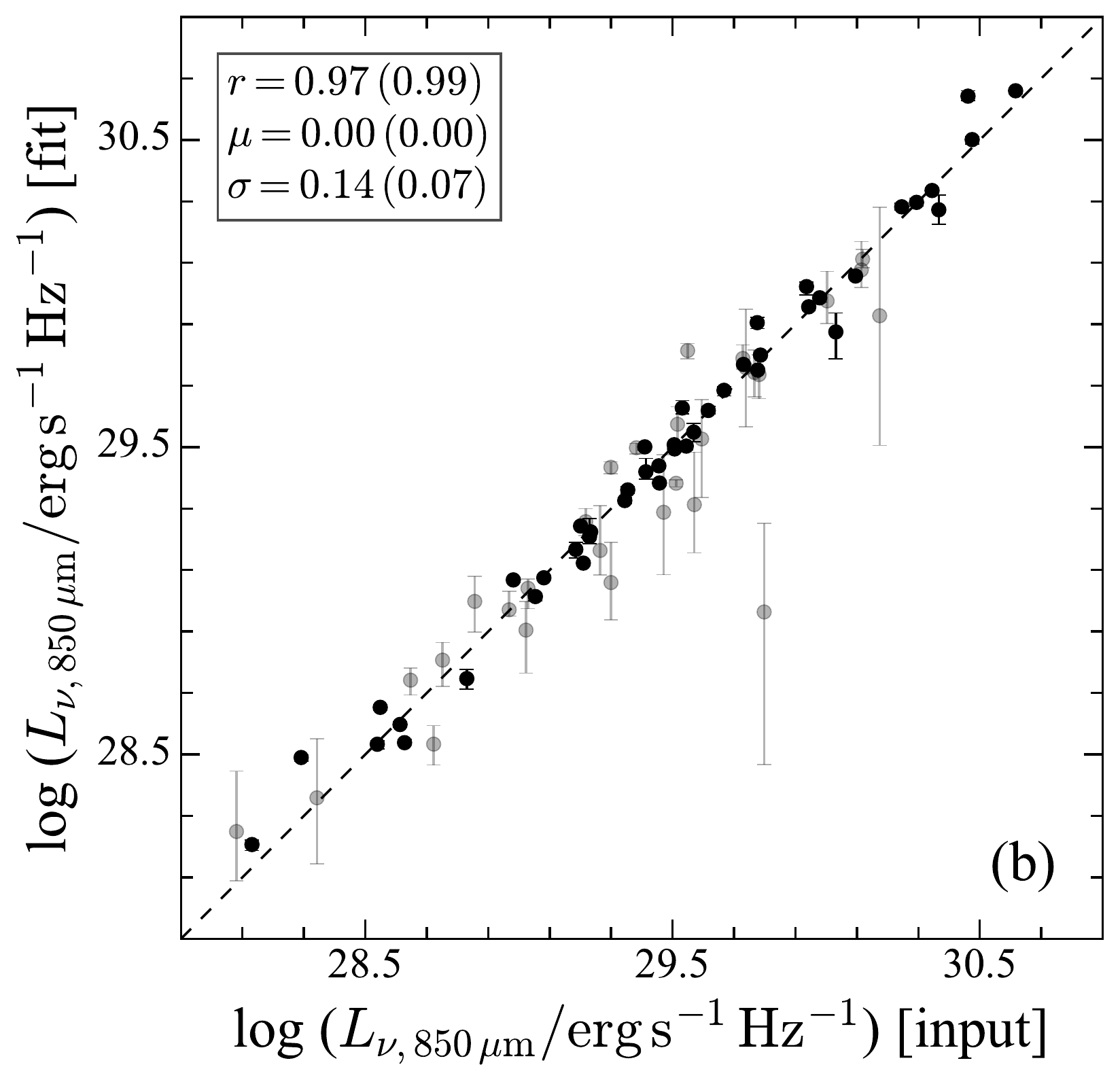}
\end{tabular}
\caption{Fits of mock SEDs to evaluate the degree to which the total IR
(8--1000 \micron) luminosity ($L_{\rm IR, host}$) and specific luminosity at 850
\micron\ ($L_\mathrm{\nu, 850\,\mu m}$) can be recovered using the DL07 model.
Symbols and conventions same as in Figure \ref{fig:relDL07}.}
\label{fig:relL}
\end{center}
\end{figure*}

Figure \ref{fig:relL} demonstrates that our fitting method can robustly
decompose the dust torus emission to yield accurate measurements of the
integrated IR (8--1000 \micron) luminosity of the galactic dust emission, as
well as the specific luminosity at 850 \micron.  Besides the DL07 parameters, 
we do not discuss other model components, which are not the primary focus 
of this study.  We simply note, in passing, that the input parameters of the 
CLUMPY torus model---apart from the optical depth $\tau_V$---are usually 
not well-reproduced by the fits, most likely because of the degeneracy with 
the ad hoc hot dust (BB) component.  A more comprehensive dust torus model 
(e.g., \citealt{GarciaGonzalez2017MNRAS,Honig2017ApJL}) is needed if we 
wish to truly study the properties of the AGN torus using the SED fitting method.  
From the posterior PDF of the parameters, we do not observe a clear degeneracy 
between the parameters of DL07 and the other components (e.g., CLUMPY).  
This is likely because 
(1) the \irs\ spectra effectively constrain the CLUMPY and DL07  models, so they 
cannot vary as freely as the fittings with photometric data only, (2) the discreteness 
of $U_\mathrm{min}$ dominates the parameter uncertainties of DL07.  As shown in 
Table \ref{tab:result}, the uncertainties of $U_\mathrm{min}$ are often unresolved.  
Unless the parameter space of the DL07 templates is refined, it is hard to analyze 
the degeneracies between the CLUMPY and DL07 parameters.  Therefore, with 
the current models, the degeneracy of the torus and DL07 parameters is dominated 
by the systematics from different torus models adopted in the fitting.  As discussed 
in Appendix \ref{apd:xl17}, this does not significantly affect $U_\mathrm{min}$ and 
$M_d$, especially when the FIR data are good enough to constrain the peak of the 
SED.  More comparisons of the SED fits with different torus models will be presented 
in a forthcoming work \citep{Zhuang2017}.

In summary: our SED fitting method can reliably measure $U_\mathrm{min}$,
$q_\mathrm{PAH}$, and $M_d$ for nearly the entire sample of PG quasars, except
for those with insufficient FIR data to constrain the DL07 model well.

\section{Systematics of Fitting the SED with MBB Model}
\label{apd:mbb}

We study the possible factors that might bias the dust mass $M_d$ and dust
temperature $T_d$ when the SED is fitted with an MBB model, a common practice
in the literature.  We generate mock SEDs using the DL07 model with different
parameters and redshift.  The SED contains the \herschel\ bands between 100
and 500 \micron.  We also consider different levels of data quality, including
the impact of upper limits.  We first summarize our conclusions:

\begin{enumerate}
\item The shortest-wavelength (100 $\mu$m) band is easily contaminated by
emission from warm dust, such that $T_{d}$ is biased toward higher values and
$M_{d}$ is underestimated.  This effect becomes serious when $\gamma$ is high,
$U_\mathrm{min}$ is low, or the redshift is high.  The parameter
$q_\mathrm{PAH}$ only marginally affects the fitting.  This bias in principle can be
mitigated by dropping the 100 \micron\ band from the fit, but in practice this
benefit is offset by the larger uncertainties or upper limits often encountered
in the 160--500 \micron\ bands.

\item The SED can hardly constrain $M_d$ and $T_d$ when there are two or more
upper limits at the longest bands.  As the data mainly constrain the
Rayleigh-Jeans tail, the fits tend to overestimate $T_{d}$ and underestimate
$M_{d}$.
\end{enumerate}

\begin{figure*}
\begin{center}
\begin{tabular}{c c}
\includegraphics[height=0.35\textheight]{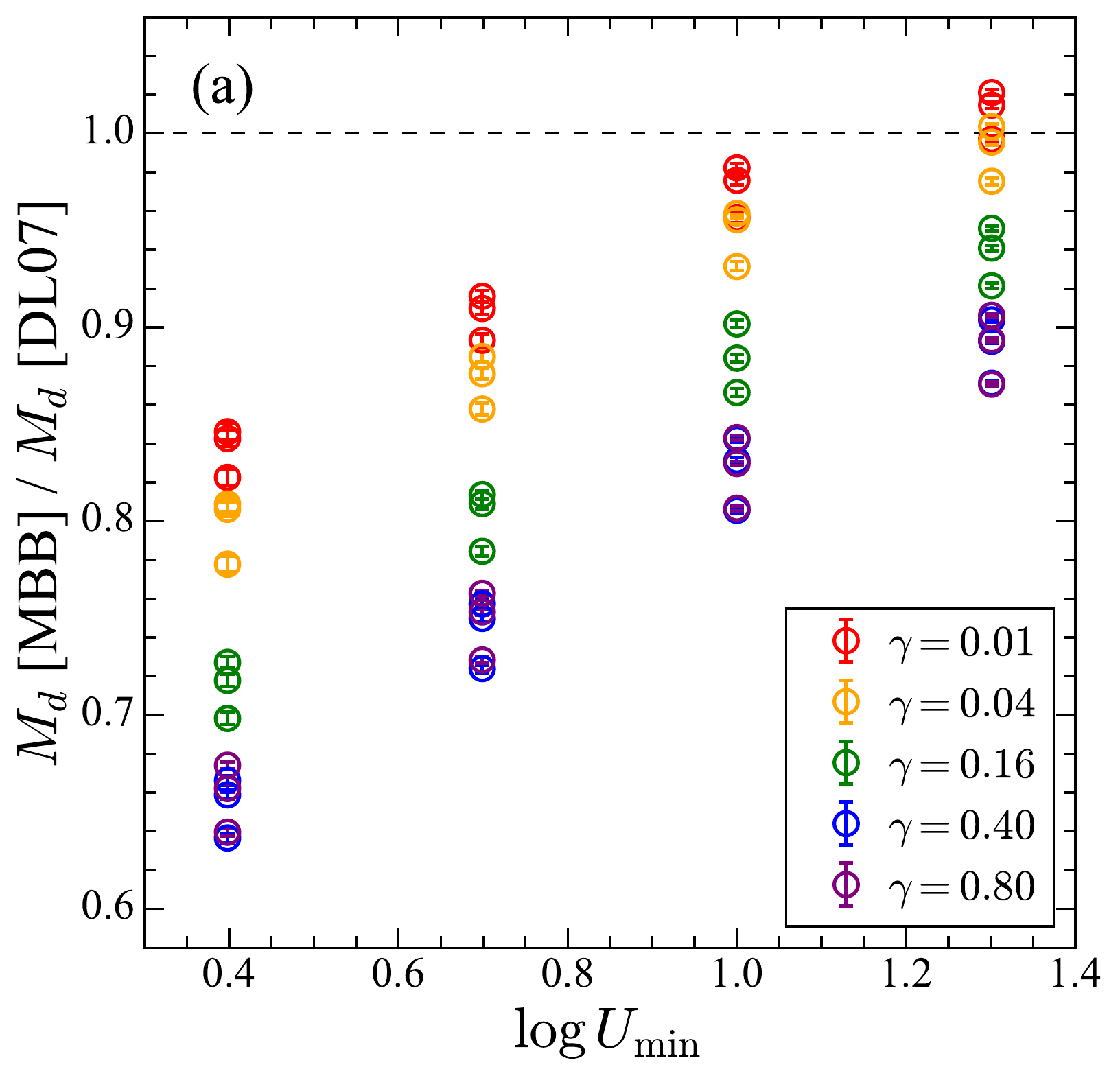} &
\includegraphics[height=0.35\textheight]{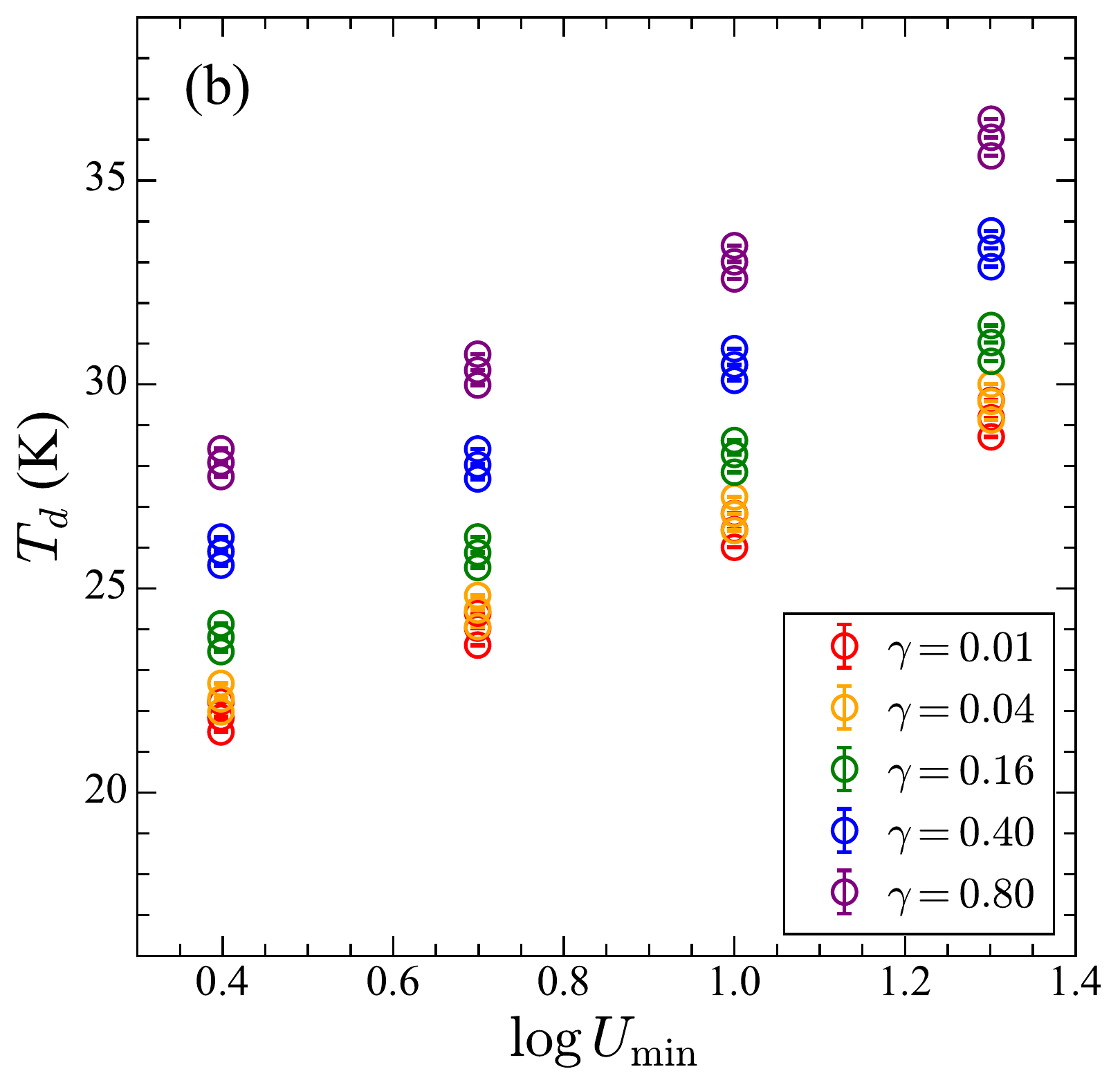}
\end{tabular}
\caption{Variation with $U_\mathrm{min}$ and $\gamma$ of (a) the dust mass 
derived from the best-fit MBB model relative to that derived from the DL07 model 
and (b) the dust temperature of the best-fit MBB model. Below the dashed line, 
the MBB model underestimates the dust mass. Each combination ($U_\mathrm{min}$, 
$\gamma$) contains three values of $q_\mathrm{PAH}$.}
\label{fig:mbb_pars}
\end{center}
\end{figure*}

We generate mock SEDs using the DL07 model with different values of
$U_\mathrm{min}$ (2.5, 5, 10, 20), $\gamma$ (0.01, 0.04, 0.16, 0.4, 0.8), and
$q_\mathrm{PAH}$ (0.47, 2.5, 4.58) and fit them with the MBB model (Figure
\ref{fig:mbb_pars}).  Fixing $\gamma$, the ratio of dust masses derived from the
MBB fit to the fiducial value from the DL07 model, $M_d$[MBB]/$M_d$[DL07],
decreases as $U_\mathrm{min}$ decreases (Figure \ref{fig:mbb_pars}(a)).
This reflects the fact that when $U_\mathrm{min}$ decreases, the FIR SED 
tends to peak at longer wavelengths ($>100\,\mu m$), such that more warm dust
emission on the shorter-wavelength side of the peak enters the 100 \micron\ band.
Note that although $T_d$ and $U_\mathrm{min}$ are positively correlated (Figure
\ref{fig:mbb_pars}(b)), as they should be, when $U_\mathrm{min}$ is low, 
the $T_d$ from the MBB fitting is biased to temperatures higher than it should 
be, due to the warm dust contamination.
On the other hand, at fixed $U_\mathrm{min}$, $M_{d}$ is increasingly
underestimated when $\gamma$ increases.  Comparing the two panels of
Figure \ref{fig:mbb_pars}, it is clear that higher values of $\gamma$ induce
more contamination by warm dust in the 100 $\mu$m band, which, in turn, leads
to a higher $T_d$ and hence lower $M_d$.  The parameter $q_\mathrm{PAH}$
has only a minor effect on $M_d$ and $T_d$.

\begin{figure*}
\begin{center}
\begin{tabular}{c c}
\includegraphics[height=0.35\textheight]{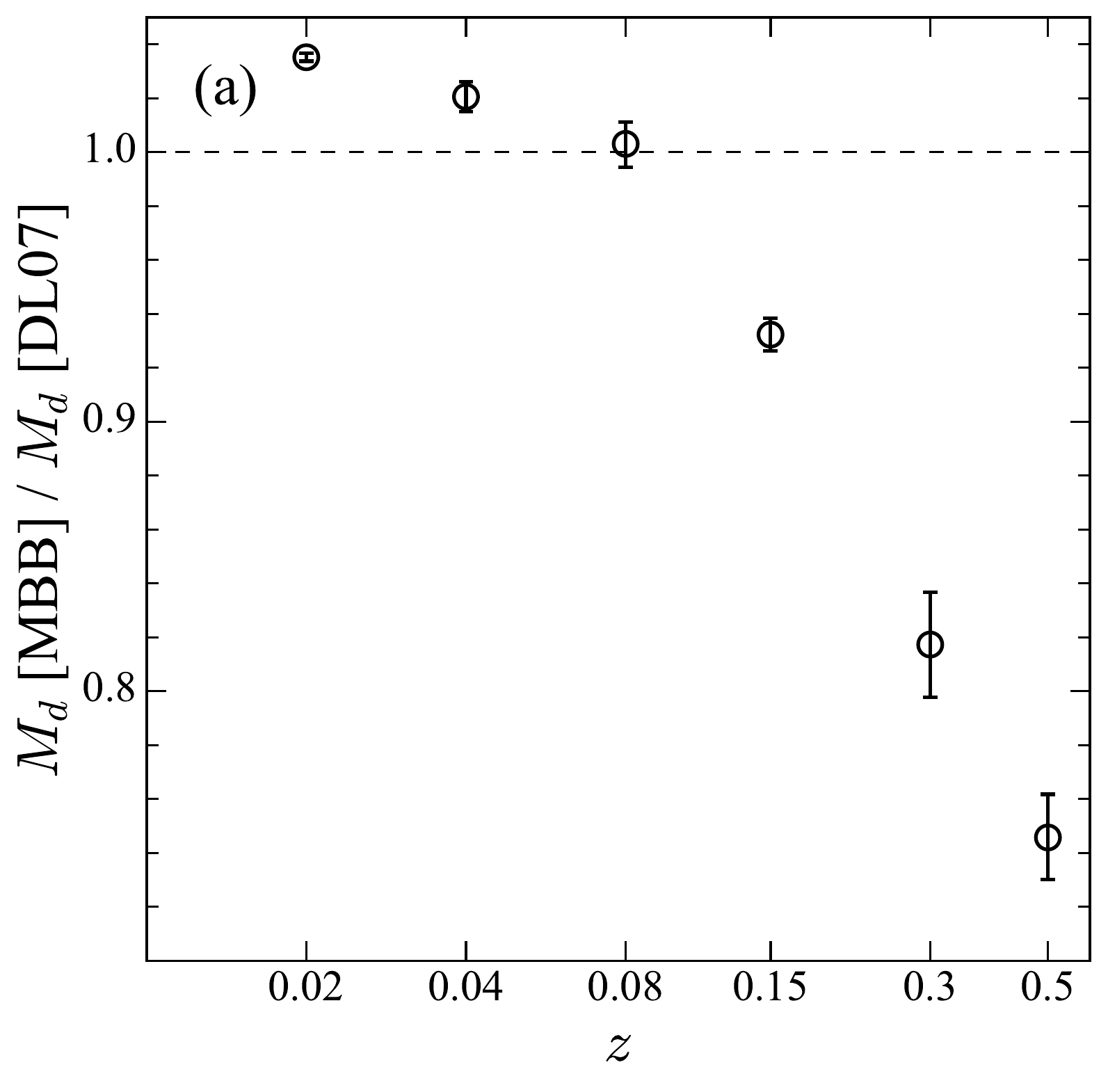} &
\includegraphics[height=0.35\textheight]{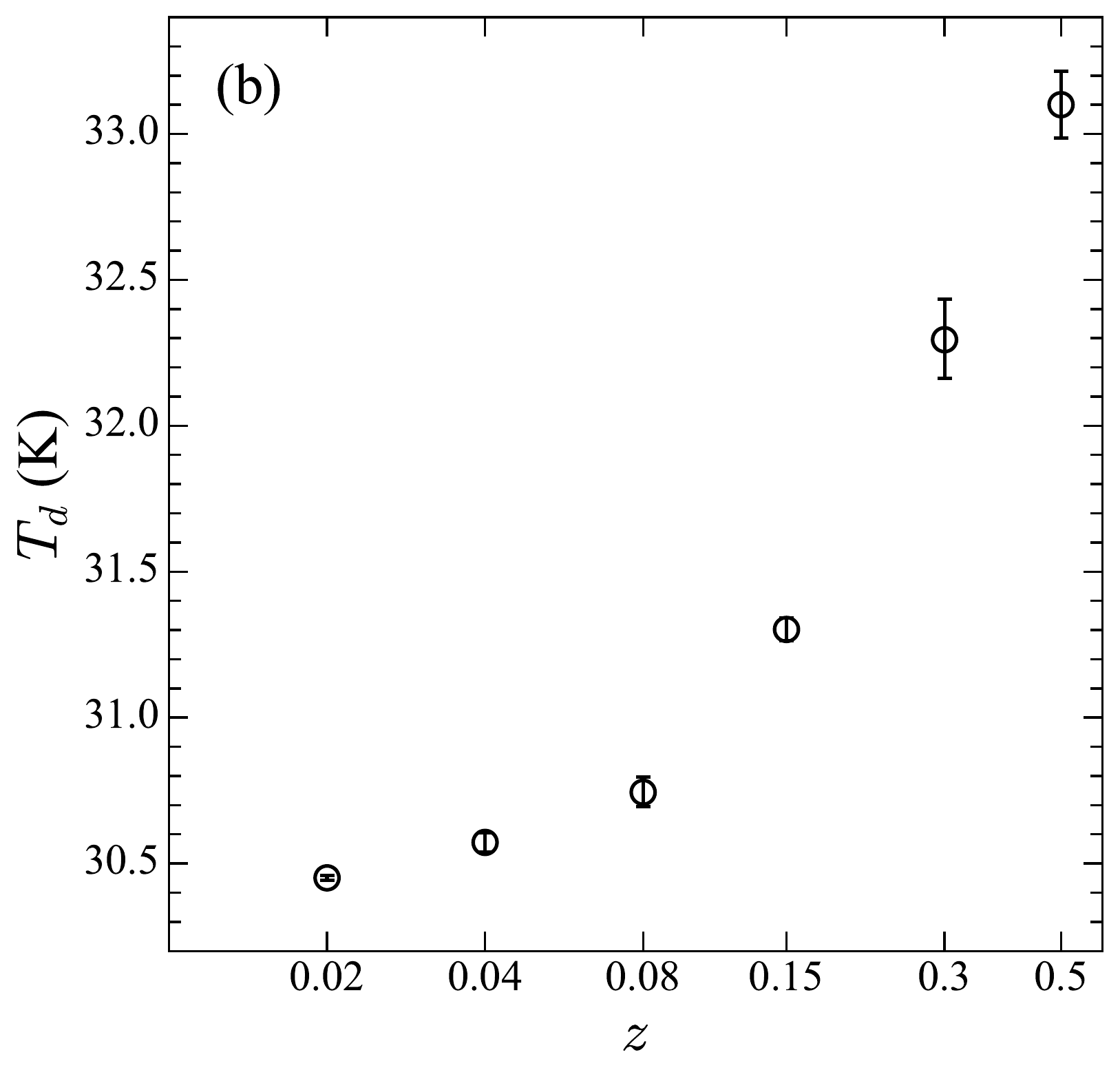}
\end{tabular}
\caption{Variation with redshift $z$ of (a) the dust mass derived from
the best-fit MBB model relative to that derived from the DL07 model and
(b) dust temperature of the best-fit MBB model.  Below the dashed line,
the MBB model underestimates the dust mass.}
\label{fig:mbb_z}
\end{center}
\end{figure*}

In order to quantify the effect of redshift, we generate mock SEDs for different
values of $z$ (0.02, 0.04, 0.08, 0.15, 0.3, 0.5) with a fixed set of fiducial
DL07 model parameters ($U_\mathrm{min} = 25$, $q_\mathrm{PAH} = 0.47$,
$\gamma = 0$, $M_d = 10^{8.1}\,M_\odot$).  Then, we fit the MBB model to the
100--500 \micron\ SED.  The best-fit $M_d$ decreases and $T_d$ increases as $z$
increases (Figure \ref{fig:mbb_z}), as a consequence of increased contamination
from warm dust emission when the peak of the SED shifts redward.

\begin{figure}
  \begin{center}
    \includegraphics[width=0.45\textwidth]{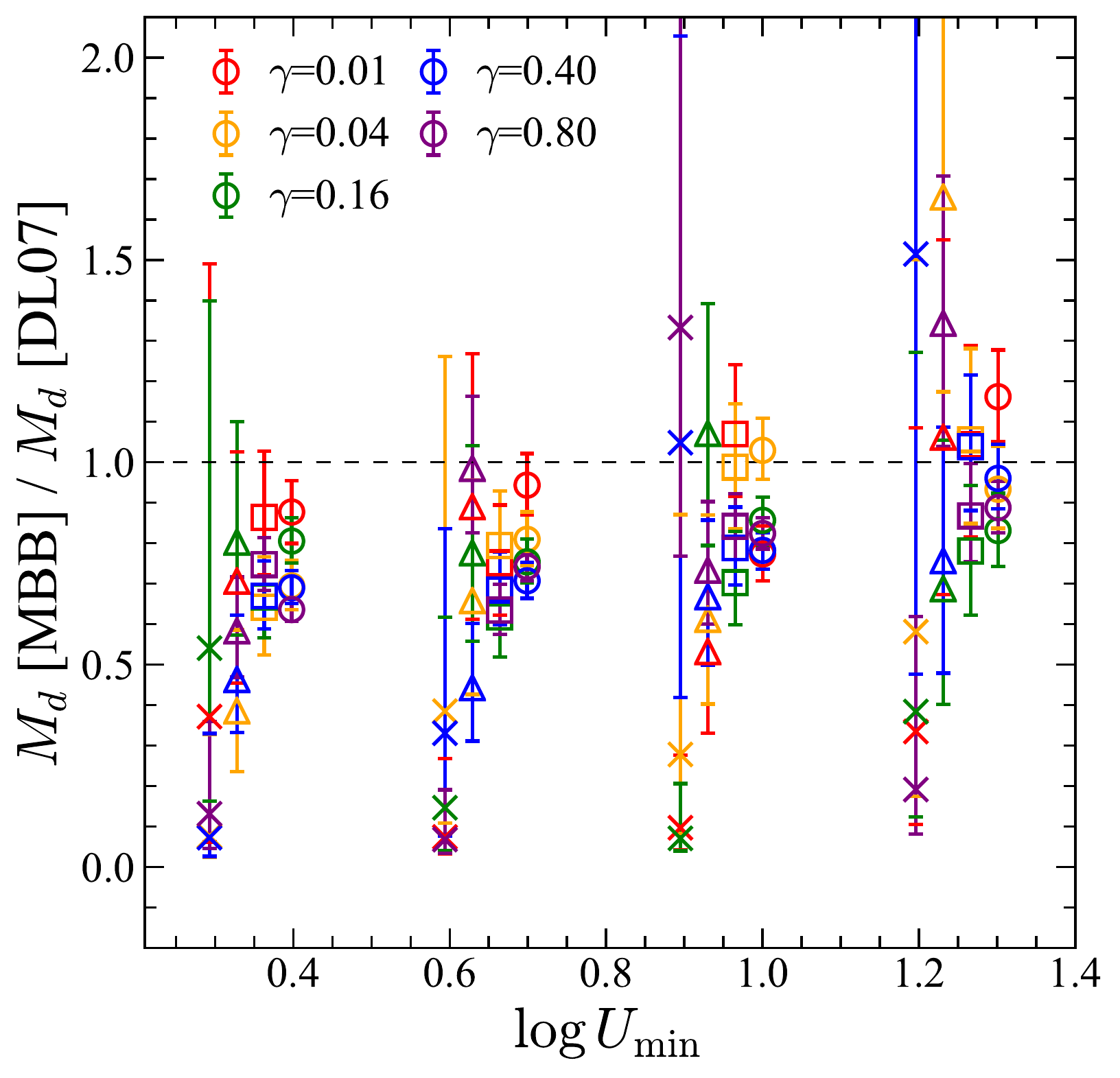}
\caption{Dust mass derived from the best-fit MBB model relative to that derived 
from the DL07 model as a function of $U_\mathrm{min}$ and $\gamma$.
The different symbols denote SEDs with different numbers of upper limits in
their spectral coverage.  Starting from the longest wavelength, circle = 0,
square = 1, triangle = 2, and cross = 3 upper limits, respectively.  For
clarity, points with the same value of $U_\mathrm{min}$ are offset slightly in
the horizontal direction.}
  \label{fig:mbb_uplim}
  \end{center}
\end{figure}

Finally, we generate mock SEDs with different numbers of upper limits (0, 1, 2, 3) 
to study the effect of non-detections on SED fits with the MBB model 
(Figure \ref{fig:mbb_uplim}).  Comparing the error bars and scatter of the
different symbols, it is clear that the uncertainty of the fits becomes larger
when there are more upper limits in the SED.  In particular, the scatter is
unacceptably large when there are as many as three upper limits.  In general, as
the number of upper limits increases toward longer wavelengths, the
Rayleigh-Jeans tail becomes more and more poorly constrained, $T_d$ is more
easily overestimated, and $M_d$ becomes systematically more underestimated.

\begin{figure*}
\centering
\includegraphics[width=0.95\textwidth]{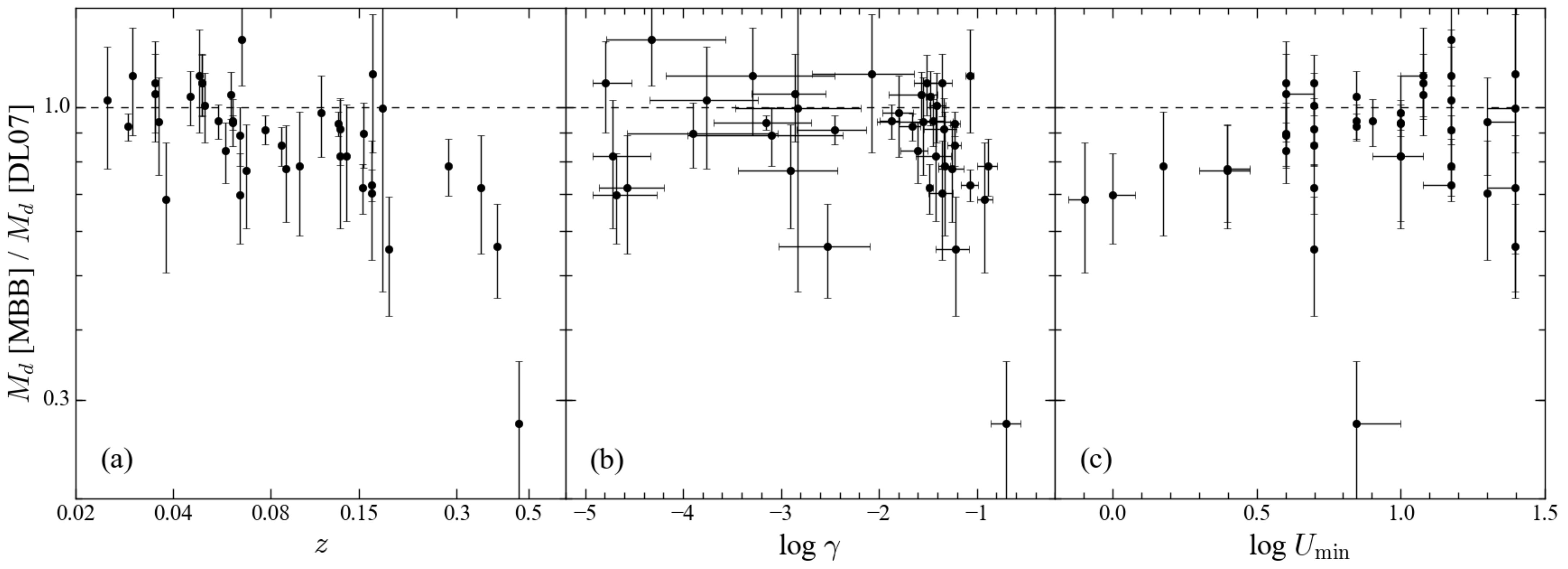}
\caption{Variation of the dust mass derived from the MBB model relative to that
derived from the DL07 model as a function of (a) $z$, (b) $\gamma$, and (c) 
$U_\mathrm{min}$.  The mass ratio is apparently decreasing toward higher 
$z$ and higher $\gamma$, for $\log \gamma > -2$.  There is no obvious trend 
with $U_\mathrm{min}$.}
\label{fig:pgmbb2}
\end{figure*}

For the PG quasars with $\geq 4$ detections in the \herschel\ bands, the 
dependence of $M_d$[MBB]/$M_d$[DL07] on DL07 parameters is shown in Figure 
\ref{fig:pgmbb2}.  It seems that $z$ and $\gamma$ are the main culprits for the 
underestimation of $M_d$[MBB].  Since $\gamma$ is easily biased in the fitting, 
we are unsure whether or not the trend with $\gamma$ is real.  The trend with $z$, 
however, is robust.  The $U_\mathrm{min}$ parameter does not significantly influence
$M_d$[MBB]/$M_d$[DL07], except that it becomes systematically lower than 1
when $\mathrm{log}\,U_\mathrm{min}<0.5$.

\end{document}

%% file: tab1.tex
\begin{longrotatetable}
\begin{deluxetable}{c c r c c c c c c c c c c r@{$\pm$}l c}
\tablecaption{Physical Properties of PG Quasars \label{tab:result}}
\tabletypesize{\scriptsize}
\tablehead{
\colhead{Object} &
\colhead{$z$} &
\colhead{$D_L$} &
\colhead{log $M_*$} &
\colhead{log $\lambda L_\lambda$(5100 \AA)} &
\colhead{FWHM$_\mathrm{H\beta}$} &
\colhead{log $M_\mathrm{BH}$} &
\colhead{log $M_\mathrm{bulge}$} &
\colhead{log $U_\mathrm{min}$} &
\colhead{$q_\mathrm{PAH}$} &
\colhead{log $\gamma$} &
\colhead{log $M_d$} &
\colhead{log $\delta_\mathrm{GDR}$} &
\mcl{2}{c}{log $M_\mathrm{gas}$} &
\colhead{Radio} \\
\colhead{} &
\colhead{} &
\colhead{(Mpc)} &
\colhead{($M_\odot$)} &
\colhead{(erg s$^{-1}$)} &
\colhead{(km s$^{-1}$)} &
\colhead{($M_\odot$)} &
\colhead{($M_\odot$)} &
\colhead{} &
\colhead{(\%)} &
\colhead{} &
\colhead{($M_\odot$)} &
\colhead{} &
\mcl{2}{c}{($M_\odot$)} &
\colhead{} \\ 
\colhead{(1)} &
\colhead{(2)} &
\colhead{(3)} &
\colhead{(4)} &
\colhead{(5)} &
\colhead{(6)} &
\colhead{(7)} &
\colhead{(8)} &
\colhead{(9)} &
\colhead{(10)} &
\colhead{(11)} &
\colhead{(12)} &
\colhead{(13)} &
\mcl{2}{c}{(14)} &
\colhead{(15)} 
}
\startdata
PG 0003$+$158 & 0.450 & 2572 & \nodata &   45.99 &              4751 &              9.45 &      11.65 &                      \nodata &                \nodata &                 \nodata &                 $<$8.9 & 2.09 & \mcl{2}{c}{$<$11.0} & S \\
PG 0003$+$199 & 0.025 &  113 & \nodata &   44.17 &              1585 &              7.52 &      10.00 & \ph{+}$1.18_{-0.00}^{+0.00}$ & $1.77_{-0.65}^{+0.00}$ & $-3.76_{-0.58}^{+0.52}$ & $6.23_{-0.03}^{+0.04}$ & 2.09 &     8.32 &     0.20 & Q \\
PG 0007$+$106 & 0.089 &  420 &   10.84 &   44.79 &              5085 &              8.87 &      11.15 & \ph{+}$0.90_{-0.06}^{+0.10}$ & $1.12_{-0.00}^{+0.00}$ & $-4.39_{-0.27}^{+0.30}$ & $7.67_{-0.13}^{+0.07}$ & 2.09 &     9.76 &     0.22 & F \\
PG 0026$+$129 & 0.142 &  693 &   10.88 &   45.07 &              1821 &              8.12 &      10.52 & \ph{+}$1.30_{-0.12}^{+0.10}$ & $0.47_{-0.00}^{+0.65}$ & $-3.06_{-0.55}^{+0.61}$ & $6.81_{-0.08}^{+0.11}$ & 2.09 &     8.90 &     0.22 & Q \\
PG 0043$+$039 & 0.384 & 2133 &   10.94 &   45.51 &              5291 &              9.28 &      11.51 &                      \nodata &                \nodata &                 \nodata &                 $<$8.6 & 2.09 & \mcl{2}{c}{$<$10.7} & Q \\
PG 0049$+$171 & 0.064 &  297 & \nodata &   43.97 &              5234 &              8.45 &      10.80 & \ph{+}$0.60_{-0.30}^{+0.24}$ & $0.47_{-0.00}^{+0.65}$ & $-4.00_{-0.69}^{+0.97}$ & $6.57_{-0.30}^{+0.30}$ & 2.09 &     8.66 &     0.36 & Q \\
PG 0050$+$124 & 0.061 &  282 &   11.12 &   44.76 &              1171 &              7.57 &      10.05 & \ph{+}$1.00_{-0.00}^{+0.00}$ & $0.47_{-0.00}^{+0.00}$ & $-3.16_{-0.53}^{+0.46}$ & $8.22_{-0.01}^{+0.01}$ & 2.08 &    10.30 &     0.20 & Q \\
PG 0052$+$251 & 0.155 &  763 &   11.05 &   45.00 &              5187 &              8.99 &      11.26 & \ph{+}$0.60_{-0.00}^{+0.00}$ & $2.50_{-0.73}^{+0.00}$ & $-3.90_{-0.68}^{+0.87}$ & $8.21_{-0.02}^{+0.02}$ & 2.08 &    10.29 &     0.20 & Q \\
PG 0157$+$001 & 0.164 &  811 &   11.53 &   44.95 &              2432 &              8.31 &      10.68 & \ph{+}$1.18_{-0.10}^{+0.00}$ & $0.47_{-0.00}^{+0.00}$ & $-1.07_{-0.09}^{+0.08}$ & $8.69_{-0.02}^{+0.03}$ & 2.11 &    10.80 &     0.20 & Q \\
PG 0804$+$761 & 0.100 &  475 &   10.64 &   45.03 &              3045 &              8.55 &      10.88 & \ph{+}$1.40_{-0.10}^{+0.00}$ & $0.47_{-0.00}^{+0.65}$ & $-1.87_{-0.56}^{+0.42}$ & $6.68_{-0.05}^{+0.06}$ & 2.11 &     8.79 &     0.21 & Q \\
PG 0838$+$770 & 0.131 &  635 &   11.14 &   44.70 &              2764 &              8.29 &      10.66 & \ph{+}$0.70_{-0.00}^{+0.00}$ & $0.47_{-0.00}^{+0.00}$ & $-1.34_{-0.20}^{+0.13}$ & $8.13_{-0.02}^{+0.02}$ & 2.08 &    10.21 &     0.20 & Q \\
PG 0844$+$349 & 0.064 &  297 &   10.69 &   44.46 &              2386 &              8.03 &      10.44 & \ph{+}$0.00_{-0.00}^{+0.08}$ & $1.77_{-0.00}^{+0.73}$ & $-4.68_{-0.24}^{+0.42}$ & $7.91_{-0.08}^{+0.03}$ & 2.11 &    10.01 &     0.21 & Q \\
PG 0921$+$525 & 0.035 &  159 & \nodata &   43.60 &              2079 &              7.45 & \ph{0}9.94 & \ph{+}$0.60_{-0.00}^{+0.10}$ & $0.47_{-0.00}^{+0.65}$ & $-2.86_{-0.44}^{+0.32}$ & $6.76_{-0.07}^{+0.02}$ & 2.09 &     8.85 &     0.21 & Q \\
PG 0923$+$201 & 0.190 &  955 &   11.09 &   45.01 &               7598 &              9.33 &      11.55 & \ph{+}$1.30_{-0.00}^{+0.10}$ & $1.77_{-0.65}^{+0.73}$ & $-3.05_{-1.03}^{+0.96}$ & $6.91_{-0.09}^{+0.10}$ & 2.08 &     9.00 &     0.22 & Q \\
PG 0923$+$129 & 0.029 &  131 & \nodata &   43.83 &              1957 &              7.52 &      10.00 & \ph{+}$0.85_{-0.00}^{+0.00}$ & $0.47_{-0.00}^{+0.65}$ & $-1.66_{-0.14}^{+0.08}$ & $7.39_{-0.01}^{+0.01}$ & 2.09 &     9.48 &     0.20 & Q \\
PG 0934$+$013 & 0.050 &  229 & \nodata &   43.85 &              1254 &              7.15 & \ph{0}9.68 & \ph{+}$0.70_{-0.00}^{+0.00}$ & $1.12_{-0.00}^{+0.00}$ & $-1.41_{-0.08}^{+0.09}$ & $7.39_{-0.02}^{+0.02}$ & 2.09 &     9.48 &     0.20 & Q \\
PG 0947$+$396 & 0.206 & 1045 &   10.73 &   44.78 &              4817 &              8.81 &      11.11 & \ph{+}$0.85_{-0.24}^{+0.23}$ & $1.12_{-0.65}^{+0.65}$ & $-4.05_{-0.66}^{+0.82}$ & $7.71_{-0.18}^{+0.21}$ & 2.10 &     9.81 &     0.28 & Q \\
PG 0953$+$414 & 0.239 & 1235 &   11.16 &   45.35 &              3111 &              8.74 &      11.04 & \ph{+}$0.60_{-0.20}^{+0.30}$ & $1.77_{-0.65}^{+0.73}$ & $-4.41_{-0.43}^{+0.74}$ & $7.97_{-0.30}^{+0.27}$ & 2.08 &    10.05 &     0.35 & Q \\
PG 1001$+$054 & 0.161 &  795 &   10.47 &   44.71 &              1700 &              7.87 &      10.30 & \ph{+}$0.90_{-0.20}^{+0.18}$ & $1.77_{-0.65}^{+0.73}$ & $-1.30_{-0.40}^{+0.26}$ & $7.37_{-0.21}^{+0.20}$ & 2.13 &     9.51 &     0.29 & Q \\
PG 1004$+$130 & 0.240 & 1240 &   11.44 &   45.51 &              6290 &              9.43 &      11.64 & \ph{+}$1.30_{-0.00}^{+0.00}$ & $0.47_{-0.00}^{+0.00}$ & $-2.51_{-0.51}^{+0.46}$ & $7.72_{-0.04}^{+0.04}$ & 2.10 &     9.82 &     0.20 & S \\
PG 1011$-$040 & 0.058 &  268 & \nodata &   44.23 &              1381 &              7.43 & \ph{0}9.93 & \ph{+}$0.60_{-0.00}^{+0.00}$ & $0.47_{-0.00}^{+0.00}$ & $-1.61_{-0.17}^{+0.11}$ & $7.56_{-0.01}^{+0.02}$ & 2.09 &     9.65 &     0.20 & Q \\
PG 1012$+$008 & 0.185 &  927 &   11.15 &   44.98 &              2615 &              8.39 &      10.74 & \ph{+}$0.70_{-0.00}^{+0.00}$ & $0.47_{-0.00}^{+0.65}$ & $-1.21_{-0.21}^{+0.13}$ & $8.12_{-0.03}^{+0.04}$ & 2.08 &    10.20 &     0.20 & Q \\
PG 1022$+$519 & 0.045 &  206 & \nodata &   43.67 &              1566 &              7.25 & \ph{0}9.77 & \ph{+}$0.85_{-0.00}^{+0.00}$ & $1.77_{-0.00}^{+0.00}$ & $-1.48_{-0.04}^{+0.03}$ & $7.25_{-0.01}^{+0.01}$ & 2.09 &     9.34 &     0.20 & Q \\
PG 1048$+$342 & 0.167 &  828 &   10.77 &   44.68 &              3581 &              8.50 &      10.84 & \ph{+}$0.40_{-0.10}^{+0.20}$ & $0.47_{-0.00}^{+0.65}$ & $-2.15_{-0.55}^{+0.32}$ & $8.04_{-0.16}^{+0.15}$ & 2.10 &    10.14 &     0.25 & Q \\
PG 1048$-$090 & 0.344 & 1875 & \nodata &   45.57 &              5611 &              9.37 &      11.58 &                      \nodata &                \nodata &                 \nodata &                 $<$8.5 & 2.09 & \mcl{2}{c}{$<$10.6} & S \\
PG 1049$-$005 & 0.357 & 1958 & \nodata &   45.60 &              5351 &              9.34 &      11.56 & \ph{+}$1.40_{-0.10}^{+0.00}$ & $1.77_{-0.00}^{+0.00}$ & $-4.57_{-0.29}^{+0.38}$ & $8.34_{-0.04}^{+0.05}$ & 2.09 &    10.44 &     0.20 & Q \\
PG 1100$+$772 & 0.313 & 1681 &   11.27 &   45.55 &              6151 &              9.44 &      11.64 & \ph{+}$0.60_{-0.12}^{+0.10}$ & $0.47_{-0.00}^{+0.00}$ & $-0.48_{-0.16}^{+0.22}$ & $8.17_{-0.17}^{+0.11}$ & 2.09 &    10.26 &     0.24 & S \\
PG 1103$-$006 & 0.425 & 2404 & \nodata &   45.64 &              6183 &              9.49 &      11.68 &                      \nodata &                \nodata &                 \nodata &                 $<$8.4 & 2.09 & \mcl{2}{c}{$<$10.5} & S \\
PG 1114$+$445 & 0.144 &  704 & \nodata &   44.70 &              4554 &              8.72 &      11.03 &      $-0.82_{-0.18}^{+0.30}$ & $4.58_{-0.68}^{+0.00}$ & $-4.57_{-0.31}^{+0.56}$ & $8.38_{-0.37}^{+0.24}$ & 2.09 &    10.47 &     0.36 & Q \\
PG 1115$+$407 & 0.154 &  757 & \nodata &   44.59 &              1679 &              7.80 &      10.24 & \ph{+}$0.70_{-0.00}^{+0.00}$ & $1.12_{-0.00}^{+0.00}$ & $-1.48_{-0.03}^{+0.04}$ & $8.44_{-0.01}^{+0.01}$ & 2.09 &    10.53 &     0.20 & Q \\
PG 1116$+$215 & 0.177 &  882 &   10.61 &   45.37 &              2897 &              8.69 &      11.00 & \ph{+}$1.40_{-0.10}^{+0.00}$ & $1.12_{-0.65}^{+0.65}$ & $-2.83_{-0.63}^{+0.65}$ & $7.26_{-0.05}^{+0.08}$ & 2.11 &     9.38 &     0.21 & Q \\
PG 1119$+$120 & 0.049 &  225 &   10.67 &   44.10 &              1773 &              7.58 &      10.05 & \ph{+}$1.08_{-0.00}^{+0.00}$ & $0.47_{-0.00}^{+0.65}$ & $-1.35_{-0.12}^{+0.10}$ & $7.16_{-0.02}^{+0.02}$ & 2.11 &     9.26 &     0.20 & Q \\
PG 1121$+$422 & 0.234 & 1205 &   10.29 &   44.85 &              2192 &              8.17 &      10.55 &                      \nodata &                \nodata &                 \nodata &                 $<$8.4 & 2.16 & \mcl{2}{c}{$<$10.6} & Q \\
PG 1126$-$041 & 0.060 &  277 &   10.85 &   44.36 &              2111 &              7.87 &      10.30 & \ph{+}$1.08_{-0.00}^{+0.00}$ & $0.47_{-0.00}^{+0.00}$ & $-1.57_{-0.33}^{+0.17}$ & $7.56_{-0.02}^{+0.01}$ & 2.09 &     9.65 &     0.20 & Q \\
PG 1149$-$110 & 0.049 &  225 & \nodata &   44.08 &              3032 &              8.04 &      10.44 & \ph{+}$0.70_{-0.00}^{+0.00}$ & $0.47_{-0.00}^{+0.00}$ & $-1.51_{-0.12}^{+0.10}$ & $7.40_{-0.02}^{+0.02}$ & 2.09 &     9.49 &     0.20 & Q \\
PG 1151$+$117 & 0.176 &  877 &   10.45 &   44.73 &              4284 &              8.68 &      10.99 &                      \nodata &                \nodata &                 \nodata &                 $<$7.9 & 2.14 & \mcl{2}{c}{$<$10.0} & Q \\
PG 1202$+$281 & 0.165 &  817 &   10.86 &   44.57 &              5036 &              8.74 &      11.04 & \ph{+}$1.40_{-0.00}^{+0.00}$ & $3.19_{-0.69}^{+0.71}$ & $-2.08_{-0.61}^{+0.44}$ & $7.41_{-0.02}^{+0.04}$ & 2.09 &     9.51 &     0.20 & Q \\
PG 1211$+$143 & 0.085 &  400 &   10.38 &   45.04 &              1817 &              8.10 &      10.50 & \ph{+}$0.30_{-0.22}^{+0.18}$ & $4.58_{-0.68}^{+0.00}$ & $-4.57_{-0.27}^{+0.49}$ & $7.36_{-0.22}^{+0.20}$ & 2.15 &     9.51 &     0.29 & Q \\
PG 1216$+$069 & 0.334 & 1812 &   10.85 &   45.69 &              5180 &              9.36 &      11.57 &                      \nodata &                \nodata &                 \nodata &                 $<$8.3 & 2.09 & \mcl{2}{c}{$<$10.4} & Q \\
PG 1226$+$023 & 0.158 &  779 &   11.51 &   45.99 &              3500 &              9.18 &      11.42 &      $-0.40_{-0.30}^{+0.30}$ & $3.90_{-0.71}^{+0.68}$ & $-2.15_{-0.66}^{+0.72}$ & $7.00_{-0.50}^{+0.56}$ & 2.10 &     9.11 &     0.57 & F \\
PG 1229$+$204 & 0.064 &  297 &   10.94 &   44.35 &              3335 &              8.26 &      10.63 & \ph{+}$0.60_{-0.00}^{+0.00}$ & $0.47_{-0.00}^{+0.65}$ & $-3.09_{-0.86}^{+0.73}$ & $7.63_{-0.02}^{+0.02}$ & 2.09 &     9.72 &     0.20 & Q \\
PG 1244$+$026 & 0.048 &  220 & \nodata &   43.77 &         \ph{0}721 &              6.62 & \ph{0}9.23 & \ph{+}$1.18_{-0.00}^{+0.00}$ & $0.47_{-0.00}^{+0.00}$ & $-1.07_{-0.04}^{+0.04}$ & $6.69_{-0.02}^{+0.02}$ & 2.09 &     8.78 &     0.20 & Q \\
PG 1259$+$593 & 0.472 & 2723 &   10.99 &   45.88 &              3377 &              9.09 &      11.34 & \ph{+}$0.40_{-0.32}^{+0.45}$ & $1.12_{-0.65}^{+0.65}$ & $-0.13_{-0.16}^{+0.09}$ & $7.80_{-0.36}^{+0.36}$ & 2.08 &     9.89 &     0.41 & Q \\
PG 1302$-$102 & 0.286 & 1515 &   11.23 &   45.80 &              3383 &              9.05 &      11.31 & \ph{+}$0.90_{-0.06}^{+0.00}$ & $0.47_{-0.00}^{+0.00}$ & $-0.45_{-0.13}^{+0.18}$ & $7.87_{-0.15}^{+0.12}$ & 2.08 &     9.96 &     0.24 & F \\
PG 1307$+$085 & 0.155 &  763 &   10.78 &   44.98 &              5307 &              9.00 &      11.27 & \ph{+}$0.90_{-0.20}^{+0.18}$ & $3.90_{-0.71}^{+0.68}$ & $-3.27_{-0.87}^{+0.80}$ & $7.46_{-0.17}^{+0.18}$ & 2.10 &     9.56 &     0.26 & Q \\
PG 1309$+$355 & 0.184 &  921 &   11.22 &   44.98 &              2917 &              8.48 &      10.82 & \ph{+}$0.40_{-0.10}^{+0.08}$ & $3.90_{-0.71}^{+0.00}$ & $-3.61_{-0.67}^{+0.72}$ & $8.32_{-0.11}^{+0.11}$ & 2.08 &    10.40 &     0.23 & F \\
PG 1310$-$108 & 0.035 &  159 & \nodata &   43.70 &              3606 &              7.99 &      10.40 & \ph{+}$0.60_{-0.00}^{+0.00}$ & $2.50_{-0.00}^{+0.69}$ & $-4.79_{-0.14}^{+0.26}$ & $6.86_{-0.02}^{+0.02}$ & 2.09 &     8.95 &     0.20 & Q \\
PG 1322$+$659 & 0.168 &  833 &   10.61 &   44.95 &              2765 &              8.42 &      10.77 & \ph{+}$1.40_{-0.00}^{+0.00}$ & $3.19_{-0.69}^{+0.00}$ & $-1.96_{-0.43}^{+0.34}$ & $7.35_{-0.03}^{+0.05}$ & 2.11 &     9.47 &     0.20 & Q \\
PG 1341$+$258 & 0.087 &  410 & \nodata &   44.31 &              3014 &              8.15 &      10.54 & \ph{+}$0.70_{-0.10}^{+0.15}$ & $1.12_{-0.00}^{+0.65}$ & $-1.28_{-0.23}^{+0.16}$ & $7.23_{-0.19}^{+0.11}$ & 2.09 &     9.32 &     0.25 & Q \\
PG 1351$+$236 & 0.055 &  253 & \nodata &   44.02 &              6527 &              8.67 &      10.98 & \ph{+}$0.85_{-0.00}^{+0.00}$ & $2.50_{-0.00}^{+0.00}$ & $-1.87_{-0.12}^{+0.08}$ & $7.68_{-0.02}^{+0.01}$ & 2.09 &     9.77 &     0.20 & Q \\
PG 1351$+$640 & 0.087 &  410 &   10.63 &   44.81 &              5646 &              8.97 &      11.24 & \ph{+}$1.40_{-0.00}^{+0.00}$ & $1.12_{-0.00}^{+0.00}$ & $-1.48_{-0.24}^{+0.17}$ & $7.56_{-0.01}^{+0.01}$ & 2.11 &     9.67 &     0.20 & Q \\
PG 1352$+$183 & 0.158 &  779 &   10.49 &   44.79 &              3581 &              8.56 &      10.89 &                      \nodata &                \nodata &                 \nodata &                 $<$7.9 & 2.13 & \mcl{2}{c}{$<$10.0} & Q \\
PG 1354$+$213 & 0.300 & 1600 &   10.97 &   44.95 &              4127 &              8.77 &      11.07 & \ph{+}$0.70_{-0.30}^{+0.38}$ & $1.12_{-0.65}^{+0.65}$ & $-3.16_{-1.10}^{+1.30}$ & $7.76_{-0.40}^{+0.39}$ & 2.09 &     9.84 &     0.44 & Q \\
PG 1402$+$261 & 0.164 &  811 &   10.86 &   44.95 &              1874 &              8.08 &      10.48 & \ph{+}$1.30_{-0.00}^{+0.00}$ & $0.47_{-0.00}^{+0.00}$ & $-1.36_{-0.13}^{+0.11}$ & $7.83_{-0.02}^{+0.02}$ & 2.09 &     9.93 &     0.20 & Q \\
PG 1404$+$226 & 0.098 &  465 & \nodata &   44.35 &         \ph{0}787 &              7.01 & \ph{0}9.56 & \ph{+}$0.18_{-0.00}^{+0.00}$ & $1.12_{-0.00}^{+0.00}$ & $-1.33_{-0.06}^{+0.06}$ & $7.89_{-0.03}^{+0.03}$ & 2.09 &     9.99 &     0.20 & Q \\
PG 1411$+$442 & 0.089 &  420 &   10.84 &   44.60 &              2640 &              8.20 &      10.58 & \ph{+}$0.40_{-0.00}^{+0.08}$ & $1.77_{-0.65}^{+0.73}$ & $-1.25_{-0.14}^{+0.12}$ & $7.81_{-0.05}^{+0.04}$ & 2.09 &     9.90 &     0.21 & Q \\
PG 1415$+$451 & 0.114 &  546 & \nodata &   44.53 &              2591 &              8.14 &      10.53 & \ph{+}$1.00_{-0.00}^{+0.00}$ & $2.50_{-0.00}^{+0.00}$ & $-1.79_{-0.17}^{+0.14}$ & $7.64_{-0.02}^{+0.02}$ & 2.09 &     9.73 &     0.20 & Q \\
PG 1416$-$129 & 0.129 &  624 & \nodata &   45.11 &              6098 &              9.19 &      11.43 & \ph{+}$1.40_{-0.10}^{+0.00}$ & $2.50_{-0.00}^{+0.69}$ & $-3.94_{-0.64}^{+0.79}$ & $7.02_{-0.04}^{+0.09}$ & 2.09 &     9.11 &     0.21 & Q \\
PG 1425$+$267 & 0.366 & 2016 &   11.15 &   45.73 &              9405 &              9.90 &      12.04 & \ph{+}$1.40_{-0.10}^{+0.00}$ & $0.47_{-0.00}^{+0.00}$ & $-1.01_{-0.28}^{+0.25}$ & $7.75_{-0.06}^{+0.06}$ & 2.08 &     9.83 &     0.21 & S \\
PG 1426$+$015 & 0.086 &  405 &   11.05 &   44.85 &              6808 &              9.15 &      11.39 & \ph{+}$0.70_{-0.00}^{+0.00}$ & $0.47_{-0.00}^{+0.00}$ & $-1.23_{-0.07}^{+0.06}$ & $7.92_{-0.01}^{+0.01}$ & 2.08 &    10.00 &     0.20 & Q \\
PG 1427$+$480 & 0.221 & 1130 &   10.77 &   44.73 &              2515 &              8.22 &      10.60 & \ph{+}$1.40_{-0.10}^{+0.00}$ & $1.77_{-0.65}^{+0.00}$ & $-2.79_{-0.80}^{+0.75}$ & $7.42_{-0.07}^{+0.09}$ & 2.10 &     9.52 &     0.22 & Q \\
PG 1435$-$067 & 0.129 &  624 &   10.51 &   44.89 &              3157 &              8.50 &      10.84 & \ph{+}$0.00_{-0.30}^{+0.18}$ & $3.19_{-0.69}^{+0.71}$ & $-1.56_{-0.44}^{+0.32}$ & $7.84_{-0.27}^{+0.31}$ & 2.13 &     9.97 &     0.35 & Q \\
PG 1440$+$356 & 0.077 &  360 &   11.05 &   44.52 &              1394 &              7.60 &      10.07 & \ph{+}$1.18_{-0.00}^{+0.00}$ & $1.77_{-0.00}^{+0.00}$ & $-2.46_{-0.38}^{+0.32}$ & $7.87_{-0.01}^{+0.01}$ & 2.08 &     9.95 &     0.20 & Q \\
PG 1444$+$407 & 0.267 & 1400 &   11.15 &   45.17 &              2457 &              8.44 &      10.78 & \ph{+}$1.30_{-0.00}^{+0.00}$ & $0.47_{-0.00}^{+0.00}$ & $-0.60_{-0.16}^{+0.17}$ & $7.43_{-0.15}^{+0.13}$ & 2.08 &     9.51 &     0.24 & Q \\
PG 1448$+$273 & 0.065 &  301 & \nodata &   44.45 &         \ph{0}815 &              7.09 & \ph{0}9.63 & \ph{+}$1.18_{-0.00}^{+0.00}$ & $2.50_{-0.73}^{+0.69}$ & $-4.33_{-0.46}^{+0.76}$ & $7.08_{-0.03}^{+0.02}$ & 2.09 &     9.17 &     0.20 & Q \\
PG 1501$+$106 & 0.036 &  164 & \nodata &   44.26 &              5454 &              8.64 &      10.96 & \ph{+}$1.30_{-0.00}^{+0.10}$ & $0.47_{-0.00}^{+0.00}$ & $-1.55_{-0.48}^{+0.30}$ & $6.60_{-0.05}^{+0.04}$ & 2.09 &     8.69 &     0.20 & Q \\
PG 1512$+$370 & 0.371 & 2048 &   11.01 &   45.57 &              6803 &              9.53 &      11.72 & \ph{+}$0.40_{-0.22}^{+0.20}$ & $1.12_{-0.65}^{+0.65}$ & $-2.50_{-0.80}^{+0.67}$ & $8.44_{-0.25}^{+0.22}$ & 2.08 &    10.52 &     0.31 & S \\
PG 1519$+$226 & 0.137 &  666 & \nodata &   44.68 &              2187 &              8.07 &      10.47 & \ph{+}$1.00_{-0.00}^{+0.00}$ & $1.12_{-0.00}^{+0.65}$ & $-1.42_{-0.20}^{+0.16}$ & $7.54_{-0.04}^{+0.03}$ & 2.09 &     9.63 &     0.20 & Q \\
PG 1534$+$580 & 0.030 &  136 & \nodata &   43.66 &              5324 &              8.30 &      10.67 & \ph{+}$1.08_{-0.08}^{+0.00}$ & $2.50_{-0.73}^{+0.00}$ & $-3.29_{-0.89}^{+0.84}$ & $6.42_{-0.04}^{+0.06}$ & 2.09 &     8.51 &     0.21 & Q \\
PG 1535$+$547 & 0.038 &  173 & \nodata &   43.93 &              1420 &              7.30 & \ph{0}9.81 &      $-0.10_{-0.06}^{+0.00}$ & $1.12_{-0.00}^{+0.00}$ & $-0.92_{-0.08}^{+0.09}$ & $7.21_{-0.07}^{+0.05}$ & 2.09 &     9.31 &     0.21 & Q \\
PG 1543$+$489 & 0.400 & 2237 &   10.93 &   45.42 &              1529 &              8.16 &      10.54 & \ph{+}$1.40_{-0.00}^{+0.00}$ & $0.47_{-0.00}^{+0.65}$ & $-2.52_{-0.50}^{+0.43}$ & $8.50_{-0.04}^{+0.04}$ & 2.09 &    10.59 &     0.20 & Q \\
PG 1545$+$210 & 0.266 & 1394 &   11.15 &   45.40 &              7022 &              9.47 &      11.67 &                      \nodata &                \nodata &                 \nodata &                 $<$8.3 & 2.08 & \mcl{2}{c}{$<$10.4} & S \\
PG 1552$+$085 & 0.119 &  572 & \nodata &   44.67 &              1377 &              7.67 &      10.12 &      $-0.30_{-0.00}^{+0.00}$ & $1.12_{-0.65}^{+0.00}$ & $-0.26_{-0.21}^{+0.17}$ & $7.64_{-0.17}^{+0.19}$ & 2.09 &     9.73 &     0.27 & Q \\
PG 1612$+$261 & 0.131 &  635 & \nodata &   44.69 &              2491 &              8.19 &      10.57 & \ph{+}$1.00_{-0.10}^{+0.08}$ & $1.12_{-0.00}^{+0.00}$ & $-4.72_{-0.21}^{+0.39}$ & $7.91_{-0.09}^{+0.07}$ & 2.09 &    10.00 &     0.22 & Q \\
PG 1613$+$658 & 0.129 &  624 &   11.46 &   44.81 &              8441 &              9.32 &      11.54 & \ph{+}$1.00_{-0.00}^{+0.00}$ & $0.47_{-0.00}^{+0.00}$ & $-1.23_{-0.05}^{+0.05}$ & $8.46_{-0.01}^{+0.01}$ & 2.10 &    10.56 &     0.20 & Q \\
PG 1617$+$175 & 0.114 &  546 &   10.47 &   44.81 &              5316 &              8.91 &      11.19 & \ph{+}$0.85_{-0.15}^{+0.23}$ & $0.47_{-0.00}^{+0.65}$ & $-0.95_{-0.36}^{+0.19}$ & $6.87_{-0.17}^{+0.14}$ & 2.13 &     9.00 &     0.25 & Q \\
PG 1626$+$554 & 0.133 &  645 &   10.84 &   44.55 &              4474 &              8.63 &      10.95 &                      \nodata &                \nodata &                 \nodata &                 $<$7.5 & 2.09 &  \mcl{2}{c}{$<$9.6} & Q \\
PG 1700$+$518 & 0.282 & 1490 &   11.39 &   45.69 &              2185 &              8.61 &      10.93 & \ph{+}$1.18_{-0.00}^{+0.00}$ & $0.47_{-0.00}^{+0.00}$ & $-0.89_{-0.11}^{+0.09}$ & $8.55_{-0.02}^{+0.02}$ & 2.09 &    10.64 &     0.20 & Q \\
PG 1704$+$608 & 0.371 & 2048 &   11.52 &   45.67 &              6552 &              9.55 &      11.74 & \ph{+}$1.40_{-0.00}^{+0.00}$ & $0.47_{-0.00}^{+0.65}$ & $-3.23_{-0.54}^{+0.52}$ & $8.16_{-0.03}^{+0.04}$ & 2.11 &    10.26 &     0.20 & S \\
PG 2112$+$059 & 0.466 & 2681 & \nodata &   46.16 &              3176 &              9.18 &      11.42 & \ph{+}$0.85_{-0.00}^{+0.15}$ & $0.47_{-0.00}^{+0.65}$ & $-0.70_{-0.16}^{+0.15}$ & $8.50_{-0.09}^{+0.09}$ & 2.09 &    10.59 &     0.22 & Q \\
PG 2130$+$099 & 0.061 &  282 &   10.85 &   44.54 &              2294 &              8.04 &      10.44 & \ph{+}$0.90_{-0.00}^{+0.00}$ & $0.47_{-0.00}^{+0.65}$ & $-1.44_{-0.13}^{+0.11}$ & $7.60_{-0.01}^{+0.02}$ & 2.09 &     9.69 &     0.20 & Q \\
PG 2209$+$184 & 0.070 &  326 & \nodata &   44.44 &              6488 &              8.89 &      11.17 & \ph{+}$0.08_{-0.00}^{+0.00}$ & $2.50_{-0.00}^{+0.69}$ & $-1.91_{-0.19}^{+0.13}$ & $7.98_{-0.05}^{+0.04}$ & 2.09 &    10.07 &     0.20 & F \\
PG 2214$+$139 & 0.067 &  311 &   10.98 &   44.63 &              4532 &              8.68 &      10.99 & \ph{+}$0.40_{-0.10}^{+0.08}$ & $3.90_{-0.71}^{+0.68}$ & $-2.90_{-0.54}^{+0.47}$ & $7.48_{-0.06}^{+0.07}$ & 2.08 &     9.56 &     0.21 & Q \\
PG 2233$+$134 & 0.325 & 1755 &   10.81 &   45.30 &              1709 &              8.19 &      10.57 & \ph{+}$0.60_{-0.20}^{+0.10}$ & $1.77_{-0.65}^{+0.73}$ & $-4.51_{-0.33}^{+0.67}$ & $8.36_{-0.16}^{+0.21}$ & 2.09 &    10.46 &     0.27 & Q \\
PG 2251$+$113 & 0.323 & 1743 &   11.05 &   45.66 &              4147 &              9.15 &      11.39 & \ph{+}$1.30_{-0.12}^{+0.10}$ & $2.50_{-0.73}^{+0.69}$ & $-3.59_{-0.74}^{+0.83}$ & $7.63_{-0.11}^{+0.15}$ & 2.08 &     9.71 &     0.24 & S \\
PG 2304$+$042 & 0.042 &  192 & \nodata &   44.04 &              6487 &              8.68 &      10.99 & \ph{+}$0.70_{-0.22}^{+0.15}$ & $1.12_{-0.65}^{+0.65}$ & $-4.40_{-0.45}^{+0.68}$ & $6.28_{-0.17}^{+0.20}$ & 2.09 &     8.37 &     0.27 & Q \\
PG 2308$+$098 & 0.432 & 2451 & \nodata &   45.75 &              7914 &              9.76 &      11.92 &                      \nodata &                \nodata &                 \nodata &                 $<$8.9 & 2.09 & \mcl{2}{c}{$<$11.0} & S \\
\enddata
\tablecomments{
(1) Object name.
(2) Redshift.
(3) The luminosity distance calculated with $\Omega_m = 0.308$,
$\Omega_\Lambda = 0.692$, and $H_{0}=67.8$ km s$^{-1}$ Mpc$^{-1}$
\citep{Planck2016AA}.
(4) The stellar mass of the quasar host galaxies from \cite{Zhang2016ApJ}.  
In order to convert from the \cite{Salpeter1955ApJ} IMF to the Kroupa-like 
IMF, we divide the stellar mass by 1.5, following \cite{Zhang2016ApJ}.
(5) The monochromatic luminosity at 5100 \AA.
(6) The FWHM of the broad \hbeta\ emission line.
(7) The mass of the BH.
(8) The stellar bulge mass of the host galaxy estimated from $M_\mathrm{BH}$.
(9) The best-fit minimum intensity of the interstellar radiation field relative
to that measured in the solar neighborhood.
(10) The best-fit mass fraction of the dust in the form of PAH molecules.
(11) The best-fit mass fraction of the dust associated with the power-law part
of the interstellar radiation field.
(12) The best-fit total dust mass.
The quoted uncertainties of the DL07 model represent the 68\% confidence 
level determined from the 16th and 84th percentile of the marginalized posterior 
PDF.  However, for $U_\mathrm{min}$ or $q_\mathrm{PAH}$, if fewer than 16\% 
of the sampled values at the discrete grids lie below (above) the best-fit value, 
the lower (upper) uncertainty of the parameter is not resolved, and it is reported 
as ``0.00'' in the table.
(13) The gas-to-dust ratio of the galaxy, estimated from the host galaxy stellar
mass.  For objects without a stellar mass measured, the median value of 
the sample is adopted, 124$\pm$6.  The value of \gdr\ has been corrected 
using Equation (\ref{eq:gdrcorr}), so that the total gas mass can be compared 
to the directly measured gas mass in an unbiased manner.
(14) The total gas mass with uncertainty, combining the uncertainties of
the dust mass and the \gdr\ (0.2 dex).
(15) The radio type of the quasar: ``Q'' for radio-quiet source, ``S'' for
steep-spectrum source, and ``F'' for flat-spectrum source.
}
\end{deluxetable}
\end{longrotatetable}

%% file: tab2.tex
\begin{table}
\begin{center}
\caption{NIR and MIR Photometry}
\label{tab:nir}
\begin{tabular}{l r@{$\pm$}l r@{$\pm$}l r@{$\pm$}l r@{$\pm$}l r@{$\pm$}l r@{$\pm$}l r@{$\pm$}l}
\hline
\hline
\multicolumn{1}{c}{Object} & 
\multicolumn{2}{c}{$F_{J}$} & 
\multicolumn{2}{c}{$F_{H}$} & 
\multicolumn{2}{c}{$F_{K_s}$} & 
\multicolumn{2}{c}{$F_{W1}$} & 
\multicolumn{2}{c}{$F_{W2}$} & 
\multicolumn{2}{c}{$F_{W3}$} & 
\multicolumn{2}{c}{$F_{W4}$} \\
\multicolumn{1}{c}{} & 
\multicolumn{2}{c}{(mJy)} & 
\multicolumn{2}{c}{(mJy)} & 
\multicolumn{2}{c}{(mJy)} & 
\multicolumn{2}{c}{(mJy)} & 
\multicolumn{2}{c}{(mJy)} & 
\multicolumn{2}{c}{(mJy)} & 
\multicolumn{2}{c}{(mJy)} \\ 
\multicolumn{1}{c}{(1)} & 
\multicolumn{2}{c}{(2)} & 
\multicolumn{2}{c}{(3)} & 
\multicolumn{2}{c}{(4)} & 
\multicolumn{2}{c}{(5)} & 
\multicolumn{2}{c}{(6)} & 
\multicolumn{2}{c}{(7)} & 
\multicolumn{2}{c}{(8)} \\ \hline
PG 0003$+$158        & 2.08  & 0.15 & 2.21 & 0.18 & 2.70 & 0.29 & 4.17 & 0.02 & 6.02 & 0.03 & 13.17 & 0.20 & 25.92 & 1.02 \\
PG 0003$+$199\tnm{a} & 20.40 & 0.68 & 27.40 & 1.12 & 45.17 & 0.99 & 71.27 & 0.05 & 100.65 & 0.07 & 178.73 & 0.41 & 290.48 & 1.31 \\
PG 0838$+$770        & 2.98  & 0.21 & 3.46 & 0.34 & 5.52 & 0.34 & 7.47 & 0.02 & 9.85 & 0.03 & 29.57 & 0.19 & 68.68 & 0.85 \\
PG 0844$+$349\tnm{a} & 8.64  & 0.67 & 8.67 & 1.14 & 12.19 & 0.98 & 16.90 & 0.03 & 22.09 & 0.03 & 52.91 & 0.35 & 96.06 & 1.24 \\
PG 0921$+$525\tnm{b} & 6.34  & 0.16 & 8.35 & 0.29 & 10.91 & 0.27 & 21.93 & 0.03 & 30.34 & 0.03 & 71.65 & 0.27 & 102.04 & 1.05 \\
PG 0923$+$201        & 3.32  & 0.14 & 4.72 & 0.28 & 9.03 & 0.24 & 21.27 & 0.03 & 26.35 & 0.05 & 39.46 & 0.29 & 56.13 & 1.22 \\
PG 1048$+$342        & 1.98  & 0.13 & 2.40 & 0.20 & 3.39 & 0.19 & 4.01 & 0.02 & 5.76 & 0.02 & 15.06 & 0.16 & 25.66 & 1.26 \\
PG 1048$-$090\tnm{c} & 0.98  & 0.15 & 1.76 & 0.23 & 1.42 & 0.32 & 5.07 & 0.02 & 7.09 & 0.04 & 11.74 & 0.26 & 22.10 & 1.48 \\
PG 1049$-$005        & 2.30  & 0.16 & 2.80 & 0.22 & 5.34 & 0.31 & 10.24 & 0.02 & 15.96 & 0.03 & 43.78 & 0.30 & 94.56 & 1.50 \\
PG 1100$+$772        & 2.49  & 0.18 & 3.32 & 0.28 & 4.35 & 0.29 & 8.58 & 0.04 & 13.19 & 0.04 & 25.80 & 0.19 & 47.85 & 0.84 \\
\hline
\end{tabular}
\end{center}
\tablecomments{
(1) Object name.
(2) $J$ band (1.235 $\mu$m) from \tmass;
(3) $H$ band (1.662 $\mu$m) from \tmass;
(4) $K_s$ band (2.159 $\mu$m) from \tmass;
(5) \w1 band (3.353 $\mu$m) from \wise;
(6) \w2 band (4.603 $\mu$m) from \wise;
(7) \w3 band (11.561 $\mu$m) from \wise;
(8) \w4 band (22.088 $\mu$m) from \wise.
The \wise\ \w3 and \w4 bands are corrected for a calibration discrepancy 
described in the text.  This table is available in its entirety in a machine-readable 
form in the on-line journal.  A portion is shown here for guidance regarding its form 
and content.
}
\tablenotetext{a}{Extended source measured with a 20\arcsec radius aperture on 
2MASS images.}
\tablenotetext{b}{There are projected companions found in the \tmass\ images.}
\tablenotetext{c}{There are projected companions found in the \wise\ images.}
\end{table}

%% file: tab3.tex
\begin{table}
\begin{center}
\caption{PACS and SPIRE Photometry}
\label{tab:herschel}
\begin{tabular}{l r@{$\pm$}l r@{$\pm$}l r@{$\pm$}l r@{$\pm$}l r@{$\pm$}l r@{$\pm$}l}
\hline
\hline
\multicolumn{1}{c}{Object} &
\multicolumn{2}{c}{$F_{70}$}  &
\multicolumn{2}{c}{$F_{100}$} &
\multicolumn{2}{c}{$F_{160}$} &
\multicolumn{2}{c}{$F_{250}$} &
\multicolumn{2}{c}{$F_{350}$} &
\multicolumn{2}{c}{$F_{500}$} \\
\multicolumn{1}{c}{}     &
\multicolumn{2}{c}{(mJy)} &
\multicolumn{2}{c}{(mJy)} &
\multicolumn{2}{c}{(mJy)} &
\multicolumn{2}{c}{(mJy)} &
\multicolumn{2}{c}{(mJy)} &
\multicolumn{2}{c}{(mJy)} \\
\multicolumn{1}{c}{(1)} &
\multicolumn{2}{c}{(2)} &
\multicolumn{2}{c}{(3)} &
\multicolumn{2}{c}{(4)} &
\multicolumn{2}{c}{(5)} &
\multicolumn{2}{c}{(6)} &
\multicolumn{2}{c}{(7)} \\ \hline
PG 0003$+$158        & 23.37  & 2.77 & 13.01   & 2.87 & \mcl{2}{c}{$<$24.30}        & \mcl{2}{c}{$<$31.23} & \mcl{2}{c}{$<$25.49}        & \mcl{2}{c}{$<$33.74} \\
PG 0026$+$129        & 29.74  & 2.31 & 27.42   & 2.59 & \mcl{2}{c}{$<$27.18}        & \mcl{2}{c}{$<$29.64} & \mcl{2}{c}{$<$29.11}        & \mcl{2}{c}{$<$32.57} \\
PG 0043$+$039\tnm{a} & 25.67  & 2.89 & 18.01   & 2.86 & \mcl{2}{c}{$<$19.86\tnm{c}} & \mcl{2}{c}{$<$32.08} & \mcl{2}{c}{$<$26.22}        & \mcl{2}{c}{$<$33.71} \\
PG 0923$+$129\tnm{b} & 811.54 & 6.06 & 1070.81 & 8.70 & 1088.33      & 34.80        & 343.19 & 14.08       & 165.62 & 10.28              & 76.14 & 11.07        \\
PG 0934$+$013        & 232.81 & 4.14 & 274.20  & 4.16 & 292.94       & 37.35        & 123.57 & 11.45       & 63.31  & 10.88              & \mcl{2}{c}{$<$43.27} \\
PG 0947$+$396\tnm{a} & 58.69  & 2.76 & 50.48   & 2.13 & 52.45\tnm{c} & 8.49         & \mcl{2}{c}{$<$30.37} & \mcl{2}{c}{$<$28.29\tnm{d}} & \mcl{2}{c}{$<$33.65} \\
PG 0953$+$414        & 35.27  & 2.25 & 32.11   & 4.42 & 52.54        & 8.45         & \mcl{2}{c}{$<$31.89} & \mcl{2}{c}{$<$25.88}        & \mcl{2}{c}{$<$31.24} \\
PG 1001$+$054        & 40.16  & 1.80 & 41.65   & 2.97 & 38.75        & 8.16         & \mcl{2}{c}{$<$28.96} & \mcl{2}{c}{$<$25.50}        & \mcl{2}{c}{$<$35.20} \\
PG 1022$+$519        & 233.97 & 4.23 & 307.07  & 6.61 & 280.75       & 18.66        & 129.82 & 9.78        & 55.03  & 8.89               & \mcl{2}{c}{$<$33.66} \\
PG 1048$+$342\tnm{a} & 30.37  & 3.27 & 45.72   & 3.80 & 79.06        & 9.71         & \mcl{2}{c}{$<$40.91} & \mcl{2}{c}{$<$26.97\tnm{d}} & \mcl{2}{c}{$<$32.51} \\
\hline
\end{tabular}
\end{center}
\tablecomments{
(1) Object name.
(2) \pacs\ 70 $\mu$m band;
(3) \pacs\ 100 $\mu$m band;
(4) \pacs\ 160 $\mu$m band;
(5) \spire\ 250 $\mu$m band;
(6) \spire\ 350 $\mu$m band;
(7) \spire\ 500 $\mu$m band.
The 5\% calibration uncertainties for \pacs\ and \spire\ photometry are not included
in the uncertainties listed in the table.  This table is available in its entirety in a
machine-readable form in the on-line journal.  A portion is shown here for guidance
regarding its form and content.
}
\tablenotetext{a}{A bright companion is found in and removed from the PACS images.}
\tablenotetext{b}{A faint companion is found but not removed in the PACS images.}
\tablenotetext{c}{The target is heavily blended with the companion in this PACS band.}
\tablenotetext{d}{The flux is likely dominated by the companion in this SPIRE band.}
\end{table}

%% file: tab4.tex
\begin{table*}
\begin{center}
\caption{Archival FIR and Radio Data}
\label{tab:arXiv}
\begin{tabular}{c r@{ }l r@{$\pm$}l r}
\hline
\hline
Object & \multicolumn{2}{c}{Band} & \multicolumn{2}{c}{$f_\nu$} & References \\
       & \multicolumn{2}{c}{}     & \multicolumn{2}{c}{(mJy)}   &           \\\hline
PG 0003+158 & 4.85 & GHz   & 327   & 45   & \cite{Gregory1991ApJS}         \\
            & 1.40 & GHz   & 805.2 & 27.0 & \cite{Condon1998AJ}            \\
            & 408  & MHz   & 2250  & 80   & \cite{Large1981MNRAS}          \\
            & 365  & MHz   & 2771  & 54   & \cite{Douglas1996AJ}           \\
            & 178  & MHz   & 4300  & 540  & \cite{Gower1967MmRAS}          \\
            & 74   & MHz   & 10480 & 1080 & \cite{Cohen2007AJ}             \\\hline
PG 0007+106 & 1.3  & mm    & 481   & 6    & \cite{Chini1989AA}             \\\hline
PG 1004+130 & 4.85 & GHz   & 427   & 59   & \cite{Gregory1991ApJS}         \\
            & 408  & MHz   & 2740  & 120  & \cite{Large1981MNRAS}          \\
            & 365  & MHz   & 1829  & 87   & \cite{Douglas1996AJ}           \\
            & 178  & MHz   & 5100  & 890  & \cite{Gower1967MmRAS}          \\
            & 74   & MHz   & 12310 & 1270 & \cite{Cohen2007AJ}             \\\hline
PG 1226+023 & 70   & $\mu$m & 488.0  & 20.2 & \cite{Shang2011ApJS} \\
            & 160  & $\mu$m & 299.0  & 29.8 & \cite{Shang2011ApJS} \\\hline
\end{tabular}
\end{center}
\tablecomments{
This table is available in its entirety in a machine-readable form in the on-line journal. 
A portion is shown here for guidance regarding its form and content.
}
\end{table*}

%% file: tab5.tex
\begin{table*}
  \begin{center}
  \caption{Model Parameters and Priors}
  \label{tab:pars}
  \begin{tabular}{ccccl}
    \hline
    \hline
    Model                        & Parameter              & Units        & Discreteness & \mcl{1}{c}{Prior}          \\
    (1)                          & (2)                    & (3)          & (4)          & \mcl{1}{c}{(5)}            \\\hline
    \multirow{2}{*}{BC03}        & $M_*$                  & $M_\odot$    & \ding{56}    & [$10^6$, $10^{14}$]        \\
                                 & $t$                    & Gyr          & \ding{52}    & 5 (fixed)                  \\\hline
    \multirow{2}{*}{BB}          & $\Omega_\mathrm{dust}$ & Sr           & \ding{56}    & [$10^{-25}$, $10^{-10}$]   \\
                                 & $T$                    & K            & \ding{56}    & [500, 1500]                \\\hline
    \multirow{7}{*}{CLUMPY}      & $i$                    & --           & \ding{52}    & [0.0, 90.0]                \\
                                 & $\tau_V$               & --           & \ding{52}    & [10.0, 300.0]              \\
                                 & $q$                    & --           & \ding{52}    & [0.0, 3.0]                 \\
                                 & $N_0$                  & --           & \ding{52}    & [1.0, 15.0]                \\
                                 & $\sigma$               & --           & \ding{52}    & [15.0, 70.0]               \\
                                 & $Y$                    & --           & \ding{52}    & [5.0, 100.0]               \\
                                 & $L$                    & erg s$^{-1}$ & \ding{56}    & [$10^{40}$, $10^{50}$]     \\\hline
    \multirow{6}{*}{DL07}        & $U_\mathrm{min}$       & --           & \ding{52}    & [0.10, 25.0]               \\
                                 & $U_\mathrm{max}$       & --           & \ding{52}    & $10^6$ (fixed)             \\
                                 & $\alpha$               & --           & \ding{56}    & 2 (fixed)                  \\
                                 & $q_\mathrm{PAH}$       & --           & \ding{52}    & [0.3, 4.8]                 \\
                                 & $\gamma$               & --           & \ding{56}    & [0.0, 1.0]                 \\
                                 & $M_d$                  & $M_\odot$    & \ding{56}    & [$10^6$, $10^\mathrm{11}$] \\\hline
    \multirow{2}{*}{Synchrotron} & $\alpha$               & --           & \ding{56}    & [0.0, 5.0]                 \\
                                 & $f_0$                  & --           & \ding{56}    & [$10^{-5}$, $10^{5}$]      \\
    \hline
  \end{tabular}
  \end{center}
\tablecomments{(1) The name of the model used in the paper.
(2) The parameters of each model.
(3) The units of the parameters.
(4) Whether the parameter is discrete and requires interpolation to implement the MCMC fitting.
(5) The prior range of the parameters.}
\end{table*}

%% file: tab6.tex
\begin{longrotatetable}
\begin{deluxetable}{ccrrclrl}
\tablecaption{PG Quasars with Gas Measurements \label{tab:gas}}
\tabletypesize{\footnotesize}
\tablehead{
\colhead{Object} &
\colhead{$S_\mathrm{CO}\Delta v$} &
\colhead{$M_\mathrm{H2}$} &
\colhead{$S_\mathrm{H\,I}\Delta v$} &
\colhead{$W_{20}$} &
\colhead{$M_\mathrm{H\,I}$} &
\colhead{$M_\mathrm{gas}$} &
\colhead{References}\\
     &
\colhead{(Jy km $\mathrm{s}^{-1}$)} &
\colhead{($10^9 M_\odot$)} &
\colhead{(Jy km $\mathrm{s}^{-1}$)} &
\colhead{(km $\mathrm{s}^{-1}$)} &
\colhead{($10^9 M_\odot$)} &
\colhead{($10^9 M_\odot$)} &
    \\
\colhead{(1)} &
\colhead{(2)} &
\colhead{(3)} &
\colhead{(4)} &
\colhead{(5)} &
\colhead{(6)} &
\colhead{(7)} &
\colhead{(8)}}
\startdata
PG 0003$+$199 & \phs    24.7 &     3.22\ph{00} & $<$0.35\ph{000} & \nodata & \phn  $<$1.42        &  $<$4.64\ph{00} & \ph{00}1, 10 \\
PG 0007$+$106 & \phn  $<$3.0 &  $<$5.10\ph{00} &    0.65\ph{000} &     800 & \phs    36.67\tnm{a} & $<$41.77\ph{00} & \ph{00}3, 9 \\
PG 0050$+$124 & \phs    30.0 &    23.67\ph{00} &    1.00\ph{000} &     420 & \phs    25.50\tnm{a} &    49.17\ph{00} & \ph{00}5, 9 \\
PG 0052$+$251 & \phs\phn 2.0 &    10.59\ph{00} & \nodata\ph{00}  & \nodata & \phs\nodata          &  \nodata\ph{0}  & \ph{00}2 \\
PG 0157$+$001 & \phs\phn 5.5 &    32.71\ph{00} & \nodata\ph{00}  & \nodata & \phs\nodata          &  \nodata\ph{0}  & \ph{00}5 \\
PG 0804$+$761 & \phs\phn 2.0 &     4.31\ph{00} & \nodata\ph{00}  & \nodata & \phs\nodata          &  \nodata\ph{0}  & \ph{00}4 \\
PG 0838$+$770 & \phs\phn 2.5 &     9.37\ph{00} & \nodata\ph{00}  & \nodata & \phs\nodata          &  \nodata\ph{0}  & \ph{00}5 \\
PG 0844$+$349 & \phn  $<$1.5 &  $<$1.30\ph{00} &    1.05\ph{000} &     488 & \phs    29.60\tnm{a} & $<$30.90\ph{00} & \ph{00}4, 10 \\
PG 0934$+$013 & \phn  $<$1.8 &  $<$0.95\ph{00} & \nodata\ph{00}  & \nodata & \phs\nodata          &  \nodata\ph{0}  & \ph{00}6 \\
PG 1011$-$040 & \phs\phn 6.8 &     4.84\ph{00} & \nodata\ph{00}  & \nodata & \phs\nodata          &  \nodata\ph{0}  & \ph{00}6 \\
PG 1119$+$120 & \phs\phn 2.7 &     1.37\ph{00} &    0.55\ph{000} &     310 & \phs\phn 8.90\tnm{a} &    10.26\ph{00} & \ph{00}5, 9 \\
PG 1126$-$041 & \phs\phn 7.5 &     5.72\ph{00} & \nodata\ph{00}  & \nodata & \nodata              &  \nodata\ph{0}  & \ph{00}6 \\
PG 1202$+$281 & \phn  $<$2.4 & $<$14.45\ph{00} & \nodata\ph{00}  & \nodata & \nodata              &  \nodata\ph{0}  & \ph{00}3 \\
PG 1211$+$143 & \phn  $<$1.5 &  $<$2.32\ph{00} & $<$0.05\ph{000} & \nodata & \phn  $<$2.56        &  $<$4.88\ph{00} & \ph{00}4, 9 \\
PG 1226$+$023 & \phs    24.2 &   133.28\ph{00} & \nodata\ph{00}  & \nodata & \phs\nodata          &  \nodata\ph{0}  & \ph{00}8 \\
PG 1229$+$204 & \phs\phn 2.4 &     2.09\ph{00} &    0.23\ph{000} &     295 & \phs\phn 6.48        &     8.57\ph{00} & \ph{00}4, 10 \\
PG 1244$+$026 & \phs \nodata &  \nodata\ph{0}  & $<$0.47\ph{000} & \nodata & \phn  $<$7.29        &  \nodata\ph{0}  & \ph{00}10 \\
PG 1309$+$355 & \phn  $<$0.6 &  $<$4.52\ph{00} & \nodata\ph{00}  & \nodata & \phs\nodata          &  \nodata\ph{0}  & \ph{00}2 \\
PG 1310$-$108 & \phn  $<$2.4 &  $<$0.62\ph{00} & \nodata\ph{00}  & \nodata & \phs\nodata          &  \nodata\ph{0}  & \ph{00}6 \\
PG 1351$+$640 & \phs\phn 2.7 &     4.38\ph{00} & \nodata\ph{00}  & \nodata & \phs\nodata          &  \nodata\ph{0}  & \ph{00}5 \\
PG 1402$+$261 & \phs\phn 2.0 &    11.89\ph{00} & \nodata\ph{00}  & \nodata & \phs\nodata          &  \nodata\ph{0}  & \ph{00}2 \\
PG 1404$+$226 & \phs\phn 2.0 &     4.14\ph{00} & \nodata\ph{00}  & \nodata & \phs\nodata          &  \nodata\ph{0}  & \ph{00}4 \\
PG 1411$+$442 & \phn  $<$1.8 &  $<$3.06\ph{00} & \nodata\ph{00}  & \nodata & \phs\nodata          &  \nodata\ph{0}  & \ph{00}4 \\
PG 1415$+$451 & \phs\phn 2.1 &     5.92\ph{00} & \nodata\ph{00}  & \nodata & \phs\nodata          &  \nodata\ph{0}  & \ph{00}5 \\
PG 1426$+$015 & \phs\phn 3.6 &     5.71\ph{00} &    0.46\ph{000} &     357 & \phs    24.13\tnm{a} &    29.84\ph{00} & \ph{00}4, 10 \\
PG 1440$+$356 & \phs\phn 6.6 &     8.36\ph{00} & \nodata\ph{00}  & \nodata & \phs\nodata          &  \nodata\ph{0}  & \ph{00}5 \\
PG 1444$+$407 & \phs\phn 0.7 &    11.39\ph{00} & \nodata\ph{00}  & \nodata & \phs\nodata          &  \nodata\ph{0}  & \ph{00}2 \\
PG 1448$+$273 & \phs \nodata &  \nodata\ph{0}  &    0.35\ph{000} &     580 & \phs    10.19\tnm{a} &  \nodata\ph{0}  & \ph{00}9 \\
PG 1501$+$106 &      $<$27.3 &  $<$7.41\ph{00} & $<$0.19\ph{000} & \nodata & \phn  $<$1.63        &  $<$9.04\ph{00} & \ph{00}1, 9 \\
PG 1545$+$210 & \phn  $<$1.0 & $<$16.15\ph{00} & \nodata\ph{00}  & \nodata & \phs\nodata          &  \nodata\ph{0}  & \ph{00}2 \\
PG 1613$+$658 & \phs\phn 8.0 &    29.05\ph{00} & \nodata\ph{00}  & \nodata & \phs\nodata          &  \nodata\ph{0}  & \ph{00}5 \\
PG 1700$+$518 & \phs\phn 3.9 &    71.08\ph{00} & \nodata\ph{00}  & \nodata & \phs\nodata          &  \nodata\ph{0}  & \ph{00}7 \\
PG 2130$+$099 & \phs\phn 3.9 &     3.08\ph{00} &    0.47\ph{000} &     506 & \phs    11.98\tnm{a} &    15.06\ph{00} & \ph{00}5, 10 \\
PG 2214$+$139 & \phs\phn 1.6 &     1.53\ph{00} &    0.77\ph{000} &     330 & \phs    23.89\tnm{a} &    25.42\ph{00} & \ph{00}4, 9 \\
\enddata
\tablenotetext{a}{The directly measured total gas mass is higher than the
dust-inferred gas mass by $>0.3$ dex.}
\tablecomments{
(1) Object name. 
(2) The CO integrated line flux.
(3) The molecular gas mass converted from the CO line emission, with
$\alpha_\mathrm{CO}=4.3\,M_\odot\,\mathrm{(K\,km\,s^{-1}\,pc^2)^{-1}}$.
(4) The \hi\ integrated line flux.
(5) Width of the line profile measured at 20\% of the peak.
(6) The \hi\ gas mass.
(7) The total gas mass, $M_\mathrm{H2}+M_\mathrm{H\,I}$.  $M_\mathrm{H2}$, 
$M_\mathrm{H\,I}$, and $M_\mathrm{gas}$ include helium and the heavier elements.
(8) References: (1) \cite{Maiolino1997ApJ},  (2) \cite{Casoli2001ASPC}, 
(3) \cite{Evans2001AJ}, (4) \cite{Scoville2003ApJ}, (5) \cite{Evans2006AJ}, 
(6) \cite{Bertram2007AA}, (7) \cite{Evans2009AJ}, (8) \cite{Xia2012ApJ}, 
(9) \cite{Hutchings1987AJ}, (10) \cite{Ho2008ApJS}.\\
Comments on the individual objects with \hi\ observations.
We search the NED and SDSS databases for the information of the nearby galaxies
within (1) 7\farcm5 radius and (2) $\pm 2 \, W_{20}$ in velocity. The host
galaxy morphologies come from the \hst\ images.
\underline{PG 0003$+$199} is isolated; no comparable-size galaxies are found nearby.
The host galaxy is a bulge-dominated disk galaxy \citep{Kim2017ApJS}.
\underline{PG 0007$+$106} resides in a dense region.  There are at least two
galaxies nearby and likely contaminating the \hi\ measurement.  The \hi\ line
profile is very broad, likely due to the confusing objects.
\underline{PG 0050$+$124} is an ongoing merger, although the companion galaxy 
is not as large.
\underline{PG 0844$+$349} is merging with a disk galaxy with comparable 
luminosity.  There are a number of faint galaxies close to the host galaxy, although 
their redshift information is lacking.  PG 0844$+$349 is likely in a dense 
environment.  The \hi\ line profile is asymmetric.
\underline{PG 1119$+$120} resides in a dense region.  There are two clear
companions.  One is inside the host galaxy envelope to the northwest and the
other is farther to the north.  There are at least five other smaller galaxies
located within a projected distance of 25$\arcsec$ of the quasar, as noted by
\cite{Surace2001AJ}.  The \hi\ profile is highly asymmetric and shows a $\sim 300
\,\mathrm{km\,s^{-1}}$ offset from the optical velocity.
\underline{PG 1211$+$143} may have several companion galaxies $>2\arcmin$ 
away but we lack line-width information to be sure.  The host galaxy 
light distribution is smooth and regular, with no sign of merger features \citep{Kim2008ApJ}.
\underline{PG 1229$+$204} is relatively isolated although there is a group of three
galaxies, $\sim 2\farcm6$ to the northwest.  There are also several relatively
small galaxies $>5\arcmin$ away.
\underline{PG 1244$+$026} resides in an isolated environment.  There are no comparable-size galaxies nearby.
\underline{PG 1426$+$015} is an ongoing merger during the final stages of 
coalescence.  There are likely a number of small galaxies nearby, although redshift information is lacking.
\underline{PG 1448$+$273} is an ongoing merger after coalescence.  The \hi\
profile is somewhat asymmetric.
\underline{PG 1501$+$106} is also relatively isolated. There is a galaxy 4\farcm6 
to the northeast with systematic velocity close to that of the quasar.
\underline{PG 2130$+$099} displays a disturbance in the NIR 
\citep{Clements2000MNRAS}.  \cite{Dunlop1993MNRAS} note that there are a 
large number of nearby companions revealed in the $K$-band image.
\underline{PG 2214$+$139} displays a shell structure in deep optical images
\citep{Hong2015ApJ}, indicating that the host galaxy is a recent merger remnant.
There is also a galaxy of similar size 4\farcm9 to the southwest.
\cite{Hutchings1987AJ} note that the \hi\ line profile is somewhat asymmetric
and broad.
}
\end{deluxetable}
\end{longrotatetable}